\documentclass[nofootinbib,twocolumn,preprintnumbers,superscriptaddress,aps]{revtex4-1}
\usepackage{amsmath}
\usepackage{amssymb}
\usepackage{graphicx}
\usepackage{booktabs}
\usepackage{color}
\usepackage{hyperref}
\usepackage{xspace}
\makeatletter
\g@addto@macro\bfseries{\boldmath}
\makeatother

\newcommand{\as}{\alpha_s}
\newcommand{\abar}{{\bar \alpha}}
\newcommand{\kT}{k_t}

\newcommand{\ptilde}{{\widetilde p}}
\newcommand{\itilde}{{\tilde \imath}}
\newcommand{\jtilde}{{\tilde \jmath}}
\newcommand{\state}{\mathcal{S}}
\newcommand{\cI}{\mathcal{I}}

\newcommand{\order}[1]{{\cal O}\!\left(#1\right)}
\newcommand{\shrink}{\hspace{-0.1em}}

\newcommand{\boldp}{\boldsymbol{p}}

\definecolor{darkgreen}{rgb}{0,0.4,0}
\definecolor{grey}{rgb}{0.5,0.5,0.5}
\definecolor{orange}{rgb}{0.9,0.5,0.0}
\newcommand{\OXaff}{Rudolf Peierls Centre for Theoretical Physics, Parks Road, Oxford OX1 3PU, UK}
\newcommand{\CERNaff}{CERN, Theoretical Physics Department, CH-1211 Geneva 23, Switzerland}
\newcommand{\UCLaff}{Department of Physics and Astronomy, University College London, London, WC1E 6BT, UK}
\newcommand{\MANaff}{Consortium for Fundamental Physics, School of
  Physics and Astronomy, University of Manchester, Manchester M13 9PL,
  United Kingdom}
\newcommand{\IPHTaff}{Institut de Physique Th\'eorique, Universit\'e
  Paris-Saclay, CNRS, CEA, F-91191, Gif-sur-Yvette, France}
\newcommand{\CNRSaff}{CNRS, UMR 7589, LPTHE, F-75005, Paris, France}

\begin{document}

\title{Parton showers beyond leading logarithmic accuracy}

\preprint{CERN-TH-2020-026}

\author{Mrinal Dasgupta}           \affiliation{\MANaff}
\author{Fr\'ed\'eric A. Dreyer}    \affiliation{\OXaff}
\author{Keith Hamilton}            \affiliation{\UCLaff}
\author{Pier Francesco Monni}      \affiliation{\CERNaff}
\author{Gavin P. Salam}            
\altaffiliation{On leave from \CNRSaff\ and \CERNaff}
\affiliation{\OXaff}
\author{Gr\'egory Soyez}           \affiliation{\IPHTaff}

\begin{abstract}
Parton showers are among the most widely used tools in collider
physics.
Despite their key importance, none so far has been able to demonstrate
accuracy beyond a basic level known as leading logarithmic (LL)
order, with ensuing limitations across a broad spectrum of physics
applications.
In this letter, we propose criteria for showers to be considered
next-to-leading logarithmic (NLL) accurate.
We then introduce new classes of shower, for final-state radiation,
that satisfy the main elements of these criteria in the widely used
large-$N_C$ limit.
As a proof of concept, we demonstrate these showers'
agreement with all-order analytical NLL calculations for a
range of observables, something never so far achieved for any parton
shower.
\end{abstract}

\maketitle

High-energy particle collisions produce complex hadronic final
states.
Understanding these final states is of crucial importance in order to
extract maximal information about the underlying
energetic scattering processes and the fundamental Lagrangian of particle physics.
To do so,
% make sense of collider data, and to plan future experiments,
there
is ubiquitous reliance on general purpose Monte Carlo (GPMC) event
generators~\cite{Buckley:2011ms}, which provide realistic simulations
of full events.
A core component of GPMCs is the parton shower, a subject of much
recent research~\cite{
Sjostrand:2004ef,
Giele:2007di,
Nagy:2007ty,
Schumann:2007mg,
Platzer:2009jq,
Jadach:2010aa,
Nagy:2012bt,
Nagy:2014mqa,
Nagy:2015hwa,
Hoche:2015sya,
Li:2016yez,
Jadach:2016zgk,
Fischer:2016vfv,
Nagy:2016pwq,
Fischer:2017htu,
Hoche:2017hno,
Hoche:2017iem,
Nagy:2017ggp,
Cabouat:2017rzi,
Dulat:2018vuy,
Platzer:2018pmd,
Martinez:2018ffw,
Isaacson:2018zdi,
Brooks:2019xso,
Forshaw:2019ver,
Nagy:2019rwb,
Hoeche:2020nsx}.
Partons refer to quarks and gluons, and a shower aims to encode the
dynamics of parton production between the high-energy scattering (e.g.\
production of electroweak or new-physics states) and the low
scale of hadronic Quantum Chromodynamics (QCD), at which experimental
observations are made.

Typically parton showers are built using a simple Markovian algorithm
that takes an $n$-parton state and stochastically maps it to an
$n+1$-parton state.
The iteration of this procedure, e.g.\ starting from a 2-parton
state, builds up events with numerous partons.
A further step, hadronisation, then maps the partons onto a
set of hadrons.
Even though this last step involves
modelling~\cite{Buckley:2009bj,Skands:2010ak}, many of the features of
the resulting events are driven by the parton shower component which
is, in principle, within the realm of calculations in perturbative
QCD.
This is because the showering occurs at momentum scales where the
strong coupling, $\as$ is small.

Much of collider physics, experimental and
theoretical~\cite{Ellis:2019qre,Azzi:2019yne,Cepeda:2019klc,CidVidal:2018eel}, is
moving towards high precision,
especially in view of large volumes of data collected so far at CERN's
Large Hadron Collider (LHC).
On the theoretical front many of the advances either involve
approximations with a small number of partons, or else are specific to
individual observables.
Parton showers, in contrast,  use a single algorithm to describe arbitrary
observables of any complexity.
This versatility comes at a cost: lesser accuracy for any specific
observable and, quite generally, at best only limited
knowledge~\cite{Marchesini:1983bm,Banfi:2006gy,Dasgupta:2018nvj,Bewick:2019rbu}
of what the accuracy even is for a given observable.
In fact there is currently no readily accepted criterion for
categorising the accuracy of parton showers.
One novel element that we introduce in this paper is therefore a
set of criteria for doing so.

The role of showers is to reproduce emissions across disparate scales.
%
% The role of showers is to reproduce the pattern of arbitrarily many
% emissions across disparate scales (distinct from
% matching/merging~\cite{Alwall:2007fs,Nason:2012pr}, which reproduce
% a small number of emissions at the hard scale).
%
Our first criterion for accuracy starts by structuring this phase space:
there are three phase space variables per emission, and two of them
(e.g.\ energy and angle) are associated with logarithmic divergences
in the product of squared matrix element and phase space. 
We define LL accuracy to include a condition
that the shower should 
generate the correct effective squared tree-level matrix element in a
limit where every pair of emissions has distinctly different values
for both logarithmic variables.
At NLL accuracy, we further require that the shower generate the correct
squared tree-level matrix element in a limit where every pair of
emissions has distinctly different values for at least one of the
logarithmic variables (or some linear combination of their
logarithms).  
Beyond NLL accuracy we would consider configurations with a pair of
emissions (or multiple pairs) both of whose logarithmic variables are
similar.

To help make this discussion concrete, let us consider showers that are
not NLL accurate according to this criterion: angular ordered
showers~\cite{Marchesini:1987cf,Corcella:2000bw,Bahr:2008pv} do not reproduce the matrix
element for configurations ordered in energy, but with commensurate angles,
and this is associated with their inability to correctly predict
$\as^n L^n$ (NLL) effects for non-global
observables~\cite{Banfi:2006gy}.
Transverse-momentum ($k_t$) ordered showers with dipole-local
recoil~\cite{Gustafson:1987rq,Lonnblad:1992tz,Sjostrand:2004ef,Giele:2007di,Schumann:2007mg,Hoche:2015sya}
do not reproduce matrix elements for configurations 
ordered in angle but with commensurate transverse momenta, because of
the way they assign transverse recoil~\cite{Dasgupta:2018nvj}.
As a result they fail to reproduce NLL effects in global observables
such as jet broadenings.
Showers that omit spin correlations fail to reproduce (the azimuthal
structure of) matrix elements for configurations ordered in angle but
with commensurate
energies~\cite{Webber:1986mc,Collins:1987cp,Catani:2010pd}, and
associated NLL terms.

Our second criterion for logarithmic accuracy tests, among other
things, the overall correctness of virtual corrections.
For showers that intertwine real and virtual corrections directly
through unitarity, once the generation of tree-level matrix elements
is set, there is only one (single-emission) degree of freedom that
remains, namely the choice of scale and scheme for the strong coupling
for each emission, as a function of its kinematics.
To claim NLL accuracy, we will require the resulting shower to
reproduce known analytical NLL resummations across
recursively infrared and collinear safe (rIRC)~\cite{Banfi:2004yd} global and non-global two-scale
observables as well as (subjet) multiplicities.

The challenge that we concentrate on here is to formulate showers that can
handle each of two regions correctly: the energy-ordered,
commensurate-angle region; and the angular-ordered, commensurate $k_t$
region.
Recall that existing $k_t$ and angular-ordered showers can each handle
one of these limits, but not both.
Strictly, full NLL accuracy also requires attention to the
angular-ordered, commensurate energy region.
However, given that general solutions for the required spin
correlations are known to
exist~\cite{Collins:1987cp,Knowles:1988hu,Nagy:2008eq,Richardson:2018pvo}, and
that they affect only a small subset of observables, we postpone their
study to future work.
For now, we also restrict our attention to final-state showers (i.e.\
lepton-lepton collisions), massless quarks and the large-$N_C$ limit.
%,
Our guiding principle will be that soft emissions should not affect,
or be affected by, subsequent emissions at disparate
rapidities.

The two classes of shower that we develop both consider emissions from
colour dipoles. 
We consider a continuous family of shower evolution variables $v$,
parameterised by a quantity $\beta$ in the range $0\le \beta < 1$,
where $\beta=0$ corresponds to transverse-momentum ordering.
%,
The phase space involves two further variables besides $v$: a
pseudorapidity-like variable within the dipole, $\bar \eta$, and an
azimuthal angle $\phi$.

We start with a shower with \emph{dipole-local recoil} (the \emph{PanLocal} shower).
Its mapping for emission of momentum $p_k$ from a dipole
$\{\ptilde_i,\ptilde_j\}$ is
\begin{subequations} \label{eq:IIc-split-map} 
  \begin{align}
    p_{k} & =a_{k}\tilde{p}_{i}+b_{k}\tilde{p}_{j}+k_{\perp}\,,\label{eq:IIc-split-map-pk}\\
    p_{i} & =a_{i}\tilde{p}_{i}+b_{i}\tilde{p}_{j}-fk_{\perp}\,,\label{eq:IIc-split-map-pi}\\
    p_{j} & =a_{j}\tilde{p}_{i}+b_{j}\tilde{p}_{j}-(1-f)k_{\perp}\,,\label{eq:IIc-split-map-pj}
  \end{align}
\end{subequations}
where $k_{\perp}=\kT\left[{n}_{\perp,1}\cos\phi+{n}_{\perp,2}\sin\phi\right]$,
with ${n}_{\perp,m}^2 = -1$,
${n}_{\perp,m}\cdot \ptilde_{i/j}
%{n}_{\perp,m}\cdot \ptilde_j
= 0$ ($m=1,2$),
${n}_{\perp,1}\cdot{n}_{\perp,2} = 0$
and
\begin{equation}
  \label{eq:evolution-variable}
  \kT = \rho v e^{\beta|\bar \eta|}\,,
  \qquad
  \rho = \left(\frac{s_{\itilde}
      s_{\jtilde}}{Q^2 s_{\itilde\jtilde}}\right)^{\frac{\beta}{2}}\,.
\end{equation}
Here $s_{\itilde \jtilde} =2\ptilde_{i}\cdot \ptilde_{j}$,
$s_{\itilde}=2\ptilde_{i} \cdot Q$, and $Q$ is the total event
momentum.
The light-cone components of $p_{k}$ are given by
\begin{equation}
  a_{k}\equiv\sqrt{\frac{s_{\jtilde}}{s_{\itilde\jtilde}s_{\itilde}}}\,
  \kT{e}^{+\bar \eta}\,,
  \qquad
  b_{k}\equiv\sqrt{\frac{s_{\itilde}}{s_{\itilde\jtilde}s_{\jtilde}}}\,
  \kT{e}^{-\bar \eta}\,,
  \label{eq:IIc-ak-bk-Sudakov-coeffs}
\end{equation}
The quantity $f$ in Eq.~(\ref{eq:IIc-split-map}) determines how
transverse recoil is shared between $p_i$ and $p_j$, cf. below.
The $a_i,b_i,a_j,b_j$ are fully specified by the requirements
$p^2_{i/j}= 0$, $(p_i+p_j+p_k)=(\ptilde_i+\ptilde_j)$ and
$p_i = \tilde p_i$ for $\kT \to 0$ and are given explicitly in
Ref.~\cite{supplement}\S\ref{sec:expl-moment-maps}.

In the event centre-of-mass frame,
$\bar \eta=0$ corresponds to a direction equidistant in angle from
$\ptilde_i$ and $\ptilde_j$.
For soft-collinear emissions, the physical pseudorapidity, $\eta =
-\ln \tan\frac{\theta}2$,
with respect to the emitter is
$\eta = |\bar \eta| + \frac{1}{\beta}\ln\rho$.
Soft-collinear emissions from distinct dipoles but with the same
$\ln v$ fall onto common contours in the Lund
plane~\cite{Andersson:1988gp}, $\kT = v e^{\beta|\eta|}$.

\begin{figure*}
  \includegraphics[width=0.32\textwidth,page=1]{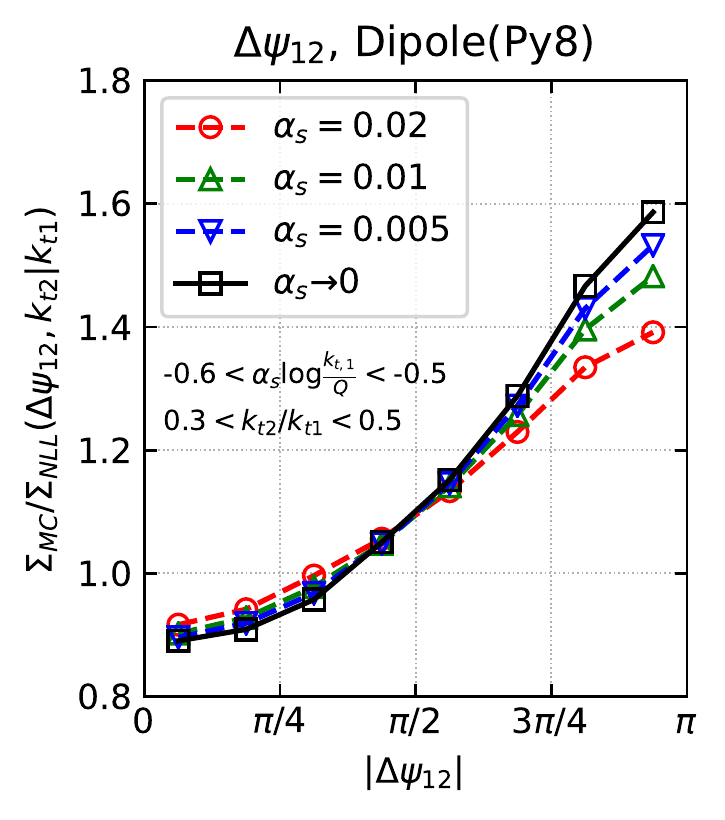}%
  \includegraphics[width=0.32\textwidth,page=2]{plots/delta-psi.pdf}%
  \includegraphics[width=0.32\textwidth,page=3]{plots/delta-psi.pdf}%
  \caption{Left: distribution for the difference in azimuthal
    angle between the two highest-$k_t$ primary Lund declusterings in
    the Pythia8 dipole shower algorithm, normalised to
    the NLL result~\cite{Banfi:2001bz}, \cite{supplement}\S\ref{sec:ee-lund};
    successively smaller $\as$ values keep fixed $\as \ln k_{t1}$.
    Middle: the same for the PanGlobal($\beta=0$) shower.
    Right: the $\as \to 0$ limit of the ratio for multiple showers.
    This observable directly tests part of our NLL (squared) matrix-element
    correctness condition.
    A unit value for the ratio signals success.
  }
  \label{fig:limits-dpsi}
\end{figure*}

For $e^+e^- \to \text{hadrons}$, the shower starts from a 2-parton
$q\bar q$ state, $\state_2$.
The probability of evolving from $\state_n \to \state_{n+1}$ in a
given slice $d\ln v$ of evolution variable is
\begin{multline}
  \label{eq:2}
  \frac{d\mathcal{P}_{n\to n+1}}{d\ln v } =
  \sum_{\text{dipoles } \{\itilde,\jtilde\}}
  \int d\bar\eta \frac{d\phi}{2\pi}
  \frac{\as(\kT) + K\as^2(\kT)}{\pi}
  \\ \times
  \left[g(\bar\eta) a_k P_{\itilde \to ik}(a_k) +
    g(-\bar\eta) b_k P_{\jtilde \to jk}(b_k)
    \right],
\end{multline}
with a function $g(\bar\eta)$ that satisfies $g(\bar\eta)+g(-\bar\eta)=1$, has
$g(\bar\eta)=0$ ($1$) for sufficiently negative (positive) $\bar\eta$,
and smoothly transitions around $\bar\eta=0$.
The $P_{\itilde \to ik}(z)$ are first-order splitting
functions~\cite{Gribov:1972ri,Altarelli:1977zs,Dokshitzer:1977sg},
normalised so 
that $\lim_{z\to0} z P_{\itilde \to ik}(z) = 2\,C$ with $C = C_F =
C_A/2=\frac43$ (our large-$N_C$
approximation, augmented~\cite{Behring:2019qufw} with $n_f=5$).
The specific choice of $g(\bar\eta)$ is not critical here, while the
splitting functions are standard.
Both are detailed in Ref.~\cite{supplement}\S\ref{sec:expl-moment-maps}.
The $\overline{\text{MS}}$ coupling, $\as(\kT)$, needs at least 2-loop
running, and $K
=\frac1{2\pi}\left[\left(\frac{67}{18}-\frac{\pi^2}{6}\right)C_A
  -\frac59 n_f\right]$~\cite{Catani:1990rr}.

The PanLocal shower comes in two variants.
In a \emph{dipole} variant, inspired by many earlier dipole
showers~\cite{Schumann:2007mg,Sjostrand:2004ef,Hoche:2015sya},
the
$P_{\itilde \to ik}(a_k)$ ($P_{\jtilde \to jk}(b_k)$) term of
Eq.~(\ref{eq:2}) is associated with the choice $f=1$ ($f=0$) in
Eq.~(\ref{eq:IIc-split-map}).
In an \emph{antenna} variant, inspired by 
Refs.~\cite{Lonnblad:1992tz,Giele:2007di}, we take a common
$f(\bar\eta)$ for both terms and set $f(\bar\eta)= g(\bar\eta)$.

A key difference relative to earlier showers is that our transition in
transverse recoil assignment between $i$ and $j$ takes place at
$\bar\eta \simeq0$, i.e.\ equal angles between the $\ptilde_i$ and
$\ptilde_j$ directions in the event centre-of-mass frame (note
similarities with Deductor~\cite{Nagy:2014mqa}).
This differs from the common choice of a transition in the middle of the
dipole centre-of-mass frame.
Our choice ensures that a given emission will not induce transverse recoil
in earlier, lower-rapidity emissions.
Additionally, we require $\beta > 0$ in the definition of the ordering
variable, Eq.~(\ref{eq:evolution-variable}).
This causes emissions at commensurate $k_t$ and widely separated in
$|\eta|$ to be effectively produced in order of increasing $|\eta|$,
so that any significant $k_t$  recoil is always taken
from the extremities of a (hard) $qg \{\ldots\}g\bar q$ dipole chain.
Together, these two elements provide a solution to the problem
observed in Ref.~\cite{Dasgupta:2018nvj}, i.e.\ that recoil assignment
in common dipole showers causes multi-gluon emission matrix elements
to be incorrect in the limit of similar $k_t$'s and disparate angles,
starting from $\as^2$, leading to incorrect NLL terms.

Note that with dipole-local recoil, NLL correctness  also requires
$\beta < 1$, because with $\beta \ge 1$ the kinematic constraint
associated with fixed dipole mass means that a first emission cuts out
regions of phase space for a second emission at similar $\ln v$.

\begin{figure*}
  \includegraphics[width=0.32\textwidth, page=1]{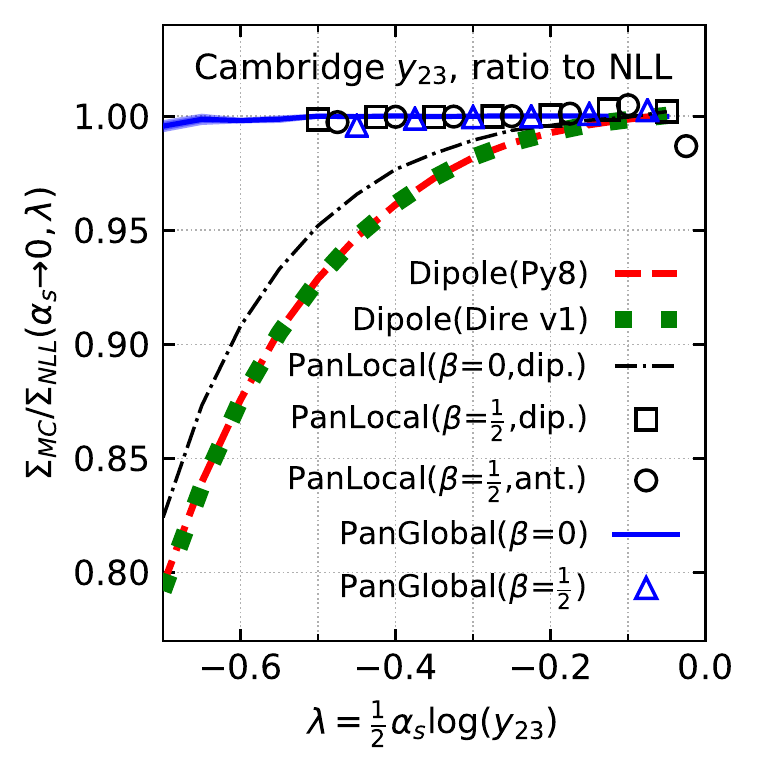}%
  \includegraphics[width=0.68\textwidth, page=1]{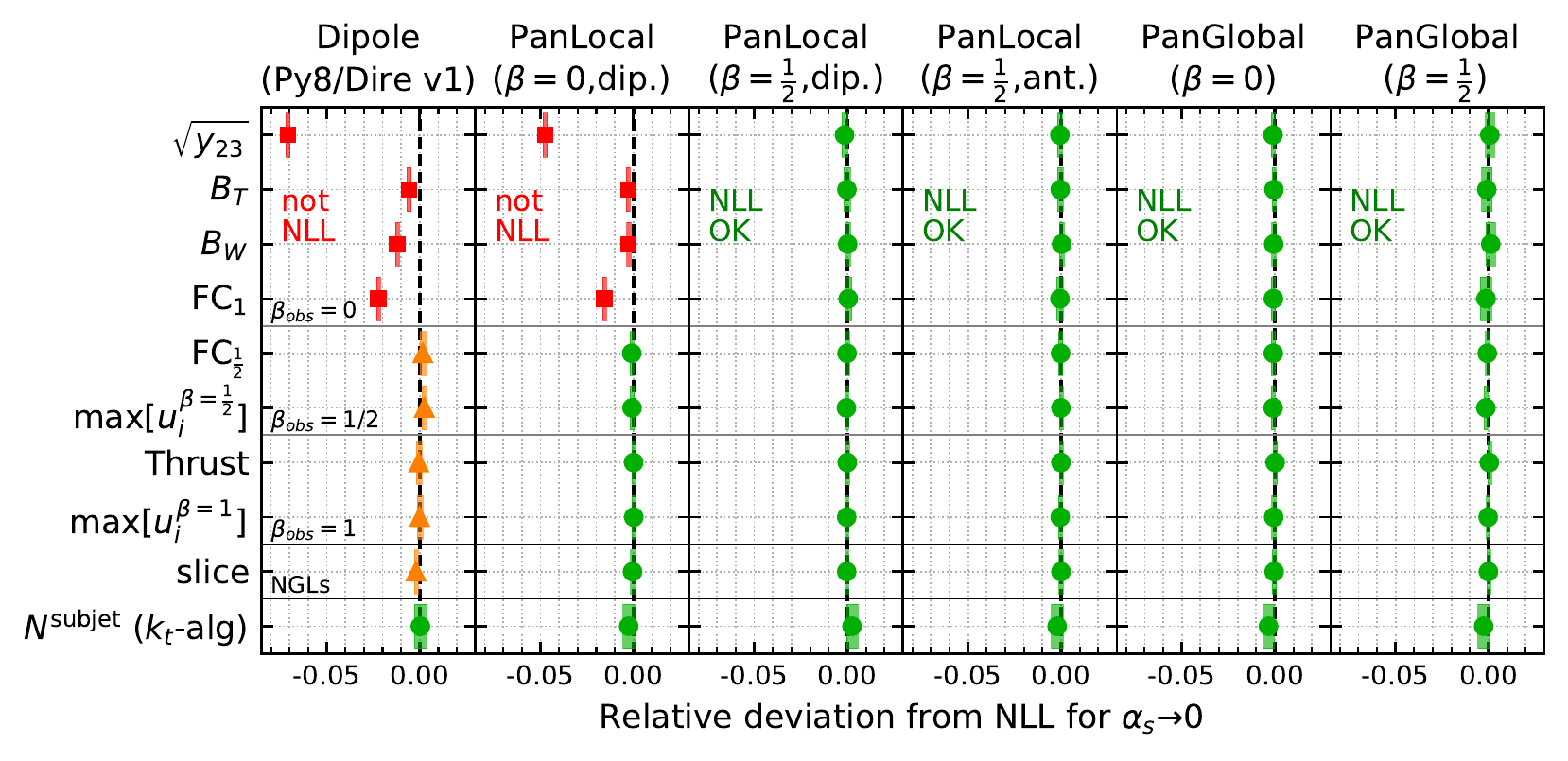}%
  \caption{Left: ratio of the cumulative $y_{23}$ distribution from several showers
      divided by the NLL answer, as a function of $\as \ln y_{23}/2$, for
      $\as \to 0$. 
      Right: summary of deviations from NLL for many shower/observable
      combinations (either
      $\Sigma_\text{shower}(\as\to 0,\as L=-0.5)/\Sigma_\text{NLL}-1$ or
      $(N^\text{subjet}_\text{shower}(\as \to 0, \as L^2 = 5 )/
      N^\text{subjet}_\text{NLL} - 1)/\sqrt{\as}$). 
      Red squares indicate clear NLL failure;
      amber triangles indicate NLL fixed-order failure that is
      masked at all orders;
      green circles indicate that all NLL tests passed.  }
  \label{fig:limits-y3}
\end{figure*}

A second class of shower can be constructed with \emph{global, i.e.\
  event-wide recoil} (the \emph{PanGlobal} shower).
It can be formulated in largely the same terms as the dipole-local
recoil shower, but with a two-step recoil procedure.
In the first step one sets
\begin{subequations} \label{eq:global-map-step-1} 
  \begin{align}
    \bar p_{k} & =a_{k}\tilde{p}_{i}+b_{k}\tilde{p}_{j}+k_{\perp}\,,\\
    \bar p_{i} & =(1-a_{k})\tilde{p}_{i}\,,\\
    \bar p_{j} & =(1-b_{k})\tilde{p}_{j}\,.
  \end{align}
\end{subequations}
The second step is to apply a boost and rescaling to the full event
(including the $\bar p_{i,j,k}$ momenta) so as to obtain final momenta
$\{p\}$ whose sum gives $Q$.
This approach assigns transverse recoil dominantly to the most energetic
particles in the event.
Thus emission from a hard $qg\{\ldots\}g \bar q$ dipole string transfers its
recoil mostly to the hard $q$ and $\bar q$ ends.
This ensures that one reproduces a pattern of independent emission for
commensurate-$k_t$ and angular-ordered gluons, while also retaining
the correct (dipole) pattern for energy-ordered, commensurate angles.
This holds even for $\beta=0$, i.e.\ with $k_t$ ordering.
Values of $\beta \ge 1$ remain problematic, however.
Note that the PanGlobal shower has power-suppressed routes to highly
collimated events.
These compete with normal Sudakov suppression, as observed also for
Pythia8~\cite{Dasgupta:2018nvj}.
We have verified that such effects are small even at the very edges of
future (FCC-hh~\cite{Benedikt:2018csr}) phenomenologically accessible
regions.
Nevertheless, ultimately one may wish to explore alternative global
recoil schemes.

The next step is to compare our showers to NLL observables.
Relative to earlier attempts at such
comparisons~\cite{Hoeche:2017jsi}, a critical novel aspect is how
we isolate the structure of NLL terms $\as^n L^n$.
For each given observable $v$, with $L = \ln v$, we consider the ratio
to the true NLL result in the limit
$\as\equiv\as(Q) \to 0$ with fixed $\as L$.
This helps us isolate the NLL terms from yet higher-order
contributions, which vanish in that limit.
Numerically, a parton shower cannot be run in the $\as \to 0$ limit
for fixed $\as L$.
However, with suitable
techniques~\cite{supplement}\S\ref{sec:taking-limits}, \cite{hida2000quad,DBLP:journals/corr/abs-1109-3627,Kleiss:2016esx},
one can run multiple small values of
$\as$ and extrapolate to
$\as = 0$.
We examine not just our showers, but also our implementations of two typical
$k_t$-ordered shower algorithms with dipole-local recoil, those of
Pythia8~\cite{Sjostrand:2004ef} and Dire v1~\cite{Hoche:2015sya}
(with the $\as+K\as^2$ choice as in Eq.~(\ref{eq:2})).

A first test concerns the multiple-emission matrix element.
We have constructed our showers specifically so that they reproduce
the squared matrix elements in the limits discussed above that are
relevant for NLL accuracy.
A simple observable for testing this is to consider the two
highest-$k_t$ Lund-plane primary
declusterings~\cite{Dreyer:2018nbf,Andrews:2018jcm} with transverse
momenta $k_{t1}$ and $k_{t2}$ (originally defined for hadronic
collisions, the $e^+e^-$ analogue is given in Ref.~\cite{supplement}\S\ref{sec:ee-lund}
and implemented with FastJet~\cite{Cacciari:2011ma}).
The $\as \to 0$ limit for fixed $\as L$ ($L = \ln k_{t1}/Q$), ensures
that the two declusterings are soft and widely separated in Lund-plane
pseudorapidity $\eta$ (which spans $|\eta| \lesssim |L| \sim
1/\as$).
In this limit the full matrix element reduces to independent
emission and so the difference of azimuthal angles between the two
emissions, $\Delta \psi_{12}$, should be uniformly distributed, for
any ratio $k_{t2}/k_{t1}$ (recall that strongly angular-ordered soft
emission is \emph{not} affected by spin correlations).
We consider the $\Delta \psi_{12}$ distribution in
Fig.~\ref{fig:limits-dpsi}.
%, normalised to the
%NLL analytic expectation.
%

The left-hand plot of Fig.~\ref{fig:limits-dpsi} shows the Pythia8
dipole algorithm (not designed as NLL accurate), while the middle plot shows our PanGlobal shower
with $\beta=0$.
The dipole result is clearly not independent of $\Delta \psi_{12}$
for $\as \to 0$, with over $60\%$ discrepancies, extending the
fixed-order conclusions of Ref.~\cite{Dasgupta:2018nvj}.
The discrepancy is only $\simeq 30\%$ for $gg$ events (not shown in
Fig.~\ref{fig:limits-dpsi}), and the difference would, e.g., skew
machine learning~\cite{Larkoski:2017jix} for quark/gluon
discrimination. 
PanGlobal is independent of $\Delta \psi_{12}$.
The right-hand plot shows the $\as \to 0$ limit for
multiple showers.
The overall pattern is as expected:
PanLocal works for $\beta=0.5$,
but not $\beta=0$,
demonstrating that with $\kT$ ordering it is not sufficient just to
change the dipole partition to get NLL accuracy.
PanGlobal works for $\beta=0$ and $\beta=0.5$.
(Showers that coincide for $\as \to 0$, e.g.\ Dire v1 and Pythia8,
typically differ at finite $\as$, reflecting NNLL differences.)

Next, we consider a range of more standard observables at NLL
accuracy.
They include the Cambridge $\sqrt{y_{23}}$ resolution
scale~\cite{Dokshitzer:1997in};
two jet
broadenings, $B_T$ and $B_W$~\cite{Catani:1992jc};
fractional moments, $\text{FC}_{1-\beta_\text{obs}}$, of the
energy-energy correlations~\cite{Banfi:2004yd}; the
thrust~\cite{Brandt:1964sa,Farhi:1977sg}, and the maximum
$u_i=k_{ti}/Q e^{-\beta_\text{obs} |\eta_i|}$ among primary Lund
declusterings $i$.
Each of these is sensitive to soft-collinear radiation as $k_t/Q
e^{-\beta_\text{obs}|\eta|}$, with the $\beta_\text{obs}$ values shown
in Fig.~\ref{fig:limits-y3} (right).
Additionally, the scalar sum of the transverse momenta in a
rapidity slice~\cite{Dasgupta:2002bw}, of full-width $2$, is useful to test
non-global logarithms (NGLs).
These observables all have the property that their
distribution at NLL can be written
as~\cite{Banfi:2001bz,Dokshitzer:1998kz,Banfi:2004yd,Catani:1992ua,Dasgupta:2002bw}
\begin{equation}
  \label{eq:NLL-master}
  \Sigma(\as\shrink, \as L) =
  \exp\! \left [
  %e^{
    \as^{\!-1} g_1(\as L) + g_2(\as L) + \order{\as^n L^{n-1}}
  %  },
  \right]\!,
\end{equation}
where $\Sigma$ is the fraction of events where the observable is
smaller than $e^{L}$ ($g_1=0$ for the rapidity slice $k_t$).
We also consider the $k_t$-algorithm~\cite{Catani:1991hj} subjet
multiplicity~\cite{Catani:1991pm}, \cite{supplement}\S\ref{sec:multiplicities}.

Fig.~\ref{fig:limits-y3} (left) illustrates our all-order tests of the
shower for one observable, $\sqrt{y_{23}}$.
It shows the ratio of the $\Sigma$ as calculated with the shower to
the NLL result, as a function of $\as \ln \sqrt{y_{23}}$ in the limit of
$\as \to 0$.
The standard dipole algorithms disagree with the NLL
result, by up to $20\%$.
This is non-negligible, though smaller than the disagreement in
Fig.~\ref{fig:limits-dpsi}, because of the azimuthally averaged nature
of the $\sqrt{y_{23}}$ observable. 
In contrast the PanGlobal and PanLocal($\beta=0.5$) showers agree with the
NLL result to within statistical uncertainties.

Fig.~\ref{fig:limits-y3} (right) shows an overall summary of our
tests.
The position of each point shows the result of
$\Sigma_\text{shower}(\as\to 0, \as L=-0.5)/\Sigma_\text{NLL}-1$
or $(N^\text{subjet}_\text{shower}(\as \to 0, \as L^2 = 5)/N^\text{subjet}_{\text{NLL}}-1)/\sqrt{\as}$.
If it differs from $0$, the point is shown as a red square.
In some cases (amber triangles) it agrees with $0$, though an
additional fixed-order analysis in a fixed-coupling toy
shower~\cite{Dasgupta:2018nvj} \cite{supplement}\S\ref{sec:toy-shower} reveals issues 
affecting NLL accuracy,
all involving hitherto undiscovered spurious super-leading
logarithmic terms.%
\footnote{%
  Such terms, $(\as L)^n (\as L^2)^p$ in $\ln \Sigma$, starting
  typically for $n=3$ (sometimes $2$), $p\ge 1$, appear for
  traditional $k_t$ ordered dipole showers for global
  ($\beta_\text{obs} > 0$) and non-global
  observables~\cite{supplement}\S\ref{sec:super-leading-logs}.
  Terms of this kind can generically exist
  \cite{Brown:1990nm,Forshaw:2006fk,Catani:2011st}, but not at leading-colour or
  for pure final-state processes with rIRC~\cite{Banfi:2004yd} safe
  observables.
  In many cases, the spurious super-leading logarithms appear to resum
  to mask any disagreement with NLL.  }

Green circles in Fig.~\ref{fig:limits-y3} (right) indicate that the
shower/observable combination passes all of our NLL tests, both at all
orders and in fixed-order expansions.
The four shower algorithms designed to be NLL accurate pass all the
tests.
These are the PanLocal shower (dipole and antenna variants) with
$\beta=\frac12$ and the PanGlobal shower with $\beta=0$ and
$\beta=\frac12$.

To conclude, we have identified two routes towards NLL parton shower
accuracy.
One involves a modification of the evolution variable and dipole
partition, while maintaining dipole-local recoil; the other replaces
dipole-local recoil with event-wide recoil.
While further work is needed towards phenomenology, the
results shown here represent the first time that individual parton
showers are demonstrated to be able to reproduce NLL accuracy
\emph{simultaneously} for both non-global and a wide set of global
observables.
It is our hope that these results, together with our NLL criteria and
validation framework, can provide the solid foundations needed for
future development of logarithmically accurate showers.

\textbf{Acknowledgements.}
We are grateful to Fabrizio Caola, Silvia Ferrario Ravasio, Basem
El-Menoufi, Alexander Karlberg, Paolo Nason, Ludovic Scyboz, Rob Verheyen, Bryan
Webber and Giulia Zanderighi for helpful discussions and comments on
the manuscript.
We thank each other's institutions for hospitality while this work was
being carried out.
This work has been funded by a Marie Sk\l{}odowska Curie Individual
Fellowship contract number 702610 Resummation4PS (PM),
by a Royal Society Research Professorship
(RP$\backslash$R1$\backslash$180112) (GPS),
by the European Research Council (ERC) under the European Union’s
Horizon 2020 research and innovation programme (grant agreement No.\
788223, PanScales) (MD, KH, GPS, GS) and
by the Science and Technology Facilities Council (STFC) under
grants ST/P000770/1 (FD), ST/P000274/1 (KH) and ST/P000800/1 (MD).

\bibliography{MC}

%merlin.mbs apsrev4-1.bst 2010-07-25 4.21a (PWD, AO, DPC) hacked
%Control: key (0)
%Control: author (72) initials jnrlst
%Control: editor formatted (1) identically to author
%Control: production of article title (-1) disabled
%Control: page (0) single
%Control: year (1) truncated
%Control: production of eprint (0) enabled
\begin{thebibliography}{79}%
\makeatletter
\providecommand \@ifxundefined [1]{%
 \@ifx{#1\undefined}
}%
\providecommand \@ifnum [1]{%
 \ifnum #1\expandafter \@firstoftwo
 \else \expandafter \@secondoftwo
 \fi
}%
\providecommand \@ifx [1]{%
 \ifx #1\expandafter \@firstoftwo
 \else \expandafter \@secondoftwo
 \fi
}%
\providecommand \natexlab [1]{#1}%
\providecommand \enquote  [1]{``#1''}%
\providecommand \bibnamefont  [1]{#1}%
\providecommand \bibfnamefont [1]{#1}%
\providecommand \citenamefont [1]{#1}%
\providecommand \href@noop [0]{\@secondoftwo}%
\providecommand \href [0]{\begingroup \@sanitize@url \@href}%
\providecommand \@href[1]{\@@startlink{#1}\@@href}%
\providecommand \@@href[1]{\endgroup#1\@@endlink}%
\providecommand \@sanitize@url [0]{\catcode `\\12\catcode `\$12\catcode
  `\&12\catcode `\#12\catcode `\^12\catcode `\_12\catcode `\%12\relax}%
\providecommand \@@startlink[1]{}%
\providecommand \@@endlink[0]{}%
\providecommand \url  [0]{\begingroup\@sanitize@url \@url }%
\providecommand \@url [1]{\endgroup\@href {#1}{\urlprefix }}%
\providecommand \urlprefix  [0]{URL }%
\providecommand \Eprint [0]{\href }%
\providecommand \doibase [0]{http://dx.doi.org/}%
\providecommand \selectlanguage [0]{\@gobble}%
\providecommand \bibinfo  [0]{\@secondoftwo}%
\providecommand \bibfield  [0]{\@secondoftwo}%
\providecommand \translation [1]{[#1]}%
\providecommand \BibitemOpen [0]{}%
\providecommand \bibitemStop [0]{}%
\providecommand \bibitemNoStop [0]{.\EOS\space}%
\providecommand \EOS [0]{\spacefactor3000\relax}%
\providecommand \BibitemShut  [1]{\csname bibitem#1\endcsname}%
\let\auto@bib@innerbib\@empty
%</preamble>
\bibitem [{\citenamefont {Buckley}\ \emph {et~al.}(2011)\citenamefont {Buckley}
  \emph {et~al.}}]{Buckley:2011ms}%
  \BibitemOpen
  \bibfield  {author} {\bibinfo {author} {\bibfnamefont {A.}~\bibnamefont
  {Buckley}} \emph {et~al.},\ }\href {\doibase 10.1016/j.physrep.2011.03.005}
  {\bibfield  {journal} {\bibinfo  {journal} {Phys. Rept.}\ }\textbf {\bibinfo
  {volume} {504}},\ \bibinfo {pages} {145} (\bibinfo {year} {2011})},\ \Eprint
  {http://arxiv.org/abs/1101.2599} {arXiv:1101.2599 [hep-ph]} \BibitemShut
  {NoStop}%
%%CITATION = ARXIV:1101.2599;%%
\bibitem [{\citenamefont {Sjostrand}\ and\ \citenamefont
  {Skands}(2005)}]{Sjostrand:2004ef}%
  \BibitemOpen
  \bibfield  {author} {\bibinfo {author} {\bibfnamefont {T.}~\bibnamefont
  {Sjostrand}}\ and\ \bibinfo {author} {\bibfnamefont {P.~Z.}\ \bibnamefont
  {Skands}},\ }\href {\doibase 10.1140/epjc/s2004-02084-y} {\bibfield
  {journal} {\bibinfo  {journal} {Eur. Phys. J.}\ }\textbf {\bibinfo {volume}
  {C39}},\ \bibinfo {pages} {129} (\bibinfo {year} {2005})},\ \Eprint
  {http://arxiv.org/abs/hep-ph/0408302} {arXiv:hep-ph/0408302 [hep-ph]}
  \BibitemShut {NoStop}%
%%CITATION = HEP-PH/0408302;%%
\bibitem [{\citenamefont {Giele}\ \emph {et~al.}(2008)\citenamefont {Giele},
  \citenamefont {Kosower},\ and\ \citenamefont {Skands}}]{Giele:2007di}%
  \BibitemOpen
  \bibfield  {author} {\bibinfo {author} {\bibfnamefont {W.~T.}\ \bibnamefont
  {Giele}}, \bibinfo {author} {\bibfnamefont {D.~A.}\ \bibnamefont {Kosower}},
  \ and\ \bibinfo {author} {\bibfnamefont {P.~Z.}\ \bibnamefont {Skands}},\
  }\href {\doibase 10.1103/PhysRevD.78.014026} {\bibfield  {journal} {\bibinfo
  {journal} {Phys. Rev.}\ }\textbf {\bibinfo {volume} {D78}},\ \bibinfo {pages}
  {014026} (\bibinfo {year} {2008})},\ \Eprint {http://arxiv.org/abs/0707.3652}
  {arXiv:0707.3652 [hep-ph]} \BibitemShut {NoStop}%
%%CITATION = ARXIV:0707.3652;%%
\bibitem [{\citenamefont {Nagy}\ and\ \citenamefont
  {Soper}(2007)}]{Nagy:2007ty}%
  \BibitemOpen
  \bibfield  {author} {\bibinfo {author} {\bibfnamefont {Z.}~\bibnamefont
  {Nagy}}\ and\ \bibinfo {author} {\bibfnamefont {D.~E.}\ \bibnamefont
  {Soper}},\ }\href {\doibase 10.1088/1126-6708/2007/09/114} {\bibfield
  {journal} {\bibinfo  {journal} {JHEP}\ }\textbf {\bibinfo {volume} {09}},\
  \bibinfo {pages} {114} (\bibinfo {year} {2007})},\ \Eprint
  {http://arxiv.org/abs/0706.0017} {arXiv:0706.0017 [hep-ph]} \BibitemShut
  {NoStop}%
%%CITATION = ARXIV:0706.0017;%%
\bibitem [{\citenamefont {Schumann}\ and\ \citenamefont
  {Krauss}(2008)}]{Schumann:2007mg}%
  \BibitemOpen
  \bibfield  {author} {\bibinfo {author} {\bibfnamefont {S.}~\bibnamefont
  {Schumann}}\ and\ \bibinfo {author} {\bibfnamefont {F.}~\bibnamefont
  {Krauss}},\ }\href {\doibase 10.1088/1126-6708/2008/03/038} {\bibfield
  {journal} {\bibinfo  {journal} {JHEP}\ }\textbf {\bibinfo {volume} {03}},\
  \bibinfo {pages} {038} (\bibinfo {year} {2008})},\ \Eprint
  {http://arxiv.org/abs/0709.1027} {arXiv:0709.1027 [hep-ph]} \BibitemShut
  {NoStop}%
%%CITATION = ARXIV:0709.1027;%%
\bibitem [{\citenamefont {Platzer}\ and\ \citenamefont
  {Gieseke}(2011)}]{Platzer:2009jq}%
  \BibitemOpen
  \bibfield  {author} {\bibinfo {author} {\bibfnamefont {S.}~\bibnamefont
  {Platzer}}\ and\ \bibinfo {author} {\bibfnamefont {S.}~\bibnamefont
  {Gieseke}},\ }\href {\doibase 10.1007/JHEP01(2011)024} {\bibfield  {journal}
  {\bibinfo  {journal} {JHEP}\ }\textbf {\bibinfo {volume} {01}},\ \bibinfo
  {pages} {024} (\bibinfo {year} {2011})},\ \Eprint
  {http://arxiv.org/abs/0909.5593} {arXiv:0909.5593 [hep-ph]} \BibitemShut
  {NoStop}%
%%CITATION = ARXIV:0909.5593;%%
\bibitem [{\citenamefont {Jadach}\ \emph {et~al.}(2010)\citenamefont {Jadach},
  \citenamefont {Kusina}, \citenamefont {Skrzypek},\ and\ \citenamefont
  {Slawinska}}]{Jadach:2010aa}%
  \BibitemOpen
  \bibfield  {author} {\bibinfo {author} {\bibfnamefont {S.}~\bibnamefont
  {Jadach}}, \bibinfo {author} {\bibfnamefont {A.}~\bibnamefont {Kusina}},
  \bibinfo {author} {\bibfnamefont {M.}~\bibnamefont {Skrzypek}}, \ and\
  \bibinfo {author} {\bibfnamefont {M.}~\bibnamefont {Slawinska}},\ }\bibfield
  {booktitle} {\emph {\bibinfo {booktitle} {{Proceedings, 10th DESY Workshop on
  Elementary Particle Theory: Loops and Legs in Quantum Field Theory: Woerlitz,
  Germany, April 25-30, 2010}}},\ }\href {\doibase
  10.1016/j.nuclphysbps.2010.09.009} {\bibfield  {journal} {\bibinfo  {journal}
  {Nucl. Phys. Proc. Suppl.}\ }\textbf {\bibinfo {volume} {205-206}},\ \bibinfo
  {pages} {295} (\bibinfo {year} {2010})},\ \Eprint
  {http://arxiv.org/abs/1007.2437} {arXiv:1007.2437 [hep-ph]} \BibitemShut
  {NoStop}%
%%CITATION = ARXIV:1007.2437;%%
\bibitem [{\citenamefont {Nagy}\ and\ \citenamefont
  {Soper}(2012)}]{Nagy:2012bt}%
  \BibitemOpen
  \bibfield  {author} {\bibinfo {author} {\bibfnamefont {Z.}~\bibnamefont
  {Nagy}}\ and\ \bibinfo {author} {\bibfnamefont {D.~E.}\ \bibnamefont
  {Soper}},\ }\href {\doibase 10.1007/JHEP06(2012)044} {\bibfield  {journal}
  {\bibinfo  {journal} {JHEP}\ }\textbf {\bibinfo {volume} {06}},\ \bibinfo
  {pages} {044} (\bibinfo {year} {2012})},\ \Eprint
  {http://arxiv.org/abs/1202.4496} {arXiv:1202.4496 [hep-ph]} \BibitemShut
  {NoStop}%
%%CITATION = ARXIV:1202.4496;%%
\bibitem [{\citenamefont {Nagy}\ and\ \citenamefont
  {Soper}(2014)}]{Nagy:2014mqa}%
  \BibitemOpen
  \bibfield  {author} {\bibinfo {author} {\bibfnamefont {Z.}~\bibnamefont
  {Nagy}}\ and\ \bibinfo {author} {\bibfnamefont {D.~E.}\ \bibnamefont
  {Soper}},\ }\href {\doibase 10.1007/JHEP06(2014)097} {\bibfield  {journal}
  {\bibinfo  {journal} {JHEP}\ }\textbf {\bibinfo {volume} {06}},\ \bibinfo
  {pages} {097} (\bibinfo {year} {2014})},\ \Eprint
  {http://arxiv.org/abs/1401.6364} {arXiv:1401.6364 [hep-ph]} \BibitemShut
  {NoStop}%
%%CITATION = ARXIV:1401.6364;%%
\bibitem [{\citenamefont {Nagy}\ and\ \citenamefont
  {Soper}(2015)}]{Nagy:2015hwa}%
  \BibitemOpen
  \bibfield  {author} {\bibinfo {author} {\bibfnamefont {Z.}~\bibnamefont
  {Nagy}}\ and\ \bibinfo {author} {\bibfnamefont {D.~E.}\ \bibnamefont
  {Soper}},\ }\href {\doibase 10.1007/JHEP07(2015)119} {\bibfield  {journal}
  {\bibinfo  {journal} {JHEP}\ }\textbf {\bibinfo {volume} {07}},\ \bibinfo
  {pages} {119} (\bibinfo {year} {2015})},\ \Eprint
  {http://arxiv.org/abs/1501.00778} {arXiv:1501.00778 [hep-ph]} \BibitemShut
  {NoStop}%
%%CITATION = ARXIV:1501.00778;%%
\bibitem [{\citenamefont {Hoeche}\ and\ \citenamefont
  {Prestel}(2015)}]{Hoche:2015sya}%
  \BibitemOpen
  \bibfield  {author} {\bibinfo {author} {\bibfnamefont {S.}~\bibnamefont
  {Hoeche}}\ and\ \bibinfo {author} {\bibfnamefont {S.}~\bibnamefont
  {Prestel}},\ }\href {\doibase 10.1140/epjc/s10052-015-3684-2} {\bibfield
  {journal} {\bibinfo  {journal} {Eur. Phys. J.}\ }\textbf {\bibinfo {volume}
  {C75}},\ \bibinfo {pages} {461} (\bibinfo {year} {2015})},\ \Eprint
  {http://arxiv.org/abs/1506.05057} {arXiv:1506.05057 [hep-ph]} \BibitemShut
  {NoStop}%
%%CITATION = ARXIV:1506.05057;%%
\bibitem [{\citenamefont {Li}\ and\ \citenamefont {Skands}(2017)}]{Li:2016yez}%
  \BibitemOpen
  \bibfield  {author} {\bibinfo {author} {\bibfnamefont {H.~T.}\ \bibnamefont
  {Li}}\ and\ \bibinfo {author} {\bibfnamefont {P.}~\bibnamefont {Skands}},\
  }\href {\doibase 10.1016/j.physletb.2017.05.011} {\bibfield  {journal}
  {\bibinfo  {journal} {Phys. Lett.}\ }\textbf {\bibinfo {volume} {B771}},\
  \bibinfo {pages} {59} (\bibinfo {year} {2017})},\ \Eprint
  {http://arxiv.org/abs/1611.00013} {arXiv:1611.00013 [hep-ph]} \BibitemShut
  {NoStop}%
%%CITATION = ARXIV:1611.00013;%%
\bibitem [{\citenamefont {Jadach}\ \emph {et~al.}(2016)\citenamefont {Jadach},
  \citenamefont {Kusina}, \citenamefont {Placzek},\ and\ \citenamefont
  {Skrzypek}}]{Jadach:2016zgk}%
  \BibitemOpen
  \bibfield  {author} {\bibinfo {author} {\bibfnamefont {S.}~\bibnamefont
  {Jadach}}, \bibinfo {author} {\bibfnamefont {A.}~\bibnamefont {Kusina}},
  \bibinfo {author} {\bibfnamefont {W.}~\bibnamefont {Placzek}}, \ and\
  \bibinfo {author} {\bibfnamefont {M.}~\bibnamefont {Skrzypek}},\ }\href
  {\doibase 10.1007/JHEP08(2016)092} {\bibfield  {journal} {\bibinfo  {journal}
  {JHEP}\ }\textbf {\bibinfo {volume} {08}},\ \bibinfo {pages} {092} (\bibinfo
  {year} {2016})},\ \Eprint {http://arxiv.org/abs/1606.01238} {arXiv:1606.01238
  [hep-ph]} \BibitemShut {NoStop}%
%%CITATION = ARXIV:1606.01238;%%
\bibitem [{\citenamefont {Fischer}\ \emph {et~al.}(2016)\citenamefont
  {Fischer}, \citenamefont {Prestel}, \citenamefont {Ritzmann},\ and\
  \citenamefont {Skands}}]{Fischer:2016vfv}%
  \BibitemOpen
  \bibfield  {author} {\bibinfo {author} {\bibfnamefont {N.}~\bibnamefont
  {Fischer}}, \bibinfo {author} {\bibfnamefont {S.}~\bibnamefont {Prestel}},
  \bibinfo {author} {\bibfnamefont {M.}~\bibnamefont {Ritzmann}}, \ and\
  \bibinfo {author} {\bibfnamefont {P.}~\bibnamefont {Skands}},\ }\href
  {\doibase 10.1140/epjc/s10052-016-4429-6} {\bibfield  {journal} {\bibinfo
  {journal} {Eur. Phys. J.}\ }\textbf {\bibinfo {volume} {C76}},\ \bibinfo
  {pages} {589} (\bibinfo {year} {2016})},\ \Eprint
  {http://arxiv.org/abs/1605.06142} {arXiv:1605.06142 [hep-ph]} \BibitemShut
  {NoStop}%
%%CITATION = ARXIV:1605.06142;%%
\bibitem [{\citenamefont {Nagy}\ and\ \citenamefont
  {Soper}(2016)}]{Nagy:2016pwq}%
  \BibitemOpen
  \bibfield  {author} {\bibinfo {author} {\bibfnamefont {Z.}~\bibnamefont
  {Nagy}}\ and\ \bibinfo {author} {\bibfnamefont {D.~E.}\ \bibnamefont
  {Soper}},\ }\href {\doibase 10.1007/JHEP10(2016)019} {\bibfield  {journal}
  {\bibinfo  {journal} {JHEP}\ }\textbf {\bibinfo {volume} {10}},\ \bibinfo
  {pages} {019} (\bibinfo {year} {2016})},\ \Eprint
  {http://arxiv.org/abs/1605.05845} {arXiv:1605.05845 [hep-ph]} \BibitemShut
  {NoStop}%
%%CITATION = ARXIV:1605.05845;%%
\bibitem [{\citenamefont {Fischer}\ \emph {et~al.}(2017)\citenamefont
  {Fischer}, \citenamefont {Lifson},\ and\ \citenamefont
  {Skands}}]{Fischer:2017htu}%
  \BibitemOpen
  \bibfield  {author} {\bibinfo {author} {\bibfnamefont {N.}~\bibnamefont
  {Fischer}}, \bibinfo {author} {\bibfnamefont {A.}~\bibnamefont {Lifson}}, \
  and\ \bibinfo {author} {\bibfnamefont {P.}~\bibnamefont {Skands}},\ }\href
  {\doibase 10.1140/epjc/s10052-017-5306-7} {\bibfield  {journal} {\bibinfo
  {journal} {Eur. Phys. J.}\ }\textbf {\bibinfo {volume} {C77}},\ \bibinfo
  {pages} {719} (\bibinfo {year} {2017})},\ \Eprint
  {http://arxiv.org/abs/1708.01736} {arXiv:1708.01736 [hep-ph]} \BibitemShut
  {NoStop}%
%%CITATION = ARXIV:1708.01736;%%
\bibitem [{\citenamefont {Hoeche}\ \emph {et~al.}(2017)\citenamefont {Hoeche},
  \citenamefont {Krauss},\ and\ \citenamefont {Prestel}}]{Hoche:2017hno}%
  \BibitemOpen
  \bibfield  {author} {\bibinfo {author} {\bibfnamefont {S.}~\bibnamefont
  {Hoeche}}, \bibinfo {author} {\bibfnamefont {F.}~\bibnamefont {Krauss}}, \
  and\ \bibinfo {author} {\bibfnamefont {S.}~\bibnamefont {Prestel}},\ }\href
  {\doibase 10.1007/JHEP10(2017)093} {\bibfield  {journal} {\bibinfo  {journal}
  {JHEP}\ }\textbf {\bibinfo {volume} {10}},\ \bibinfo {pages} {093} (\bibinfo
  {year} {2017})},\ \Eprint {http://arxiv.org/abs/1705.00982} {arXiv:1705.00982
  [hep-ph]} \BibitemShut {NoStop}%
%%CITATION = ARXIV:1705.00982;%%
\bibitem [{\citenamefont {Hoeche}\ and\ \citenamefont
  {Prestel}(2017)}]{Hoche:2017iem}%
  \BibitemOpen
  \bibfield  {author} {\bibinfo {author} {\bibfnamefont {S.}~\bibnamefont
  {Hoeche}}\ and\ \bibinfo {author} {\bibfnamefont {S.}~\bibnamefont
  {Prestel}},\ }\href {\doibase 10.1103/PhysRevD.96.074017} {\bibfield
  {journal} {\bibinfo  {journal} {Phys. Rev.}\ }\textbf {\bibinfo {volume}
  {D96}},\ \bibinfo {pages} {074017} (\bibinfo {year} {2017})},\ \Eprint
  {http://arxiv.org/abs/1705.00742} {arXiv:1705.00742 [hep-ph]} \BibitemShut
  {NoStop}%
%%CITATION = ARXIV:1705.00742;%%
\bibitem [{\citenamefont {Nagy}\ and\ \citenamefont
  {Soper}(2018)}]{Nagy:2017ggp}%
  \BibitemOpen
  \bibfield  {author} {\bibinfo {author} {\bibfnamefont {Z.}~\bibnamefont
  {Nagy}}\ and\ \bibinfo {author} {\bibfnamefont {D.~E.}\ \bibnamefont
  {Soper}},\ }\href {\doibase 10.1103/PhysRevD.98.014034} {\bibfield  {journal}
  {\bibinfo  {journal} {Phys. Rev.}\ }\textbf {\bibinfo {volume} {D98}},\
  \bibinfo {pages} {014034} (\bibinfo {year} {2018})},\ \Eprint
  {http://arxiv.org/abs/1705.08093} {arXiv:1705.08093 [hep-ph]} \BibitemShut
  {NoStop}%
%%CITATION = ARXIV:1705.08093;%%
\bibitem [{\citenamefont {Cabouat}\ and\ \citenamefont
  {Sjöstrand}(2018)}]{Cabouat:2017rzi}%
  \BibitemOpen
  \bibfield  {author} {\bibinfo {author} {\bibfnamefont {B.}~\bibnamefont
  {Cabouat}}\ and\ \bibinfo {author} {\bibfnamefont {T.}~\bibnamefont
  {Sjöstrand}},\ }\href {\doibase 10.1140/epjc/s10052-018-5645-z,
  10.1140/s10052-018-5645-z} {\bibfield  {journal} {\bibinfo  {journal} {Eur.
  Phys. J.}\ }\textbf {\bibinfo {volume} {C78}},\ \bibinfo {pages} {226}
  (\bibinfo {year} {2018})},\ \Eprint {http://arxiv.org/abs/1710.00391}
  {arXiv:1710.00391 [hep-ph]} \BibitemShut {NoStop}%
%%CITATION = ARXIV:1710.00391;%%
\bibitem [{\citenamefont {Dulat}\ \emph {et~al.}(2018)\citenamefont {Dulat},
  \citenamefont {Hoeche},\ and\ \citenamefont {Prestel}}]{Dulat:2018vuy}%
  \BibitemOpen
  \bibfield  {author} {\bibinfo {author} {\bibfnamefont {F.}~\bibnamefont
  {Dulat}}, \bibinfo {author} {\bibfnamefont {S.}~\bibnamefont {Hoeche}}, \
  and\ \bibinfo {author} {\bibfnamefont {S.}~\bibnamefont {Prestel}},\ }\href
  {\doibase 10.1103/PhysRevD.98.074013} {\bibfield  {journal} {\bibinfo
  {journal} {Phys. Rev.}\ }\textbf {\bibinfo {volume} {D98}},\ \bibinfo {pages}
  {074013} (\bibinfo {year} {2018})},\ \Eprint
  {http://arxiv.org/abs/1805.03757} {arXiv:1805.03757 [hep-ph]} \BibitemShut
  {NoStop}%
%%CITATION = ARXIV:1805.03757;%%
\bibitem [{\citenamefont {Plaetzer}\ \emph {et~al.}(2018)\citenamefont
  {Plaetzer}, \citenamefont {Sjodahl},\ and\ \citenamefont
  {Thorén}}]{Platzer:2018pmd}%
  \BibitemOpen
  \bibfield  {author} {\bibinfo {author} {\bibfnamefont {S.}~\bibnamefont
  {Plaetzer}}, \bibinfo {author} {\bibfnamefont {M.}~\bibnamefont {Sjodahl}}, \
  and\ \bibinfo {author} {\bibfnamefont {J.}~\bibnamefont {Thorén}},\ }\href
  {\doibase 10.1007/JHEP11(2018)009} {\bibfield  {journal} {\bibinfo  {journal}
  {JHEP}\ }\textbf {\bibinfo {volume} {11}},\ \bibinfo {pages} {009} (\bibinfo
  {year} {2018})},\ \Eprint {http://arxiv.org/abs/1808.00332} {arXiv:1808.00332
  [hep-ph]} \BibitemShut {NoStop}%
%%CITATION = ARXIV:1808.00332;%%
\bibitem [{\citenamefont {\'Angeles~Mart\'inez}\ \emph
  {et~al.}(2018)\citenamefont {\'Angeles~Mart\'inez}, \citenamefont
  {De~Angelis}, \citenamefont {Forshaw}, \citenamefont {Plaetzer},\ and\
  \citenamefont {Seymour}}]{Martinez:2018ffw}%
  \BibitemOpen
  \bibfield  {author} {\bibinfo {author} {\bibfnamefont {R.}~\bibnamefont
  {\'Angeles~Mart\'inez}}, \bibinfo {author} {\bibfnamefont {M.}~\bibnamefont
  {De~Angelis}}, \bibinfo {author} {\bibfnamefont {J.~R.}\ \bibnamefont
  {Forshaw}}, \bibinfo {author} {\bibfnamefont {S.}~\bibnamefont {Plaetzer}}, \
  and\ \bibinfo {author} {\bibfnamefont {M.~H.}\ \bibnamefont {Seymour}},\
  }\href {\doibase 10.1007/JHEP05(2018)044} {\bibfield  {journal} {\bibinfo
  {journal} {JHEP}\ }\textbf {\bibinfo {volume} {05}},\ \bibinfo {pages} {044}
  (\bibinfo {year} {2018})},\ \Eprint {http://arxiv.org/abs/1802.08531}
  {arXiv:1802.08531 [hep-ph]} \BibitemShut {NoStop}%
%%CITATION = ARXIV:1802.08531;%%
\bibitem [{\citenamefont {Isaacson}\ and\ \citenamefont
  {Prestel}(2019)}]{Isaacson:2018zdi}%
  \BibitemOpen
  \bibfield  {author} {\bibinfo {author} {\bibfnamefont {J.}~\bibnamefont
  {Isaacson}}\ and\ \bibinfo {author} {\bibfnamefont {S.}~\bibnamefont
  {Prestel}},\ }\href {\doibase 10.1103/PhysRevD.99.014021} {\bibfield
  {journal} {\bibinfo  {journal} {Phys. Rev.}\ }\textbf {\bibinfo {volume}
  {D99}},\ \bibinfo {pages} {014021} (\bibinfo {year} {2019})},\ \Eprint
  {http://arxiv.org/abs/1806.10102} {arXiv:1806.10102 [hep-ph]} \BibitemShut
  {NoStop}%
%%CITATION = ARXIV:1806.10102;%%
\bibitem [{\citenamefont {Brooks}\ and\ \citenamefont
  {Skands}(2019)}]{Brooks:2019xso}%
  \BibitemOpen
  \bibfield  {author} {\bibinfo {author} {\bibfnamefont {H.}~\bibnamefont
  {Brooks}}\ and\ \bibinfo {author} {\bibfnamefont {P.}~\bibnamefont
  {Skands}},\ }\href {\doibase 10.1103/PhysRevD.100.076006} {\bibfield
  {journal} {\bibinfo  {journal} {Phys.Rev.D}\ }\textbf {\bibinfo {volume}
  {100}},\ \bibinfo {pages} {076006} (\bibinfo {year} {2019})},\ \Eprint
  {http://arxiv.org/abs/1907.08980} {arXiv:1907.08980 [hep-ph]} \BibitemShut
  {NoStop}%
\bibitem [{\citenamefont {Forshaw}\ \emph {et~al.}(2019)\citenamefont
  {Forshaw}, \citenamefont {Holguin},\ and\ \citenamefont
  {Plaetzer}}]{Forshaw:2019ver}%
  \BibitemOpen
  \bibfield  {author} {\bibinfo {author} {\bibfnamefont {J.~R.}\ \bibnamefont
  {Forshaw}}, \bibinfo {author} {\bibfnamefont {J.}~\bibnamefont {Holguin}}, \
  and\ \bibinfo {author} {\bibfnamefont {S.}~\bibnamefont {Plaetzer}},\ }\href
  {\doibase 10.1007/JHEP08(2019)145} {\  (\bibinfo {year} {2019}),\
  10.1007/JHEP08(2019)145},\ \bibinfo {note} {[JHEP08,145(2019)]},\ \Eprint
  {http://arxiv.org/abs/1905.08686} {arXiv:1905.08686 [hep-ph]} \BibitemShut
  {NoStop}%
%%CITATION = ARXIV1905.08686;%%
\bibitem [{\citenamefont {Nagy}\ and\ \citenamefont
  {Soper}(2019)}]{Nagy:2019rwb}%
  \BibitemOpen
  \bibfield  {author} {\bibinfo {author} {\bibfnamefont {Z.}~\bibnamefont
  {Nagy}}\ and\ \bibinfo {author} {\bibfnamefont {D.~E.}\ \bibnamefont
  {Soper}},\ }\href {\doibase 10.1103/PhysRevD.100.074005} {\bibfield
  {journal} {\bibinfo  {journal} {Phys. Rev.}\ }\textbf {\bibinfo {volume}
  {D100}},\ \bibinfo {pages} {074005} (\bibinfo {year} {2019})},\ \Eprint
  {http://arxiv.org/abs/1908.11420} {arXiv:1908.11420 [hep-ph]} \BibitemShut
  {NoStop}%
%%CITATION = ARXIV:1908.11420;%%
\bibitem [{\citenamefont {Hoeche}\ and\ \citenamefont
  {Reichelt}(2020)}]{Hoeche:2020nsx}%
  \BibitemOpen
  \bibfield  {author} {\bibinfo {author} {\bibfnamefont {S.}~\bibnamefont
  {Hoeche}}\ and\ \bibinfo {author} {\bibfnamefont {D.}~\bibnamefont
  {Reichelt}},\ }\href@noop {} {\  (\bibinfo {year} {2020})},\ \Eprint
  {http://arxiv.org/abs/2001.11492} {arXiv:2001.11492 [hep-ph]} \BibitemShut
  {NoStop}%
%%CITATION = ARXIV:2001.11492;%%
\bibitem [{\citenamefont {Buckley}\ \emph {et~al.}(2010)\citenamefont
  {Buckley}, \citenamefont {Hoeth}, \citenamefont {Lacker}, \citenamefont
  {Schulz},\ and\ \citenamefont {von Seggern}}]{Buckley:2009bj}%
  \BibitemOpen
  \bibfield  {author} {\bibinfo {author} {\bibfnamefont {A.}~\bibnamefont
  {Buckley}}, \bibinfo {author} {\bibfnamefont {H.}~\bibnamefont {Hoeth}},
  \bibinfo {author} {\bibfnamefont {H.}~\bibnamefont {Lacker}}, \bibinfo
  {author} {\bibfnamefont {H.}~\bibnamefont {Schulz}}, \ and\ \bibinfo {author}
  {\bibfnamefont {J.~E.}\ \bibnamefont {von Seggern}},\ }\href {\doibase
  10.1140/epjc/s10052-009-1196-7} {\bibfield  {journal} {\bibinfo  {journal}
  {Eur. Phys. J.}\ }\textbf {\bibinfo {volume} {C65}},\ \bibinfo {pages} {331}
  (\bibinfo {year} {2010})},\ \Eprint {http://arxiv.org/abs/0907.2973}
  {arXiv:0907.2973 [hep-ph]} \BibitemShut {NoStop}%
%%CITATION = ARXIV:0907.2973;%%
\bibitem [{\citenamefont {Skands}(2010)}]{Skands:2010ak}%
  \BibitemOpen
  \bibfield  {author} {\bibinfo {author} {\bibfnamefont {P.~Z.}\ \bibnamefont
  {Skands}},\ }\href {\doibase 10.1103/PhysRevD.82.074018} {\bibfield
  {journal} {\bibinfo  {journal} {Phys. Rev.}\ }\textbf {\bibinfo {volume}
  {D82}},\ \bibinfo {pages} {074018} (\bibinfo {year} {2010})},\ \Eprint
  {http://arxiv.org/abs/1005.3457} {arXiv:1005.3457 [hep-ph]} \BibitemShut
  {NoStop}%
%%CITATION = ARXIV:1005.3457;%%
\bibitem [{\citenamefont {Keith~Ellis}\ and\ \citenamefont
  {Zanderighi}(2019)}]{Ellis:2019qre}%
  \BibitemOpen
  \bibfield  {author} {\bibinfo {author} {\bibfnamefont {R.}~\bibnamefont
  {Keith~Ellis}}\ and\ \bibinfo {author} {\bibfnamefont {G.}~\bibnamefont
  {Zanderighi}},\ }in\ \href {\doibase 10.1142/9789813238053_0004} {\emph
  {\bibinfo {booktitle} {From My Vast Repertoire ...: Guido Altarelli's
  Legacy}}},\ \bibinfo {editor} {edited by\ \bibinfo {editor} {\bibfnamefont
  {A.}~\bibnamefont {Levy}}, \bibinfo {editor} {\bibfnamefont {S.}~\bibnamefont
  {Forte}}, \ and\ \bibinfo {editor} {\bibfnamefont {G.}~\bibnamefont
  {Ridolfi}}}\ (\bibinfo  {publisher} {World Scientific},\ \bibinfo {address}
  {Singapore},\ \bibinfo {year} {2019})\ pp.\ \bibinfo {pages}
  {31--52}\BibitemShut {NoStop}%
%%CITATION = INSPIRE-1700916;%%
\bibitem [{\citenamefont {Azzi}\ \emph {et~al.}(2019)\citenamefont {Azzi} \emph
  {et~al.}}]{Azzi:2019yne}%
  \BibitemOpen
  \bibfield  {author} {\bibinfo {author} {\bibfnamefont {P.}~\bibnamefont
  {Azzi}} \emph {et~al.},\ }\href {\doibase 10.23731/CYRM-2019-007.1}
  {\bibfield  {journal} {\bibinfo  {journal} {CERN Yellow Rep. Monogr.}\
  }\textbf {\bibinfo {volume} {7}},\ \bibinfo {pages} {1} (\bibinfo {year}
  {2019})},\ \Eprint {http://arxiv.org/abs/1902.04070} {arXiv:1902.04070
  [hep-ph]} \BibitemShut {NoStop}%
%%CITATION = ARXIV:1902.04070;%%
\bibitem [{\citenamefont {Cepeda}\ \emph {et~al.}(2019)\citenamefont {Cepeda}
  \emph {et~al.}}]{Cepeda:2019klc}%
  \BibitemOpen
  \bibfield  {author} {\bibinfo {author} {\bibfnamefont {M.}~\bibnamefont
  {Cepeda}} \emph {et~al.},\ }\href {\doibase 10.23731/CYRM-2019-007.221}
  {\bibfield  {journal} {\bibinfo  {journal} {CERN Yellow Rep. Monogr.}\
  }\textbf {\bibinfo {volume} {7}},\ \bibinfo {pages} {221} (\bibinfo {year}
  {2019})},\ \Eprint {http://arxiv.org/abs/1902.00134} {arXiv:1902.00134
  [hep-ph]} \BibitemShut {NoStop}%
%%CITATION = ARXIV:1902.00134;%%
\bibitem [{\citenamefont {Cid~Vidal}\ \emph {et~al.}(2019)\citenamefont
  {Cid~Vidal} \emph {et~al.}}]{CidVidal:2018eel}%
  \BibitemOpen
  \bibfield  {author} {\bibinfo {author} {\bibfnamefont {X.}~\bibnamefont
  {Cid~Vidal}} \emph {et~al.},\ }\href {\doibase 10.23731/CYRM-2019-007.585}
  {\bibfield  {journal} {\bibinfo  {journal} {CERN Yellow Rep. Monogr.}\
  }\textbf {\bibinfo {volume} {7}},\ \bibinfo {pages} {585} (\bibinfo {year}
  {2019})},\ \Eprint {http://arxiv.org/abs/1812.07831} {arXiv:1812.07831
  [hep-ph]} \BibitemShut {NoStop}%
%%CITATION = ARXIV:1812.07831;%%
\bibitem [{\citenamefont {Marchesini}\ and\ \citenamefont
  {Webber}(1984)}]{Marchesini:1983bm}%
  \BibitemOpen
  \bibfield  {author} {\bibinfo {author} {\bibfnamefont {G.}~\bibnamefont
  {Marchesini}}\ and\ \bibinfo {author} {\bibfnamefont {B.~R.}\ \bibnamefont
  {Webber}},\ }\href {\doibase 10.1016/0550-3213(84)90463-2} {\bibfield
  {journal} {\bibinfo  {journal} {Nucl. Phys.}\ }\textbf {\bibinfo {volume}
  {B238}},\ \bibinfo {pages} {1} (\bibinfo {year} {1984})}\BibitemShut
  {NoStop}%
%%CITATION = NUPHA,B238,1;%%
\bibitem [{\citenamefont {Banfi}\ \emph {et~al.}(2007)\citenamefont {Banfi},
  \citenamefont {Corcella},\ and\ \citenamefont {Dasgupta}}]{Banfi:2006gy}%
  \BibitemOpen
  \bibfield  {author} {\bibinfo {author} {\bibfnamefont {A.}~\bibnamefont
  {Banfi}}, \bibinfo {author} {\bibfnamefont {G.}~\bibnamefont {Corcella}}, \
  and\ \bibinfo {author} {\bibfnamefont {M.}~\bibnamefont {Dasgupta}},\ }\href
  {\doibase 10.1088/1126-6708/2007/03/050} {\bibfield  {journal} {\bibinfo
  {journal} {JHEP}\ }\textbf {\bibinfo {volume} {03}},\ \bibinfo {pages} {050}
  (\bibinfo {year} {2007})},\ \Eprint {http://arxiv.org/abs/hep-ph/0612282}
  {arXiv:hep-ph/0612282 [hep-ph]} \BibitemShut {NoStop}%
%%CITATION = HEP-PH/0612282;%%
\bibitem [{\citenamefont {Dasgupta}\ \emph {et~al.}(2018)\citenamefont
  {Dasgupta}, \citenamefont {Dreyer}, \citenamefont {Hamilton}, \citenamefont
  {Monni},\ and\ \citenamefont {Salam}}]{Dasgupta:2018nvj}%
  \BibitemOpen
  \bibfield  {author} {\bibinfo {author} {\bibfnamefont {M.}~\bibnamefont
  {Dasgupta}}, \bibinfo {author} {\bibfnamefont {F.~A.}\ \bibnamefont
  {Dreyer}}, \bibinfo {author} {\bibfnamefont {K.}~\bibnamefont {Hamilton}},
  \bibinfo {author} {\bibfnamefont {P.~F.}\ \bibnamefont {Monni}}, \ and\
  \bibinfo {author} {\bibfnamefont {G.~P.}\ \bibnamefont {Salam}},\ }\href
  {\doibase 10.1007/JHEP09(2018)033} {\bibfield  {journal} {\bibinfo  {journal}
  {JHEP}\ }\textbf {\bibinfo {volume} {09}},\ \bibinfo {pages} {033} (\bibinfo
  {year} {2018})},\ \Eprint {http://arxiv.org/abs/1805.09327} {arXiv:1805.09327
  [hep-ph]} \BibitemShut {NoStop}%
%%CITATION = ARXIV:1805.09327;%%
\bibitem [{\citenamefont {Bewick}\ \emph {et~al.}(2019)\citenamefont {Bewick},
  \citenamefont {Ferrario~Ravasio}, \citenamefont {Richardson},\ and\
  \citenamefont {Seymour}}]{Bewick:2019rbu}%
  \BibitemOpen
  \bibfield  {author} {\bibinfo {author} {\bibfnamefont {G.}~\bibnamefont
  {Bewick}}, \bibinfo {author} {\bibfnamefont {S.}~\bibnamefont
  {Ferrario~Ravasio}}, \bibinfo {author} {\bibfnamefont {P.}~\bibnamefont
  {Richardson}}, \ and\ \bibinfo {author} {\bibfnamefont {M.~H.}\ \bibnamefont
  {Seymour}},\ }\href@noop {} {\  (\bibinfo {year} {2019})},\ \Eprint
  {http://arxiv.org/abs/1904.11866} {arXiv:1904.11866 [hep-ph]} \BibitemShut
  {NoStop}%
%%CITATION = ARXIV:1904.11866;%%
\bibitem [{\citenamefont {Marchesini}\ and\ \citenamefont
  {Webber}(1988)}]{Marchesini:1987cf}%
  \BibitemOpen
  \bibfield  {author} {\bibinfo {author} {\bibfnamefont {G.}~\bibnamefont
  {Marchesini}}\ and\ \bibinfo {author} {\bibfnamefont {B.~R.}\ \bibnamefont
  {Webber}},\ }\href {\doibase 10.1016/0550-3213(88)90089-2} {\bibfield
  {journal} {\bibinfo  {journal} {Nucl. Phys.}\ }\textbf {\bibinfo {volume}
  {B310}},\ \bibinfo {pages} {461} (\bibinfo {year} {1988})}\BibitemShut
  {NoStop}%
%%CITATION = NUPHA,B310,461;%%
\bibitem [{\citenamefont {Corcella}\ \emph {et~al.}(2001)\citenamefont
  {Corcella}, \citenamefont {Knowles}, \citenamefont {Marchesini},
  \citenamefont {Moretti}, \citenamefont {Odagiri}, \citenamefont {Richardson},
  \citenamefont {Seymour},\ and\ \citenamefont {Webber}}]{Corcella:2000bw}%
  \BibitemOpen
  \bibfield  {author} {\bibinfo {author} {\bibfnamefont {G.}~\bibnamefont
  {Corcella}}, \bibinfo {author} {\bibfnamefont {I.~G.}\ \bibnamefont
  {Knowles}}, \bibinfo {author} {\bibfnamefont {G.}~\bibnamefont {Marchesini}},
  \bibinfo {author} {\bibfnamefont {S.}~\bibnamefont {Moretti}}, \bibinfo
  {author} {\bibfnamefont {K.}~\bibnamefont {Odagiri}}, \bibinfo {author}
  {\bibfnamefont {P.}~\bibnamefont {Richardson}}, \bibinfo {author}
  {\bibfnamefont {M.~H.}\ \bibnamefont {Seymour}}, \ and\ \bibinfo {author}
  {\bibfnamefont {B.~R.}\ \bibnamefont {Webber}},\ }\href {\doibase
  10.1088/1126-6708/2001/01/010} {\bibfield  {journal} {\bibinfo  {journal}
  {JHEP}\ }\textbf {\bibinfo {volume} {01}},\ \bibinfo {pages} {010} (\bibinfo
  {year} {2001})},\ \Eprint {http://arxiv.org/abs/hep-ph/0011363}
  {arXiv:hep-ph/0011363 [hep-ph]} \BibitemShut {NoStop}%
%%CITATION = HEP-PH/0011363;%%
\bibitem [{\citenamefont {Bahr}\ \emph {et~al.}(2008)\citenamefont {Bahr} \emph
  {et~al.}}]{Bahr:2008pv}%
  \BibitemOpen
  \bibfield  {author} {\bibinfo {author} {\bibfnamefont {M.}~\bibnamefont
  {Bahr}} \emph {et~al.},\ }\href {\doibase 10.1140/epjc/s10052-008-0798-9}
  {\bibfield  {journal} {\bibinfo  {journal} {Eur. Phys. J.}\ }\textbf
  {\bibinfo {volume} {C58}},\ \bibinfo {pages} {639} (\bibinfo {year}
  {2008})},\ \Eprint {http://arxiv.org/abs/0803.0883} {arXiv:0803.0883
  [hep-ph]} \BibitemShut {NoStop}%
%%CITATION = ARXIV:0803.0883;%%
\bibitem [{\citenamefont {Gustafson}\ and\ \citenamefont
  {Pettersson}(1988)}]{Gustafson:1987rq}%
  \BibitemOpen
  \bibfield  {author} {\bibinfo {author} {\bibfnamefont {G.}~\bibnamefont
  {Gustafson}}\ and\ \bibinfo {author} {\bibfnamefont {U.}~\bibnamefont
  {Pettersson}},\ }\href {\doibase 10.1016/0550-3213(88)90441-5} {\bibfield
  {journal} {\bibinfo  {journal} {Nucl. Phys.}\ }\textbf {\bibinfo {volume}
  {B306}},\ \bibinfo {pages} {746} (\bibinfo {year} {1988})}\BibitemShut
  {NoStop}%
%%CITATION = NUPHA,B306,746;%%
\bibitem [{\citenamefont {Lonnblad}(1992)}]{Lonnblad:1992tz}%
  \BibitemOpen
  \bibfield  {author} {\bibinfo {author} {\bibfnamefont {L.}~\bibnamefont
  {Lonnblad}},\ }\href {\doibase 10.1016/0010-4655(92)90068-A} {\bibfield
  {journal} {\bibinfo  {journal} {Comput. Phys. Commun.}\ }\textbf {\bibinfo
  {volume} {71}},\ \bibinfo {pages} {15} (\bibinfo {year} {1992})}\BibitemShut
  {NoStop}%
%%CITATION = CPHCB,71,15;%%
\bibitem [{\citenamefont {Webber}(1986)}]{Webber:1986mc}%
  \BibitemOpen
  \bibfield  {author} {\bibinfo {author} {\bibfnamefont {B.~R.}\ \bibnamefont
  {Webber}},\ }\href {\doibase 10.1146/annurev.ns.36.120186.001345} {\bibfield
  {journal} {\bibinfo  {journal} {Ann. Rev. Nucl. Part. Sci.}\ }\textbf
  {\bibinfo {volume} {36}},\ \bibinfo {pages} {253} (\bibinfo {year}
  {1986})}\BibitemShut {NoStop}%
%%CITATION = ARNUA,36,253;%%
\bibitem [{\citenamefont {Collins}(1988)}]{Collins:1987cp}%
  \BibitemOpen
  \bibfield  {author} {\bibinfo {author} {\bibfnamefont {J.~C.}\ \bibnamefont
  {Collins}},\ }\href {\doibase 10.1016/0550-3213(88)90654-2} {\bibfield
  {journal} {\bibinfo  {journal} {Nucl. Phys.}\ }\textbf {\bibinfo {volume}
  {B304}},\ \bibinfo {pages} {794} (\bibinfo {year} {1988})}\BibitemShut
  {NoStop}%
%%CITATION = NUPHA,B304,794;%%
\bibitem [{\citenamefont {Catani}\ and\ \citenamefont
  {Grazzini}(2011)}]{Catani:2010pd}%
  \BibitemOpen
  \bibfield  {author} {\bibinfo {author} {\bibfnamefont {S.}~\bibnamefont
  {Catani}}\ and\ \bibinfo {author} {\bibfnamefont {M.}~\bibnamefont
  {Grazzini}},\ }\href {\doibase 10.1016/j.nuclphysb.2010.12.007} {\bibfield
  {journal} {\bibinfo  {journal} {Nucl. Phys.}\ }\textbf {\bibinfo {volume}
  {B845}},\ \bibinfo {pages} {297} (\bibinfo {year} {2011})},\ \Eprint
  {http://arxiv.org/abs/1011.3918} {arXiv:1011.3918 [hep-ph]} \BibitemShut
  {NoStop}%
%%CITATION = ARXIV:1011.3918;%%
\bibitem [{\citenamefont {Banfi}\ \emph {et~al.}(2005)\citenamefont {Banfi},
  \citenamefont {Salam},\ and\ \citenamefont {Zanderighi}}]{Banfi:2004yd}%
  \BibitemOpen
  \bibfield  {author} {\bibinfo {author} {\bibfnamefont {A.}~\bibnamefont
  {Banfi}}, \bibinfo {author} {\bibfnamefont {G.~P.}\ \bibnamefont {Salam}}, \
  and\ \bibinfo {author} {\bibfnamefont {G.}~\bibnamefont {Zanderighi}},\
  }\href {\doibase 10.1088/1126-6708/2005/03/073} {\bibfield  {journal}
  {\bibinfo  {journal} {JHEP}\ }\textbf {\bibinfo {volume} {03}},\ \bibinfo
  {pages} {073} (\bibinfo {year} {2005})},\ \Eprint
  {http://arxiv.org/abs/hep-ph/0407286} {arXiv:hep-ph/0407286 [hep-ph]}
  \BibitemShut {NoStop}%
%%CITATION = HEP-PH/0407286;%%
\bibitem [{\citenamefont {Knowles}(1990)}]{Knowles:1988hu}%
  \BibitemOpen
  \bibfield  {author} {\bibinfo {author} {\bibfnamefont {I.~G.}\ \bibnamefont
  {Knowles}},\ }\href {\doibase 10.1016/0010-4655(90)90063-7} {\bibfield
  {journal} {\bibinfo  {journal} {Comput. Phys. Commun.}\ }\textbf {\bibinfo
  {volume} {58}},\ \bibinfo {pages} {271} (\bibinfo {year} {1990})}\BibitemShut
  {NoStop}%
%%CITATION = CPHCB,58,271;%%
\bibitem [{\citenamefont {Nagy}\ and\ \citenamefont
  {Soper}(2008)}]{Nagy:2008eq}%
  \BibitemOpen
  \bibfield  {author} {\bibinfo {author} {\bibfnamefont {Z.}~\bibnamefont
  {Nagy}}\ and\ \bibinfo {author} {\bibfnamefont {D.~E.}\ \bibnamefont
  {Soper}},\ }\href {\doibase 10.1088/1126-6708/2008/07/025} {\bibfield
  {journal} {\bibinfo  {journal} {JHEP}\ }\textbf {\bibinfo {volume} {07}},\
  \bibinfo {pages} {025} (\bibinfo {year} {2008})},\ \Eprint
  {http://arxiv.org/abs/0805.0216} {arXiv:0805.0216 [hep-ph]} \BibitemShut
  {NoStop}%
%%CITATION = ARXIV:0805.0216;%%
\bibitem [{\citenamefont {Richardson}\ and\ \citenamefont
  {Webster}(2020)}]{Richardson:2018pvo}%
  \BibitemOpen
  \bibfield  {author} {\bibinfo {author} {\bibfnamefont {P.}~\bibnamefont
  {Richardson}}\ and\ \bibinfo {author} {\bibfnamefont {S.}~\bibnamefont
  {Webster}},\ }\href {\doibase 10.1140/epjc/s10052-019-7429-5} {\bibfield
  {journal} {\bibinfo  {journal} {Eur. Phys. J.}\ }\textbf {\bibinfo {volume}
  {C80}},\ \bibinfo {pages} {83} (\bibinfo {year} {2020})},\ \Eprint
  {http://arxiv.org/abs/1807.01955} {arXiv:1807.01955 [hep-ph]} \BibitemShut
  {NoStop}%
%%CITATION = ARXIV:1807.01955;%%
\bibitem [{\citenamefont {Dasgupta}\ \emph {et~al.}(2020)\citenamefont
  {Dasgupta}, \citenamefont {Dreyer}, \citenamefont {Hamilton}, \citenamefont
  {Monni}, \citenamefont {Salam},\ and\ \citenamefont {Soyez}}]{supplement}%
  \BibitemOpen
  \bibfield  {author} {\bibinfo {author} {\bibfnamefont {M.}~\bibnamefont
  {Dasgupta}}, \bibinfo {author} {\bibfnamefont {F.~A.}\ \bibnamefont
  {Dreyer}}, \bibinfo {author} {\bibfnamefont {K.}~\bibnamefont {Hamilton}},
  \bibinfo {author} {\bibfnamefont {P.~F.}\ \bibnamefont {Monni}}, \bibinfo
  {author} {\bibfnamefont {G.~P.}\ \bibnamefont {Salam}}, \ and\ \bibinfo
  {author} {\bibfnamefont {G.}~\bibnamefont {Soyez}},\ }\href@noop {}
  {\bibfield  {journal} {\bibinfo  {journal} {Supplemental material to this
  letter}\ } (\bibinfo {year} {2020})}\BibitemShut {NoStop}%
\bibitem [{\citenamefont {Andersson}\ \emph {et~al.}(1989)\citenamefont
  {Andersson}, \citenamefont {Gustafson}, \citenamefont {Lonnblad},\ and\
  \citenamefont {Pettersson}}]{Andersson:1988gp}%
  \BibitemOpen
  \bibfield  {author} {\bibinfo {author} {\bibfnamefont {B.}~\bibnamefont
  {Andersson}}, \bibinfo {author} {\bibfnamefont {G.}~\bibnamefont
  {Gustafson}}, \bibinfo {author} {\bibfnamefont {L.}~\bibnamefont {Lonnblad}},
  \ and\ \bibinfo {author} {\bibfnamefont {U.}~\bibnamefont {Pettersson}},\
  }\href {\doibase 10.1007/BF01550942} {\bibfield  {journal} {\bibinfo
  {journal} {Z. Phys.}\ }\textbf {\bibinfo {volume} {C43}},\ \bibinfo {pages}
  {625} (\bibinfo {year} {1989})}\BibitemShut {NoStop}%
%%CITATION = ZEPYA,C43,625;%%
\bibitem [{\citenamefont {Banfi}\ \emph {et~al.}(2002)\citenamefont {Banfi},
  \citenamefont {Salam},\ and\ \citenamefont {Zanderighi}}]{Banfi:2001bz}%
  \BibitemOpen
  \bibfield  {author} {\bibinfo {author} {\bibfnamefont {A.}~\bibnamefont
  {Banfi}}, \bibinfo {author} {\bibfnamefont {G.~P.}\ \bibnamefont {Salam}}, \
  and\ \bibinfo {author} {\bibfnamefont {G.}~\bibnamefont {Zanderighi}},\
  }\href {\doibase 10.1088/1126-6708/2002/01/018} {\bibfield  {journal}
  {\bibinfo  {journal} {JHEP}\ }\textbf {\bibinfo {volume} {01}},\ \bibinfo
  {pages} {018} (\bibinfo {year} {2002})},\ \Eprint
  {http://arxiv.org/abs/hep-ph/0112156} {arXiv:hep-ph/0112156 [hep-ph]}
  \BibitemShut {NoStop}%
%%CITATION = HEP-PH/0112156;%%
\bibitem [{\citenamefont {Gribov}\ and\ \citenamefont
  {Lipatov}(1972)}]{Gribov:1972ri}%
  \BibitemOpen
  \bibfield  {author} {\bibinfo {author} {\bibfnamefont {V.~N.}\ \bibnamefont
  {Gribov}}\ and\ \bibinfo {author} {\bibfnamefont {L.~N.}\ \bibnamefont
  {Lipatov}},\ }\href@noop {} {\bibfield  {journal} {\bibinfo  {journal} {Sov.
  J. Nucl. Phys.}\ }\textbf {\bibinfo {volume} {15}},\ \bibinfo {pages} {438}
  (\bibinfo {year} {1972})},\ \bibinfo {note} {[Yad.
  Fiz.15,781(1972)]}\BibitemShut {NoStop}%
%%CITATION = SJNCA,15,438;%%
\bibitem [{\citenamefont {Altarelli}\ and\ \citenamefont
  {Parisi}(1977)}]{Altarelli:1977zs}%
  \BibitemOpen
  \bibfield  {author} {\bibinfo {author} {\bibfnamefont {G.}~\bibnamefont
  {Altarelli}}\ and\ \bibinfo {author} {\bibfnamefont {G.}~\bibnamefont
  {Parisi}},\ }\href {\doibase 10.1016/0550-3213(77)90384-4} {\bibfield
  {journal} {\bibinfo  {journal} {Nucl. Phys.}\ }\textbf {\bibinfo {volume}
  {B126}},\ \bibinfo {pages} {298} (\bibinfo {year} {1977})}\BibitemShut
  {NoStop}%
%%CITATION = NUPHA,B126,298;%%
\bibitem [{\citenamefont {Dokshitzer}(1977)}]{Dokshitzer:1977sg}%
  \BibitemOpen
  \bibfield  {author} {\bibinfo {author} {\bibfnamefont {Y.~L.}\ \bibnamefont
  {Dokshitzer}},\ }\href@noop {} {\bibfield  {journal} {\bibinfo  {journal}
  {Sov. Phys. JETP}\ }\textbf {\bibinfo {volume} {46}},\ \bibinfo {pages} {641}
  (\bibinfo {year} {1977})},\ \bibinfo {note} {[Zh. Eksp. Teor.
  Fiz.73,1216(1977)]}\BibitemShut {NoStop}%
%%CITATION = SPHJA,46,641;%%
\bibitem [{\citenamefont {Behring}\ \emph {et~al.}(2019)\citenamefont
  {Behring}, \citenamefont {Melnikov}, \citenamefont {Rietkerk}, \citenamefont
  {Tancredi},\ and\ \citenamefont {Wever}}]{Behring:2019qufw}%
  \BibitemOpen
  \bibfield  {author} {\bibinfo {author} {\bibfnamefont {A.}~\bibnamefont
  {Behring}}, \bibinfo {author} {\bibfnamefont {K.}~\bibnamefont {Melnikov}},
  \bibinfo {author} {\bibfnamefont {R.}~\bibnamefont {Rietkerk}}, \bibinfo
  {author} {\bibfnamefont {L.}~\bibnamefont {Tancredi}}, \ and\ \bibinfo
  {author} {\bibfnamefont {C.}~\bibnamefont {Wever}},\ }\href {\doibase
  10.1103/PhysRevD.100.114034} {\bibfield  {journal} {\bibinfo  {journal}
  {Phys.Rev.D}\ }\textbf {\bibinfo {volume} {100}},\ \bibinfo {pages} {114034}
  (\bibinfo {year} {2019})},\ \Eprint {http://arxiv.org/abs/1910.10059}
  {arXiv:1910.10059 [hep-ph]} \BibitemShut {NoStop}%
\bibitem [{\citenamefont {Catani}\ \emph
  {et~al.}(1991{\natexlab{a}})\citenamefont {Catani}, \citenamefont {Webber},\
  and\ \citenamefont {Marchesini}}]{Catani:1990rr}%
  \BibitemOpen
  \bibfield  {author} {\bibinfo {author} {\bibfnamefont {S.}~\bibnamefont
  {Catani}}, \bibinfo {author} {\bibfnamefont {B.~R.}\ \bibnamefont {Webber}},
  \ and\ \bibinfo {author} {\bibfnamefont {G.}~\bibnamefont {Marchesini}},\
  }\href {\doibase 10.1016/0550-3213(91)90390-J} {\bibfield  {journal}
  {\bibinfo  {journal} {Nucl. Phys.}\ }\textbf {\bibinfo {volume} {B349}},\
  \bibinfo {pages} {635} (\bibinfo {year} {1991}{\natexlab{a}})}\BibitemShut
  {NoStop}%
%%CITATION = NUPHA,B349,635;%%
\bibitem [{\citenamefont {Abada}\ \emph {et~al.}(2019)\citenamefont {Abada}
  \emph {et~al.}}]{Benedikt:2018csr}%
  \BibitemOpen
  \bibfield  {author} {\bibinfo {author} {\bibfnamefont {A.}~\bibnamefont
  {Abada}} \emph {et~al.} (\bibinfo {collaboration} {FCC}),\ }\href {\doibase
  10.1140/epjst/e2019-900087-0} {\bibfield  {journal} {\bibinfo  {journal}
  {Eur. Phys. J. ST}\ }\textbf {\bibinfo {volume} {228}},\ \bibinfo {pages}
  {755} (\bibinfo {year} {2019})}\BibitemShut {NoStop}%
%%CITATION = 00619,228,755;%%
\bibitem [{\citenamefont {Hoeche}\ \emph {et~al.}(2018)\citenamefont {Hoeche},
  \citenamefont {Reichelt},\ and\ \citenamefont {Siegert}}]{Hoeche:2017jsi}%
  \BibitemOpen
  \bibfield  {author} {\bibinfo {author} {\bibfnamefont {S.}~\bibnamefont
  {Hoeche}}, \bibinfo {author} {\bibfnamefont {D.}~\bibnamefont {Reichelt}}, \
  and\ \bibinfo {author} {\bibfnamefont {F.}~\bibnamefont {Siegert}},\ }\href
  {\doibase 10.1007/JHEP01(2018)118} {\bibfield  {journal} {\bibinfo  {journal}
  {JHEP}\ }\textbf {\bibinfo {volume} {01}},\ \bibinfo {pages} {118} (\bibinfo
  {year} {2018})},\ \Eprint {http://arxiv.org/abs/1711.03497} {arXiv:1711.03497
  [hep-ph]} \BibitemShut {NoStop}%
%%CITATION = ARXIV:1711.03497;%%
\bibitem [{\citenamefont {Hida}\ \emph {et~al.}(2000)\citenamefont {Hida},
  \citenamefont {Li},\ and\ \citenamefont {Bailey}}]{hida2000quad}%
  \BibitemOpen
  \bibfield  {author} {\bibinfo {author} {\bibfnamefont {Y.}~\bibnamefont
  {Hida}}, \bibinfo {author} {\bibfnamefont {X.~S.}\ \bibnamefont {Li}}, \ and\
  \bibinfo {author} {\bibfnamefont {D.~H.}\ \bibnamefont {Bailey}},\ }in\
  \href@noop {} {\emph {\bibinfo {booktitle} {15th IEEE Symposium on Computer
  Arithmetic}}}\ (\bibinfo {year} {2000})\ pp.\ \bibinfo {pages}
  {155--162}\BibitemShut {NoStop}%
\bibitem [{\citenamefont {Lipowski}\ and\ \citenamefont
  {Lipowska}(2011)}]{DBLP:journals/corr/abs-1109-3627}%
  \BibitemOpen
  \bibfield  {author} {\bibinfo {author} {\bibfnamefont {A.}~\bibnamefont
  {Lipowski}}\ and\ \bibinfo {author} {\bibfnamefont {D.}~\bibnamefont
  {Lipowska}},\ }\href {http://arxiv.org/abs/1109.3627} {\bibfield  {journal}
  {\bibinfo  {journal} {CoRR}\ }\textbf {\bibinfo {volume} {abs/1109.3627}}
  (\bibinfo {year} {2011})},\ \Eprint {http://arxiv.org/abs/1109.3627}
  {arXiv:1109.3627} \BibitemShut {NoStop}%
\bibitem [{\citenamefont {Kleiss}\ and\ \citenamefont
  {Verheyen}(2016)}]{Kleiss:2016esx}%
  \BibitemOpen
  \bibfield  {author} {\bibinfo {author} {\bibfnamefont {R.}~\bibnamefont
  {Kleiss}}\ and\ \bibinfo {author} {\bibfnamefont {R.}~\bibnamefont
  {Verheyen}},\ }\href {\doibase 10.1140/epjc/s10052-016-4231-5} {\bibfield
  {journal} {\bibinfo  {journal} {Eur. Phys. J.}\ }\textbf {\bibinfo {volume}
  {C76}},\ \bibinfo {pages} {359} (\bibinfo {year} {2016})},\ \Eprint
  {http://arxiv.org/abs/1605.09246} {arXiv:1605.09246 [hep-ph]} \BibitemShut
  {NoStop}%
%%CITATION = ARXIV:1605.09246;%%
\bibitem [{\citenamefont {Dreyer}\ \emph {et~al.}(2018)\citenamefont {Dreyer},
  \citenamefont {Salam},\ and\ \citenamefont {Soyez}}]{Dreyer:2018nbf}%
  \BibitemOpen
  \bibfield  {author} {\bibinfo {author} {\bibfnamefont {F.~A.}\ \bibnamefont
  {Dreyer}}, \bibinfo {author} {\bibfnamefont {G.~P.}\ \bibnamefont {Salam}}, \
  and\ \bibinfo {author} {\bibfnamefont {G.}~\bibnamefont {Soyez}},\ }\href
  {\doibase 10.1007/JHEP12(2018)064} {\bibfield  {journal} {\bibinfo  {journal}
  {JHEP}\ }\textbf {\bibinfo {volume} {12}},\ \bibinfo {pages} {064} (\bibinfo
  {year} {2018})},\ \Eprint {http://arxiv.org/abs/1807.04758} {arXiv:1807.04758
  [hep-ph]} \BibitemShut {NoStop}%
%%CITATION = ARXIV:1807.04758;%%
\bibitem [{\citenamefont {Andrews}\ \emph {et~al.}(2018)\citenamefont {Andrews}
  \emph {et~al.}}]{Andrews:2018jcm}%
  \BibitemOpen
  \bibfield  {author} {\bibinfo {author} {\bibfnamefont {H.~A.}\ \bibnamefont
  {Andrews}} \emph {et~al.},\ }\href@noop {} {\  (\bibinfo {year} {2018})},\
  \Eprint {http://arxiv.org/abs/1808.03689} {arXiv:1808.03689 [hep-ph]}
  \BibitemShut {NoStop}%
%%CITATION = ARXIV:1808.03689;%%
\bibitem [{\citenamefont {Cacciari}\ \emph {et~al.}(2012)\citenamefont
  {Cacciari}, \citenamefont {Salam},\ and\ \citenamefont
  {Soyez}}]{Cacciari:2011ma}%
  \BibitemOpen
  \bibfield  {author} {\bibinfo {author} {\bibfnamefont {M.}~\bibnamefont
  {Cacciari}}, \bibinfo {author} {\bibfnamefont {G.~P.}\ \bibnamefont {Salam}},
  \ and\ \bibinfo {author} {\bibfnamefont {G.}~\bibnamefont {Soyez}},\ }\href
  {\doibase 10.1140/epjc/s10052-012-1896-2} {\bibfield  {journal} {\bibinfo
  {journal} {Eur. Phys. J.}\ }\textbf {\bibinfo {volume} {C72}},\ \bibinfo
  {pages} {1896} (\bibinfo {year} {2012})},\ \Eprint
  {http://arxiv.org/abs/1111.6097} {arXiv:1111.6097 [hep-ph]} \BibitemShut
  {NoStop}%
%%CITATION = ARXIV:1111.6097;%%
\bibitem [{\citenamefont {Larkoski}\ \emph {et~al.}(2020)\citenamefont
  {Larkoski}, \citenamefont {Moult},\ and\ \citenamefont
  {Nachman}}]{Larkoski:2017jix}%
  \BibitemOpen
  \bibfield  {author} {\bibinfo {author} {\bibfnamefont {A.~J.}\ \bibnamefont
  {Larkoski}}, \bibinfo {author} {\bibfnamefont {I.}~\bibnamefont {Moult}}, \
  and\ \bibinfo {author} {\bibfnamefont {B.}~\bibnamefont {Nachman}},\ }\href
  {\doibase 10.1016/j.physrep.2019.11.001} {\bibfield  {journal} {\bibinfo
  {journal} {Phys. Rept.}\ }\textbf {\bibinfo {volume} {841}},\ \bibinfo
  {pages} {1} (\bibinfo {year} {2020})},\ \Eprint
  {http://arxiv.org/abs/1709.04464} {arXiv:1709.04464 [hep-ph]} \BibitemShut
  {NoStop}%
%%CITATION = ARXIV:1709.04464;%%
\bibitem [{\citenamefont {Dokshitzer}\ \emph {et~al.}(1997)\citenamefont
  {Dokshitzer}, \citenamefont {Leder}, \citenamefont {Moretti},\ and\
  \citenamefont {Webber}}]{Dokshitzer:1997in}%
  \BibitemOpen
  \bibfield  {author} {\bibinfo {author} {\bibfnamefont {Y.~L.}\ \bibnamefont
  {Dokshitzer}}, \bibinfo {author} {\bibfnamefont {G.~D.}\ \bibnamefont
  {Leder}}, \bibinfo {author} {\bibfnamefont {S.}~\bibnamefont {Moretti}}, \
  and\ \bibinfo {author} {\bibfnamefont {B.~R.}\ \bibnamefont {Webber}},\
  }\href {\doibase 10.1088/1126-6708/1997/08/001} {\bibfield  {journal}
  {\bibinfo  {journal} {JHEP}\ }\textbf {\bibinfo {volume} {08}},\ \bibinfo
  {pages} {001} (\bibinfo {year} {1997})},\ \Eprint
  {http://arxiv.org/abs/hep-ph/9707323} {arXiv:hep-ph/9707323 [hep-ph]}
  \BibitemShut {NoStop}%
%%CITATION = HEP-PH/9707323;%%
\bibitem [{\citenamefont {Catani}\ \emph
  {et~al.}(1992{\natexlab{a}})\citenamefont {Catani}, \citenamefont {Turnock},\
  and\ \citenamefont {Webber}}]{Catani:1992jc}%
  \BibitemOpen
  \bibfield  {author} {\bibinfo {author} {\bibfnamefont {S.}~\bibnamefont
  {Catani}}, \bibinfo {author} {\bibfnamefont {G.}~\bibnamefont {Turnock}}, \
  and\ \bibinfo {author} {\bibfnamefont {B.~R.}\ \bibnamefont {Webber}},\
  }\href {\doibase 10.1016/0370-2693(92)91565-Q} {\bibfield  {journal}
  {\bibinfo  {journal} {Phys. Lett.}\ }\textbf {\bibinfo {volume} {B295}},\
  \bibinfo {pages} {269} (\bibinfo {year} {1992}{\natexlab{a}})}\BibitemShut
  {NoStop}%
%%CITATION = PHLTA,B295,269;%%
\bibitem [{\citenamefont {Brandt}\ \emph {et~al.}(1964)\citenamefont {Brandt},
  \citenamefont {Peyrou}, \citenamefont {Sosnowski},\ and\ \citenamefont
  {Wroblewski}}]{Brandt:1964sa}%
  \BibitemOpen
  \bibfield  {author} {\bibinfo {author} {\bibfnamefont {S.}~\bibnamefont
  {Brandt}}, \bibinfo {author} {\bibfnamefont {C.}~\bibnamefont {Peyrou}},
  \bibinfo {author} {\bibfnamefont {R.}~\bibnamefont {Sosnowski}}, \ and\
  \bibinfo {author} {\bibfnamefont {A.}~\bibnamefont {Wroblewski}},\ }\href
  {\doibase 10.1016/0031-9163(64)91176-X} {\bibfield  {journal} {\bibinfo
  {journal} {Phys. Lett.}\ }\textbf {\bibinfo {volume} {12}},\ \bibinfo {pages}
  {57} (\bibinfo {year} {1964})}\BibitemShut {NoStop}%
%%CITATION = PHLTA,12,57;%%
\bibitem [{\citenamefont {Farhi}(1977)}]{Farhi:1977sg}%
  \BibitemOpen
  \bibfield  {author} {\bibinfo {author} {\bibfnamefont {E.}~\bibnamefont
  {Farhi}},\ }\href {\doibase 10.1103/PhysRevLett.39.1587} {\bibfield
  {journal} {\bibinfo  {journal} {Phys. Rev. Lett.}\ }\textbf {\bibinfo
  {volume} {39}},\ \bibinfo {pages} {1587} (\bibinfo {year}
  {1977})}\BibitemShut {NoStop}%
%%CITATION = PRLTA,39,1587;%%
\bibitem [{\citenamefont {Dasgupta}\ and\ \citenamefont
  {Salam}(2002)}]{Dasgupta:2002bw}%
  \BibitemOpen
  \bibfield  {author} {\bibinfo {author} {\bibfnamefont {M.}~\bibnamefont
  {Dasgupta}}\ and\ \bibinfo {author} {\bibfnamefont {G.~P.}\ \bibnamefont
  {Salam}},\ }\href {\doibase 10.1088/1126-6708/2002/03/017} {\bibfield
  {journal} {\bibinfo  {journal} {JHEP}\ }\textbf {\bibinfo {volume} {03}},\
  \bibinfo {pages} {017} (\bibinfo {year} {2002})},\ \Eprint
  {http://arxiv.org/abs/hep-ph/0203009} {arXiv:hep-ph/0203009 [hep-ph]}
  \BibitemShut {NoStop}%
%%CITATION = HEP-PH/0203009;%%
\bibitem [{\citenamefont {Dokshitzer}\ \emph {et~al.}(1998)\citenamefont
  {Dokshitzer}, \citenamefont {Lucenti}, \citenamefont {Marchesini},\ and\
  \citenamefont {Salam}}]{Dokshitzer:1998kz}%
  \BibitemOpen
  \bibfield  {author} {\bibinfo {author} {\bibfnamefont {Y.~L.}\ \bibnamefont
  {Dokshitzer}}, \bibinfo {author} {\bibfnamefont {A.}~\bibnamefont {Lucenti}},
  \bibinfo {author} {\bibfnamefont {G.}~\bibnamefont {Marchesini}}, \ and\
  \bibinfo {author} {\bibfnamefont {G.~P.}\ \bibnamefont {Salam}},\ }\href
  {\doibase 10.1088/1126-6708/1998/01/011} {\bibfield  {journal} {\bibinfo
  {journal} {JHEP}\ }\textbf {\bibinfo {volume} {01}},\ \bibinfo {pages} {011}
  (\bibinfo {year} {1998})},\ \Eprint {http://arxiv.org/abs/hep-ph/9801324}
  {arXiv:hep-ph/9801324 [hep-ph]} \BibitemShut {NoStop}%
%%CITATION = HEP-PH/9801324;%%
\bibitem [{\citenamefont {Catani}\ \emph {et~al.}(1993)\citenamefont {Catani},
  \citenamefont {Trentadue}, \citenamefont {Turnock},\ and\ \citenamefont
  {Webber}}]{Catani:1992ua}%
  \BibitemOpen
  \bibfield  {author} {\bibinfo {author} {\bibfnamefont {S.}~\bibnamefont
  {Catani}}, \bibinfo {author} {\bibfnamefont {L.}~\bibnamefont {Trentadue}},
  \bibinfo {author} {\bibfnamefont {G.}~\bibnamefont {Turnock}}, \ and\
  \bibinfo {author} {\bibfnamefont {B.~R.}\ \bibnamefont {Webber}},\ }\href
  {\doibase 10.1016/0550-3213(93)90271-P} {\bibfield  {journal} {\bibinfo
  {journal} {Nucl. Phys.}\ }\textbf {\bibinfo {volume} {B407}},\ \bibinfo
  {pages} {3} (\bibinfo {year} {1993})}\BibitemShut {NoStop}%
%%CITATION = NUPHA,B407,3;%%
\bibitem [{\citenamefont {Catani}\ \emph
  {et~al.}(1991{\natexlab{b}})\citenamefont {Catani}, \citenamefont
  {Dokshitzer}, \citenamefont {Olsson}, \citenamefont {Turnock},\ and\
  \citenamefont {Webber}}]{Catani:1991hj}%
  \BibitemOpen
  \bibfield  {author} {\bibinfo {author} {\bibfnamefont {S.}~\bibnamefont
  {Catani}}, \bibinfo {author} {\bibfnamefont {Y.~L.}\ \bibnamefont
  {Dokshitzer}}, \bibinfo {author} {\bibfnamefont {M.}~\bibnamefont {Olsson}},
  \bibinfo {author} {\bibfnamefont {G.}~\bibnamefont {Turnock}}, \ and\
  \bibinfo {author} {\bibfnamefont {B.~R.}\ \bibnamefont {Webber}},\ }\href
  {\doibase 10.1016/0370-2693(91)90196-W} {\bibfield  {journal} {\bibinfo
  {journal} {Phys. Lett.}\ }\textbf {\bibinfo {volume} {B269}},\ \bibinfo
  {pages} {432} (\bibinfo {year} {1991}{\natexlab{b}})}\BibitemShut {NoStop}%
%%CITATION = PHLTA,B269,432;%%
\bibitem [{\citenamefont {Catani}\ \emph
  {et~al.}(1992{\natexlab{b}})\citenamefont {Catani}, \citenamefont
  {Dokshitzer}, \citenamefont {Fiorani},\ and\ \citenamefont
  {Webber}}]{Catani:1991pm}%
  \BibitemOpen
  \bibfield  {author} {\bibinfo {author} {\bibfnamefont {S.}~\bibnamefont
  {Catani}}, \bibinfo {author} {\bibfnamefont {Y.~L.}\ \bibnamefont
  {Dokshitzer}}, \bibinfo {author} {\bibfnamefont {F.}~\bibnamefont {Fiorani}},
  \ and\ \bibinfo {author} {\bibfnamefont {B.~R.}\ \bibnamefont {Webber}},\
  }\href {\doibase 10.1016/0550-3213(92)90296-N} {\bibfield  {journal}
  {\bibinfo  {journal} {Nucl. Phys.}\ }\textbf {\bibinfo {volume} {B377}},\
  \bibinfo {pages} {445} (\bibinfo {year} {1992}{\natexlab{b}})}\BibitemShut
  {NoStop}%
%%CITATION = NUPHA,B377,445;%%
\bibitem [{\citenamefont {Brown}\ and\ \citenamefont
  {Stirling}(1990)}]{Brown:1990nm}%
  \BibitemOpen
  \bibfield  {author} {\bibinfo {author} {\bibfnamefont {N.}~\bibnamefont
  {Brown}}\ and\ \bibinfo {author} {\bibfnamefont {W.~J.}\ \bibnamefont
  {Stirling}},\ }\href {\doibase 10.1016/0370-2693(90)90502-W} {\bibfield
  {journal} {\bibinfo  {journal} {Phys. Lett.}\ }\textbf {\bibinfo {volume}
  {B252}},\ \bibinfo {pages} {657} (\bibinfo {year} {1990})}\BibitemShut
  {NoStop}%
%%CITATION = PHLTA,B252,657;%%
\bibitem [{\citenamefont {Forshaw}\ \emph {et~al.}(2006)\citenamefont
  {Forshaw}, \citenamefont {Kyrieleis},\ and\ \citenamefont
  {Seymour}}]{Forshaw:2006fk}%
  \BibitemOpen
  \bibfield  {author} {\bibinfo {author} {\bibfnamefont {J.~R.}\ \bibnamefont
  {Forshaw}}, \bibinfo {author} {\bibfnamefont {A.}~\bibnamefont {Kyrieleis}},
  \ and\ \bibinfo {author} {\bibfnamefont {M.~H.}\ \bibnamefont {Seymour}},\
  }\href {\doibase 10.1088/1126-6708/2006/08/059} {\bibfield  {journal}
  {\bibinfo  {journal} {JHEP}\ }\textbf {\bibinfo {volume} {08}},\ \bibinfo
  {pages} {059} (\bibinfo {year} {2006})},\ \Eprint
  {http://arxiv.org/abs/hep-ph/0604094} {arXiv:hep-ph/0604094 [hep-ph]}
  \BibitemShut {NoStop}%
%%CITATION = HEP-PH/0604094;%%
\bibitem [{\citenamefont {Catani}\ \emph {et~al.}(2012)\citenamefont {Catani},
  \citenamefont {de~Florian},\ and\ \citenamefont {Rodrigo}}]{Catani:2011st}%
  \BibitemOpen
  \bibfield  {author} {\bibinfo {author} {\bibfnamefont {S.}~\bibnamefont
  {Catani}}, \bibinfo {author} {\bibfnamefont {D.}~\bibnamefont {de~Florian}},
  \ and\ \bibinfo {author} {\bibfnamefont {G.}~\bibnamefont {Rodrigo}},\ }\href
  {\doibase 10.1007/JHEP07(2012)026} {\bibfield  {journal} {\bibinfo  {journal}
  {JHEP}\ }\textbf {\bibinfo {volume} {07}},\ \bibinfo {pages} {026} (\bibinfo
  {year} {2012})},\ \Eprint {http://arxiv.org/abs/1112.4405} {arXiv:1112.4405
  [hep-ph]} \BibitemShut {NoStop}%
%%CITATION = ARXIV:1112.4405;%%
\end{thebibliography}%

%======================================================================
%======================================================================
%======================================================================
%======================================================================
%======================================================================
%======================================================================
%======================================================================
%
% Include supplemental material
%

\newpage

\onecolumngrid
\newpage
\appendix

\section*{Supplemental material}

\makeatletter
\renewcommand\@biblabel[1]{[#1S]}
\makeatother

The main purpose of this supplement is to summarise material that,
while not critical to the message and understanding of the results of
the manuscript, may be of benefit to those wishing to reproduce our
results or consult additional evidence beyond the summary plot,
Fig.~\ref{fig:limits-y3}.

Section~\ref{sec:expl-moment-maps} gives the fully worked out
kinematic maps and the specific choices we made for the transition
functions $g(\bar\eta)$ in Eq.~(\ref{eq:2}).
Section~\ref{sec:toy-shower} outlines a toy version of the (standard)
dipole and new parton showers. It provides a useful additional diagnostic 
tool for understanding transverse recoil effects and their impact on
logarithmic accuracy.  Note that the results presented in the main
text have been obtained with full showers, not the toy versions. The
toy versions serve mainly to demonstrate that we have fully understood
the origin of the standard dipole-shower deviations from NLL. The toy
versions are directly relevant to the results presented in the main
text only insofar as they provide the basis for the amber colouring of
5 entries in Fig.~\ref{fig:limits-y3} right.
%
% Section~\ref{sec:toy-shower} explains the nature of our toy shower, a
% useful intermediate step for understanding transverse recoil effects
% and their impact on logarithmic accuracy.
%
Section~\ref{sec:super-leading-logs} explains the origin of the
super-leading logarithmic terms found with that toy shower for
standard dipole recoil.
Section~\ref{sec:ee-lund} outlines how we have adapted the
hadron-collider definition of the Lund
declustering~\cite{Dreyer:2018nbf} to the $e^+e^-$ case, including the
expected result for the normalisation of the $\Delta \psi_{12}$
distribution. 
One of our test observables, the subjet multiplicity, differs in the
structure of the logarithms from other observables, so that is
summarised in Section~\ref{sec:multiplicities}.
Finally, section~\ref{sec:taking-limits} summarises technical
challenges that arise when taking $\as$ towards zero with finite
$\as L$ and solutions we adopted, which may be of interest to others who
wish to embark on a similar path.

%----------------------------------------------------------------------
\subsection{Explicit formulae for the PanLocal and PanGlobal showers}
\label{sec:expl-moment-maps}
In this section we report the explicit equations for the kinematic
maps and the branching probabilities of the two NLL showers given in
the letter. 
All Jacobian factors associated with the kinematic maps given in this
section have the property that they tend to unity in singular regions
of the phase space. For this reason, we omit them in our phase space
parametrisation.

\subsubsection{The PanLocal dipole shower}

The kinematic map for the emission of $p_k$ off the dipole
$\{\ptilde_i,\ptilde_j\}$ is defined in Eq.~\eqref{eq:IIc-split-map},
where $f=1\,(0)$ for the $P_{\itilde \to ik}(a_k)$ ($P_{\jtilde \to
  jk}(b_k)$) term in Eq.~(\ref{eq:2}). 
The difference with the
common dipole-local recoil scheme is that the transition takes place
at $\bar\eta=0$, which corresponds to equal angles between $\ptilde_i$ and
$\ptilde_j$ in the event centre-of-mass frame.

For simplicity, we give explicit formulas just for the $f=1$ case,
corresponding physically to a situation where $p_k$ is emitted from
$\ptilde_i$.
The extension to $f=0$ is straightforward.
The coefficients $a_{i,j}$ and $b_{i,j}$ are given by
\begin{align}
a_{i} 
        =1-a_{k}\,;\quad b_{i} =\frac{a_{k}b_{k}}{a_i}\,;\quad
a_{j} =0\,;\quad b_{j} =\frac{a_{i}-b_{k}}{a_i}\,,\label{eq:IIc-ai-bi-aj-bj}
\end{align}
where the coefficients $a_k$ and $b_k$ are defined in
Eq.~\eqref{eq:IIc-ak-bk-Sudakov-coeffs}.
The phase space boundaries,
requiring the conservation of the dipole invariant mass and
that all partons remain on shell, map to the following condition,
\begin{equation}
a_k + b_k < 1\,.
\end{equation}
The splitting functions $P_{\itilde \to ik}(z)$ of the evolution
equation~\eqref{eq:2} describe the splitting of the parton $\ptilde_i$
into the partons $p_i$ and $p_k$. If $\ptilde_i$ is a gluon, in the
large-$N_C$ limit its splitting function is shared between two
contiguous dipoles such that
\begin{equation}
\label{eq:half-splitting}
P_{g\rightarrow \text{X}}(z) = \frac12 P^\text{asym.}_{g\rightarrow
  \text{X}}(z) + \frac12 P^\text{asym.}_{g\rightarrow \text{X}}(1-z)\,,
\end{equation}
and $\ptilde_i$ emits according to $\frac12 P^\text{asym.}_{g\rightarrow
  \text{X}}$ in each of the two dipoles.
The splitting functions are given by
\begin{align}
\label{eq:splitting-functions-dipole}
  P_{q\rightarrow q+g}(z) = C_F \frac{1+(1-z)^2}{z}\,;\quad
\frac12 P^\text{asym.}_{g\rightarrow
  q+\overline{q}}(z) = n_F T_R\, (1-z)^2\,;\quad
\frac12 P^\text{asym.}_{g\rightarrow g+g}(z) =\frac{C_A}{2}\frac{1+(1-z)^3}{z}\,,
\end{align}
where $z$ is the momentum fraction (i.e. $a_k$ or $b_k$) of the
emitted parton.
The asymmetric versions of the splitting functions that we adopt are
one of a continuous class of choices that we could have made that are
positive definite and satisfy Eq.~(\ref{eq:half-splitting}).

We finally comment on the function $g(\bar\eta)$ in
Eq.~\eqref{eq:2}.
The main constraint that one has is that
$g(\bar\eta)+g(-\bar\eta)=1$, and $g(\bar\eta)=0$ ($1$) for sufficiently negative
(positive) $\bar\eta$, while smoothly transitioning around
$\bar\eta=0$.
Several choices are of course possible, and we adopt the
following one,
\begin{subequations}
  \label{eq:IIc-g}
  \begin{align}
    \bar\eta < -1:~ g(\bar\eta) = 0\,;\qquad
%    \\
    -1 < \bar\eta < 1:~ g(\bar\eta) = \frac{15}{16} \left(\frac{\bar\eta ^5}{5}-\frac{2 \bar\eta ^3}{3}+\bar\eta +\frac{8}{15}\right)\,;\qquad
%    \\
    \bar\eta > +1:~ g(\bar\eta) = 1\,.
  \end{align}
\end{subequations}
We have in particular opted for a function that saturates (smoothly)
at $|\bar\eta|=1$, so as to avoid unphysical recoil assignments leaking
far in rapidity. 

\subsubsection{The PanLocal antenna shower}

The antenna variant of the PanLocal shower differs from the dipole
version just described in the distribution of the transverse recoil
across the dipole, parametrised by the function $f$ in
Eq.~\eqref{eq:IIc-split-map}. Specifically, we adopt the choice
\begin{equation}
\label{eq:fofeta}
f \equiv f(\bar\eta) = \frac{e^{2\bar\eta}}{1+e^{2\bar\eta}}\,.
\end{equation}
As a result, the explicit form of the mapping is considerably more
involved than that for the PanLocal dipole shower. For an emission
$p_k$ off the dipole $\{\ptilde_i,\ptilde_j\}$, the coefficients
$a_{i,j}$ and $b_{i,j}$ of Eq.~\eqref{eq:IIc-split-map} are
\begin{subequations} \label{eq:IIb-ai-bi-aj-bj-solns}
\begin{align}
a_{i} & =\frac{(\sqrt{\lambda_{1}}+\sqrt{\lambda_{2}})^{2}+4f^{2}}{4(1-b_{k})}\,a_{k}b_{k}\,,\label{eq:IIb-ai-bi-aj-bj-solns-ai}\\
b_{i} & =\frac{(\sqrt{\lambda_{1}}-\sqrt{\lambda_{2}})^{2}+4f^{2}}{4(1-a_{k})}\,a_{k}b_{k}\,,\label{eq:IIb-ai-bi-aj-bj-solns-bi}\\
a_{j} & =\frac{(\sqrt{\lambda_{1}}-\sqrt{\lambda_{2}})^{2}+4(1-f)^{2}}{4(1-b_{k})}\,a_{k}b_{k}\,,\label{eq:IIb-ai-bi-aj-bj-solns-aj}\\
b_{j} & =\frac{(\sqrt{\lambda_{1}}+\sqrt{\lambda_{2}})^{2}+4(1-f)^{2}}{4(1-a_{k})}\,a_{k}b_{k}\,,\label{eq:IIb-ai-bi-aj-bj-solns-bj}
\end{align}
\end{subequations} 
where $a_{k}$, $b_{k}$ are given in
Eq.~\eqref{eq:IIc-ak-bk-Sudakov-coeffs}, and
\begin{subequations} \label{eq:IIb-lambda-lambda1-lambda2}
\begin{align}
\label{eq:IIb-lambda}
\lambda_{1}  =(1-a_k-b_k)/(a_k b_k)\,;\qquad
\lambda_{2}  = \lambda_{1}  + 4 f(1-f)\,.
\end{align}
\end{subequations}
For $f=1$ the above equations reproduce the map adopted for the
PanLocal dipole shower~\eqref{eq:IIc-ai-bi-aj-bj}.
Moreover, analogously to the PanLocal dipole shower, the phase space
boundaries are obtained by requiring
\begin{equation}
a_k + b_k < 1\,.
\end{equation}
The splitting functions are those defined in the previous section for
the PanLocal dipole shower, while the function $g(\bar\eta)$ for the
PanLocal antenna shower is chosen as
$g(\bar\eta) = 1 - g(-\bar\eta) = f(\bar\eta)$.

\subsubsection{The PanGlobal shower}
The PanGlobal shower differs from the PanLocal design in that only the
longitudinal recoil is distributed locally among the dipole ends as in
Eq.~\eqref{eq:global-map-step-1}. The transverse recoil is
handled as follows.
Having constructed the intermediate post-branching momenta as in
Eq.~\eqref{eq:global-map-step-1}, we evaluate their sum, 
\begin{equation}
  \bar{Q}
  \,
  =
  \,
  \sum_{i\in\text{event}}\bar{p}_{i}
  \,
  =
  \,
  Q+k_{\perp}
  \,
  ,\label{eq:IIIa-barQ-def}
\end{equation}
and then rescale each momentum in the event by a factor
\begin{equation}
r=\sqrt{\frac{Q^{2}}{\bar{Q}^{2}}}\,,\label{eq:IIIa-r-defn} 
\end{equation}
to yield \begin{subequations} \label{eq:IIIa-split-map-step-2} 
\begin{align}
  p_{k}^{\prime} & =ra_{k}\tilde{p}_{i}+rb_{k}\tilde{p}_{j}+rk_{\perp}\,,\\
  p_{i}^{\prime} & =r(1-a_{k})\tilde{p}_{i}\,,\\
  p_{j}^{\prime} & =r(1-b_{k})\tilde{p}_{j}\,,\\
  p_{l}^{\prime} & =r\tilde{p}_{l}
  \text{ }
  \forall
  \text{ }
  \tilde{p}_{l}\notin\{\tilde{p}_{i},\tilde{p}_{j}\}
  \,.
\end{align}
\end{subequations} Finally we construct the following Lorentz boost
\begin{equation}
  \mathbb{B}_{\text{ }\nu}^{\mu}
  =
  \left[
    g_{\text{ }\nu}^{\mu}
    +
    \frac{2Q^{\mu}Q_{\nu}^{\prime}}{Q^{2}}
    -
    \frac{2\left(Q+Q^{\prime}\right)^{\mu}\left(Q+Q^{\prime}\right)_{\nu}}
         {\left(Q+Q^{\prime}\right)^{2}}\right]
  \,,
  \qquad 
  Q^{\prime}
  =
  r\,\bar{Q}
  \,
  ,\label{eq:IIIa-Lambda-defn-in-terms-of-Q-Qbar-rho}
\end{equation}
and apply it to the latter, primed momenta to obtain the post
branching kinematics:
\begin{subequations}
\label{eq:IIIa-split-map-step-3}
\begin{align}
  p_{k} & =
  \mathbb{B}\left[ra_{k}\tilde{p}_{i}+rb_{k}\tilde{p}_{j}+rk_{\perp}\right]\,,\\
  p_{i} & =
  \mathbb{B}\left[r(1-a_{k})\tilde{p}_{i}\right]\,,\\
  p_{j} & =
  \mathbb{B}\left[r(1-b_{k})\tilde{p}_{j}\right]\,,\\
  p_{l} & =
  \mathbb{B}\left[r\tilde{p}_{l}\right]
  \text{ }
  \forall
  \text{ }
  \tilde{p}_{l}\notin\{\tilde{p}_{i},\tilde{p}_{j}\}
  \,.
\end{align}
\end{subequations}
The phase space boundaries are simply set by imposing
\begin{equation}
a_k < 1\,,\quad b_k < 1\,.
\end{equation}
The splitting functions are given in
Eq.~\eqref{eq:splitting-functions-dipole}, while for the $g(\bar\eta)$
function we adopt the same choice as for the PanLocal antenna shower,
namely $g(\bar\eta) = 1 - g(-\bar\eta) = f(\bar\eta)$, with
$f(\bar\eta)$ given in Eq.~\eqref{eq:fofeta}.

%----------------------------------------------------------------------
\subsection{Toy shower formulation and results}
\label{sec:toy-shower}

The toy shower is intended to provide a simple way to determine
expectations for logarithmic effects in showers with dipole-local
recoil.
The shower functions in a soft approximation, and uses a fixed
coupling.
Moreover, it only considers the emission of primary radiation, while
neglecting any secondary branchings.
These approximations help make the toy model sufficiently simple that
it becomes a powerful tool for examining and understanding the
interplay between event-shape type observables and parton showers.

Rather than storing 4-momenta, for each particle one stores a
2-dimensional transverse momentum and a rapidity and determines
observables directly from those variables.
The rapidity is to be understood in a primary Lund-declustering sense, i.e.\
as related to the opening angle of the emission from its
emitter. 

\subsubsection{Shower definitions}
Our notation is as follows: $\boldp_{\perp,i}$ and $\eta_i$ are,
respectively, the (two-dimensional) vector transverse momentum and the
rapidity of particle $i$ at a given stage of the evolution.
We will use quantities with a hat, $\boldsymbol{\hat
  p}_{\perp,i}$ and $\hat \eta_i$, to denote the values when the
particle was first created.
When discussing the effect of a specific branching on other
emissions, $\widetilde{\boldp}_{\perp, i}$ and
$\widetilde{\eta}_i$ will be the values before the branching, while
$\boldp_{\perp,i}$ and $\eta_i$ will be those after the
branching.
Emissions are maintained in a dipole chain that is ordered in
increasing $\hat \eta_i$.
When we write a dipole as $[ab]$ it is implicit that $\hat \eta_a <
\hat \eta_b$.

We start from a $q\bar q$ system, where $q$ ($\bar q$) has infinite
positive (negative) rapidity and zero transverse momentum, and an
evolution scale $\ln v = 0$ (we implicitly have a centre of mass
energy set to $Q=1$).
The evolution variable $v$ maps to $\ln \hat p_{\perp}$ and $\hat\eta$ as
$\ln v = \ln \hat p_{\perp}- \beta |\hat\eta|$,
where the
$\beta$-dependent term is relevant only for the PanLocal and PanGlobal
showers. 
This form matches the kinematic pattern of primary-emission generation
for the full showers, both dipole and PanLocal / PanGlobal.
The probability of evolving from $\state_n \to \state_{n+1}$ in a
given slice $d\ln v$ of evolution variable is
\begin{equation}
  \label{eq:toy-splitting}
  \frac{d\mathcal{P}_{n\to n+1}}{d\ln v } =
  \abar \int_{-\eta_\text{max}}^{+\eta_\text{max}} d\hat\eta_{n+1} \frac{d\hat\phi_{n+1}}{2\pi}\,,
  \qquad \eta_{\max} = -\frac{\ln v}{1+\beta}\,,
  \qquad \abar \equiv \frac{2 \as C_F}{\pi}.
\end{equation}
Given an $\ln v$ for emission $n+1$, the same phase space allows one to
select $\hat\eta_{n+1}$ and $\hat\phi_{n+1}$ and then one inserts the emission
into the dipole chain, by identifying the partons that have
$\hat \eta$ immediately below and above $\hat\eta_{n+1}$.
This structure applies to all our toy showers and they differ only
through the recoil prescription that is adopted when inserting an
emission.

\paragraph{Recoil for dipole showers.} For $k_t$ ordered showers with
dipole-local recoil (the toy analogue is as given already in
Appendix~B of Ref.\cite{Dasgupta:2018nvj}), like Dire and Pythia8,
emission $n+1$ will induce recoil in either the left or right-hand
member of the dipole into which it is inserted.
Generically we encode the recoil as $(k,s)$ where $k$ is the index of
the recoiling parton and $s$ the sign of its rapidity modification.
The recoiling parton is modified as
\begin{equation}
  \label{eq:generic-recoil}
  \boldp_{\perp, k}=
  \widetilde{\boldp}_{\perp, k} - \hat{\boldp}_{\perp, n+1};
  \qquad
  \eta_k = \widetilde{\eta}_k + s
  \ln\frac{|\widetilde{\boldp}_{\perp, k}
    - \hat{\boldp}_{\perp, n+1}|}{|\widetilde{\boldp}_{\perp, k} |}\,.\\
\end{equation}
If $k$ is quark or anti-quark, the rapidities are infinite and so the
rapidity modification is irrelevant.
For any given dipole, we define a mid-rapidity point
%\gps{KH comment: stay with mid?}
\begin{subequations}
  \begin{align}
    \text{$[\bar qg_i]$ dipole:}
    &\quad \eta_\text{mid} = \frac{1}{2}\left(\ln \hat{p}_{\perp, i}+\hat{\eta}_i\right),
    \\
    \text{$[g_i g_j]$ dipole:}&\quad \eta_\text{mid} =
                                \frac{1}{2}\left(\ln \frac{{\hat p}_{\perp,
                                j}}{{\hat p}_{\perp,
                                i}}+{\hat \eta}_i + {\hat \eta}_j\right),
    \\
    \text{$[g_i q]$ dipole:}&\quad \eta_\text{mid} =
                              \frac{1}{2}\left(-\ln {\hat p}_{\perp, i}+{\hat \eta}_i\right) .
  \end{align}%
  \label{eq:midpoints}%
\end{subequations}%
If $\hat\eta_{n+1}$ is below (above) $\eta_\text{mid}$ then we take $k$ in
Eq.~(\ref{eq:generic-recoil}) to be the index of the dipole member at
smaller (larger) rapidity and $s=+1$ ($s=-1$).
Note that we have made an explicit choice to use the $\hat p$
variables in defining the midpoints in Eq.~(\ref{eq:midpoints}).
The full shower is arguably closer to choosing the $\tilde p$
variables instead.
However in terms of probing single-logarithmic effects, both choices
should give equivalent answers (they differ only in a region of
logarithmic width $\order{1}$ close to the midpoint), and $\hat p$
variables are a little simpler computationally.
Note also that we have a discrete transition in the recoil being
assigned to one parton or the other, whereas full dipole showers have
a smooth transition (though the Pythia8 and Dire algorithms implement
that smooth transition differently).
Again the difference is a sub-leading logarithmic effect, because it
affects a region of rapidity width of order $1$, i.e.\ without
logarithmic enhancements.
\medskip
\paragraph{Recoil for the PanLocal showers.}
If the emission has positive (negative) $\hat \eta$, the recoil is assigned to
the dipole member with more positive (negative) $\hat \eta$, and with
$s=-1$ ($s=+1$).
In the full shower this is equivalent to saying that if the emission
is forward (backward) in the event frame, the recoil is assigned to the
more forward (backward) of the two dipole members.
This is the same for both the dipole and antenna variants of the
PanLocal shower.
\medskip

\paragraph{Recoil for the PanGlobal shower.}
In the soft limit of the PanGlobal shower, transverse recoil from any
given emission is assigned in equal fractions to the quark and
anti-quark.\medskip

\paragraph{Recoil for the CAESAR reference.}
It will be useful to compare results to known resummations from the
CAESAR approach~\cite{Banfi:2004yd}.
The CAESAR reference results for recursively infrared and collinear
safe global observables can be obtained by setting $k$ to be the
anti-quark (quark) for $\hat\eta_{n+1} <0$ ($\hat\eta_{n+1} >0$).

%......................................................................
\subsubsection{Observables}

We have two main classes of observables, each parameterised by a
quantity $\beta$ which determines their angular dependence.
(In the main text we have used $\beta_\text{obs}$, to distinguish this
parameter from the $\beta$ used in the shower evolution variable; in
this section, to keep the notation more compact, we drop the
``$\text{obs}$'' subscript, but with the understanding that $\beta$
here, when used in the context of an observable, always means
$\beta_\text{obs}$.)
One class, $S_\beta$, is additive in the contributions from different
emissions and at NLL accuracy it is equivalent to the
energy-correlation moments $FC_{1-\beta}$ (Appendix~I of
Ref.~\cite{Banfi:2004yd})
\begin{equation}
  \label{eq:FC-S-toy}
  FC_{1-\beta} = S_\beta = \sum_{i\notin q \bar q} p_{\perp,i}
  e^{-\beta |\eta_i|}\,.
\end{equation}
The thrust is equivalent to $S_{\beta=1}$ at NLL accuracy.
The other class takes the maximum across all emissions
\begin{equation}
  \label{eq:M-toy}
  M_\beta = \max_{i\notin q \bar q} \left\{p_{\perp,i} e^{-\beta |\eta_i|} \right\},
\end{equation}
and at NLL accuracy the angular-ordered Durham (or Cambridge)
$\sqrt{y_{23}} = M_{\beta=0}$.
We have also defined left (right) hemisphere versions of these
observables, $S_{L,\beta}$ and $M_{L,\beta}$ ($S_{R,\beta}$ and
$M_{R,\beta}$), where we restrict the sum or maximum to particles with
$\eta_i <0$ ($\eta_i >0$).
Additionally, defining a hemisphere vector transverse sum $V_{L,0}$
\begin{equation}
  \label{eq:V-toy}
  V_{L,0} = \left|\sum_{i\notin q \bar q} \boldp_{\perp,i}\,\Theta(-\eta_i)\right|\,,
\end{equation}
and analogously for $V_{R,0}$, we can write the jet
broadenings~\cite{Catani:1992jc} as 
\begin{equation}
  \label{eq:B-toy}
  B_{X} = \frac12\left(S_{X,0} + V_{X,0}\right)
  \text{ [for }X=L,R]\,,
  \qquad
  B_T = B_L + B_R\,,
  \qquad
  B_W = \max\{B_L,B_R\}\,.
\end{equation}
Finally we define the scalar transverse momentum sum in a slice of
half-width $\Delta$ centred at zero rapidity
\begin{equation}
  \label{eq:slice-toy}
  S^\text{slice}_{\beta,\Delta} = \sum_{i\notin q \bar q} p_{\perp,i}
  e^{-\beta |\eta_i|} \Theta(\Delta - |\eta_i|)\,,
\end{equation}
evaluated with $\beta = 0$.

Note that the toy shower does not capture non-global logarithms for slice
and hemisphere observables, since the non-global logarithms are
induced by secondary emissions (which, we recall, are not included in
the toy shower).
However, it is still of interest for these observables, because it can
diagnose the presence of at least some classes of super-leading
logarithms in showers with dipole-local recoil, as discussed below.

%......................................................................
\subsubsection{All-order results}
\begin{figure}
  \centering
  \includegraphics[width=0.3\textwidth,page=1]{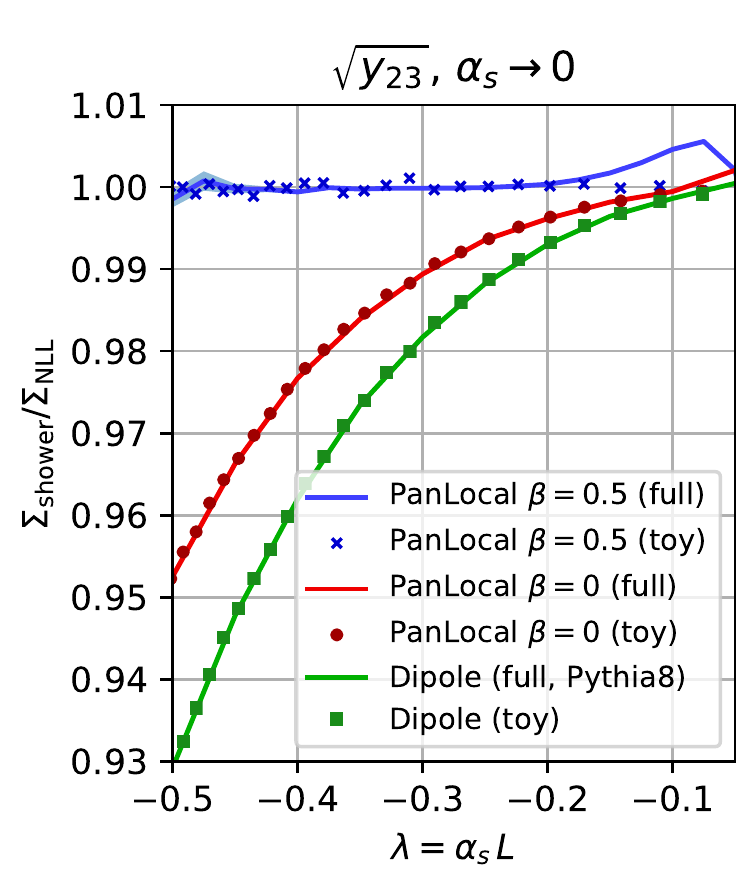}%
  \includegraphics[width=0.3\textwidth,page=3]{plots/toy-full-allorder-comparison.pdf}%
  \includegraphics[width=0.3\textwidth,page=4]{plots/toy-full-allorder-comparison.pdf}
  \caption{Comparison of the ratio
    $\Sigma_\text{shower}/\Sigma_\text{NLL}$ between the toy shower
    and the full shower for three reference observables ($\sqrt{y_{23}}$,
    $B_W$ and $\text{FC}_1$), in the limit $\as \to 0$, as a function
    of $\as L$.  %
    For the full showers the figure shows the ratio of the shower
    prediction to the full NLL result, while for the toy shower it
    shows the ratio to the CAESAR-like toy shower.
    Three full showers are shown in each plot, each compared to the
    corresponding toy shower.
    The PanLocal full showers are shown in their dipole variants
    (identical conclusions hold for the antenna variant).
    Small $(0.5\%$) issues at $\lambda \gtrsim -0.1$ are a consequence
    of the fact that for the largest of the $\as$ values used in the
    extrapolation, the corresponding $L$ values do not quite satisfy
    $e^{L} \ll 1$.  }
  \label{fig:toy-allorder-summary}
\end{figure}

The toy showers can be used as simplified models of the full showers, to
derive quantitative expectations for their logarithmic accuracy.
As an example, in
Fig.~\ref{fig:toy-allorder-summary} we show a comparison between some
toy showers (the PanLocal shower with $\beta
= 0$ and
$\beta=1/2$,
and the Dipole shower) and the corresponding full showers for three
observables: the angular-ordered Durham
$\sqrt{y_{23}}$
resolution scale (left plot), the wide jet broadening $B_W$
(middle plot), and a moment of energy-energy correlation
$\text{FC}_{1}$ (right plot).
Specifically, the three plots show the ratio
$\Sigma_\text{shower}/\Sigma_\text{NLL}$ in the limit $\as\to 0$ as a
function of $\lambda=\as L$.
$\Sigma_\text{NLL}$
stands for the correct NLL prediction for each
observable.
In the toy shower case, NLL is to be understood in a sense that 
includes neither running coupling, nor hard-collinear effects, while
it does include transverse recoil effects.
The NLL prediction is obtained with the toy shower itself using the
CAESAR-type recoil described above, though we have also checked that
it is consistent with analytical calculations that use the same
simplifications.

Given that the toy shower doesn't include running coupling, a
translation is needed between $\alpha_s L$ in the full shower and that
in the toy shower.
For global observables with $\beta_\text{obs}=0$, as shown in
Fig.~\ref{fig:toy-allorder-summary}, that translation is unambiguous
and is given by
\begin{equation}
  \label{eq:7}
  \as L^\text{toy} = \frac{\as L^\text{full}}{1 - 2 b_0\as L^\text{full}}\,,
\end{equation}
where $b_0 = \frac{11 C_A - 2n_f}{12\pi}$.

In all cases in Fig.~\ref{fig:toy-allorder-summary} we observe that
the prediction of the toy shower agrees well with the full shower
result.
An exception is given by the small
$|\lambda|$ region where some small ($0.5\%$) discrepancies between
the toy and full shower are due to the fact that, for some of the
$\as$ values used in the $\as \to 0$ extrapolation, the value of $|L|$
is not sufficiently large for non-logarithmic effects to be entirely
negligible in this region (e.g.\ relative $e^{L}$
contributions). 

We have carried out analogous tests for all of the observable/shower
combinations analysed in this article.
For the $\beta_\text{obs}=0$ cases, one can perform a direct
comparison between the full shower and the toy shower using
Eq.~(\ref{eq:7}) and one obtains good agreement.
For $\beta_\text{obs}>0$, there is no direct translation for
running-coupling effects, however these cases all show agreement with
NLL, both in the full and toy showers.

%......................................................................
\subsubsection{Fixed-order results}

In all-order results, whether for the full shower or the toy shower,
a key difficulty is to unambiguously sort through different
logarithmic orders.
The limit $\as \to 0$ that we use is essential in this respect, but
the accessible range of $\as$ values remains limited, and a
fixed-order expansion provides an alternative way of identifying
parametrically dominant logarithmic terms that might come with a small
coefficient.

The expansion of the cross section $\Sigma(L) = \sum_{n=0} \abar^n
\Sigma_n(L)$ for an observable $V$ to be below
some threshold $e^{L}$ can be written as follows
\begin{equation}
  \label{eq:toy-expansion-master}
  \Sigma(L) = 
  \sum_{m,n}
  \left[\frac{1}{m!} \prod_{i=1}^m
    \left(\abar\, \int d\Phi_i^\text{real}\right)\right]
  \left[\frac{1}{n!} \prod_{i=1}^n
    \left(-\abar\, \int d\Phi_i^\text{virt}\right)\right]
  \Theta(L - \ln V(\mathcal{I}(p_1^\text{real},\ldots, p_m^\text{real})))\,,
\end{equation}
where $d\Phi_i$ is given by
\begin{equation}
  \label{eq:toy-expansion-master-phasespace}
  d\Phi_i \equiv
  d\hat \eta_i
  \frac{1}{2\pi}
  \frac{d^2 \hat \boldp_{\perp,i}}{{\hat p}_{\perp,i}^2}
  \Theta(-\ln \hat p_{\perp,i} > |\hat \eta_i| )\,.
\end{equation}
The insertion operator
$\mathcal{I}(p_1^\text{real},\ldots, p_m^\text{real})$ in
Eq.~(\ref{eq:toy-expansion-master}) inserts the emissions in order of
decreasing $\ln v = \ln \hat p_{\perp,i} - \beta |\hat \eta_i|$ with
the appropriate recoil prescription for the given shower, e.g.\
Eq.~(\ref{eq:generic-recoil}) for dipole showers.

A direct evaluation of Eq.~(\ref{eq:toy-expansion-master}) leads to
terms with up to $2n$ logarithms for the coefficient of $\abar^n$,
from the exponentiation of $\abar L^2$ structures.
For observables that exponentiate, and at fixed coupling, these terms
disappear in $\ln \Sigma(L)$, leaving at most terms $\abar^n L^m$ with
$m \le n$.
In a numerical (Monte Carlo) calculation of $\ln \Sigma$, one could
evaluate individual terms at different orders in the $\abar$ expansion
of $\Sigma(L)$ and then combine them to obtain the expansion of
$\ln \Sigma(L)$.
However this would lead to large cancellations between
$\abar^n L^{2n}$ terms coming from Monte Carlo calculations at
distinct orders, with uncorrelated statistical errors.

Instead, we take the approach of directly evaluating the expansion of
\begin{equation}
  \label{eq:F}
  F \equiv \exp(- \abar \Sigma_1(L)) \Sigma(L)\,,
\end{equation}
where $\Sigma_1(L)$ is the coefficient of $\abar$ in the $\abar$
series expansion of $\Sigma(L)$.
A necessary
(but not sufficient) condition for an NLL-correct shower, in the
fixed-coupling approximation of our toy model, is that $F$
should only have terms $\abar^n L^m$ with
$m \le n$, like $\ln \Sigma(L)$.
Note that with running coupling there would be terms $\abar^n
L^{n+1}$, and it would make more sense to use an analogue of $F$ in
which the exponential pre-factor was adjusted for running coupling
effects.
The non-trivial result starts at second order, and writing
$F = \sum_n \abar^n F_n$, the first few terms are
\begin{subequations}
  \begin{align}
    \label{eq:toy-expansion-master-2nd}
    F_2(L) \equiv [\exp(- \abar \Sigma_1(L)) \Sigma(L)]_2
    &= \frac1{2!} \int d\Phi_1 d\Phi_2
      \left(\Theta_{12} - \Theta_1 \Theta_2\right),
    \\
    F_3(L) \equiv [\exp(- \abar \Sigma_1(L)) \Sigma(L)]_3
    &= \frac1{3!} \int d\Phi_1 d\Phi_2 d\Phi_3 \left(\Theta_{123}
      -  \Theta_{12} \Theta_3 
      -  \Theta_{23} \Theta_1 
      -  \Theta_{13} \Theta_2
      + 2 \Theta_1 \Theta_2 \Theta_3
      \right),
    \\
    F_4(L) \equiv [\exp(- \abar \Sigma_1(L)) \Sigma(L)]_4
    &= \frac1{4!} \int d\Phi_1 d\Phi_2 d\Phi_3  d\Phi_4
      (\Theta_{1234}
      -  \Theta_{123} \Theta_4
      -  \Theta_{124} \Theta_3
      -  \Theta_{134} \Theta_2
      -  \Theta_{234} \Theta_1
    \nonumber \\[-6pt] &\qquad\qquad\qquad\qquad\qquad\quad
      +  \Theta_{12} \Theta_3 \Theta_4
      +  \Theta_{13} \Theta_2 \Theta_4
      +  \Theta_{14} \Theta_2 \Theta_3
      +  \Theta_{23} \Theta_1 \Theta_4
    \nonumber \\ &\qquad\qquad\qquad\qquad\qquad\quad
      +  \Theta_{24} \Theta_1 \Theta_3
      +  \Theta_{34} \Theta_1 \Theta_2
         %%% now the 
      - 3 \Theta_1 \Theta_2 \Theta_3 \Theta_4
         ),
  \end{align}
\end{subequations}
where we have introduced the shorthand
\begin{equation}
  \label{eq:Theta-shorthand}
  \Theta_{i\ldots n} \equiv \Theta[L - \ln V(\cI(p_i,\ldots, p_n))]\,.
\end{equation}
In practice we evaluate the difference between a given shower and the
CAESAR result
$\delta F_n(L) \equiv F_n^\text{shower}(L) - F_n^\text{CAESAR}(L)$,
so as to remove known NLL terms.
In a NLL-correct shower, the $\delta F$ should at most have
contributions $\abar^n L^m$ with $m < n$.
It will be convenient to study $\delta F_n/L^n$, which should go to
zero for NLL-accurate showers for large negative $L$.
If it tends to a non-zero constant, that will signal NLL failure.

\begin{figure}
  \centering
  \includegraphics[width=0.32\textwidth,page=1]{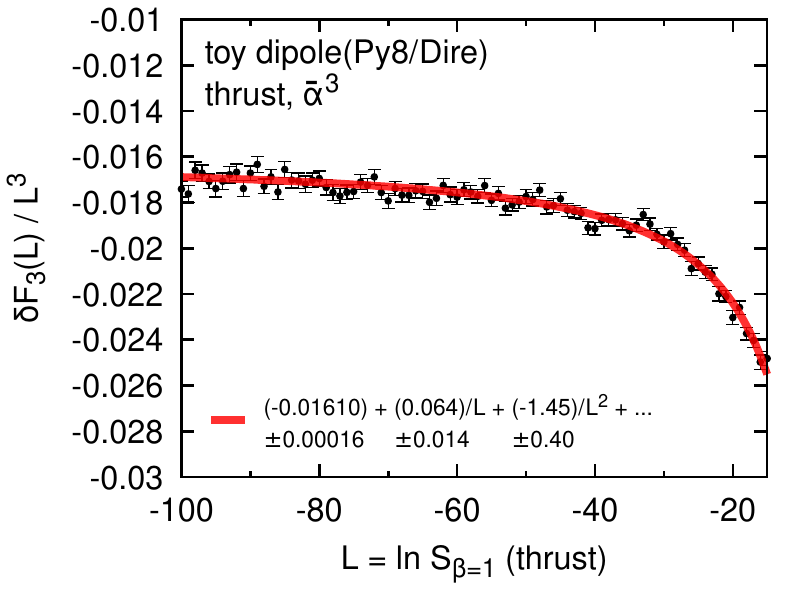}
  \includegraphics[width=0.32\textwidth,page=2]{plots/toy-FO-paper.pdf}
  \includegraphics[width=0.32\textwidth,page=3]{plots/toy-FO-paper.pdf}
  \\
  \includegraphics[width=0.32\textwidth,page=4]{plots/toy-FO-paper.pdf}
  \includegraphics[width=0.32\textwidth,page=5]{plots/toy-FO-paper.pdf}
  \includegraphics[width=0.32\textwidth,page=6]{plots/toy-FO-paper.pdf}
  \caption{Fixed order results from the toy implementation of the
    standard dipole showers.
    The plots show the difference between the toy dipole shower and
    the (NLL-correct) CAESAR results for the $F_n$ coefficient of
    $\abar^n$ in the expansion of Eq.~(\ref{eq:F}),
    divided by $L^n$.
    For an NLL-correct shower, the results should
    tend to zero for large negative $L$.
    The first row shows the result of $n=3$, the second row that of $n=4$. 
    The columns correspond to different observables (thrust, slice
    transverse momentum and hemisphere $\sqrt{y_{23}}$).
    Observe how the results tend to constants (NLL discrepancy) or
    demonstrate a linear or even quadratic dependence on $L$
    (super-leading logarithms).
    The coefficients have been fitted taking into account correlations
    between points, and we include powers down to $L^{-3}$ in the fit
    of $\delta F_n/L^n$.
    The fit range is from $-100$ to $-5$ and the quoted error includes
    both the (statistical) fit uncertainty and the difference in
    coefficients obtained with the range $[-100, -10]$ (added in
    quadrature).
  }
  \label{fig:super-leading-logs}
\end{figure}

A subset of results for dipole showers is shown in
Fig.~\ref{fig:super-leading-logs}.
The left-hand column of plots is for the thrust observable and shows
$\delta F_3(L)/L^3$ (top) and $\delta F_4(L)/L^4$ (bottom)
as a function of $L$.
At order $\as^3$ one sees that the result tends to a
constant, signifying an $\as^3 L^3$ term and NLL failure (as reported in the revised
version of Ref.~\cite{Dasgupta:2018nvj}).
Rather surprising, however, is that $\delta F_4/L^4$ (lower-left plot
of Fig.~\ref{fig:super-leading-logs}) appears to have a linear
behaviour at large negative $L$, signalling a term $\as^4 L^5$.
Such terms are super-leading, in the sense they are larger than any
term that should be present in $F(L)$ (or in $\ln \Sigma$) for rIRC
safe~\cite{Banfi:2004yd} observables in our
fixed-coupling approximation.\footnote{The one context where such terms
  are believed to exist is in association with coherence-violating
  effects~\cite{Forshaw:2006fk,Catani:2011st}.
  However the corresponding cases always involve hadronic systems
  in the initial state as well as the final state, and the
  super-leading logarithms are sub-leading in colour.
  The terms observed here arise at leading colour in a purely
  final-state context.}
We have found such terms to be present for dipole showers for all
global observables with $\beta_\text{obs} > 0$ and there is strong reason to
believe that they are related to sub-leading terms $\as^2 L$ being
enhanced by powers of $\as L^2$, giving $\as^n L^{2n-3}$.
The toy shower cannot accurately predict the coefficients of all such
terms in the full shower, because they are affected also by secondary
radiation (while the toy model has only primary radiation).
However the conclusion that there are such terms is, we believe,
robust.

Let us now examine two non-global observables: the transverse momentum in
a slice ($S^\text{slice}_{\beta=0,\Delta=1}$) and a hemisphere $\sqrt{y_{23}}$ (angular-ordered Durham) jet
resolution parameter ($M_{R,\beta=0}$).
The toy-shower $F_n(L)/L^n$ results are shown in the middle and
right-hand columns, respectively, of
Fig.~\ref{fig:super-leading-logs}.
The slice looks similar to the thrust case, with $\as^3 L^3$ and
$\as^4 L^5$ terms.
The hemisphere $\sqrt{y_{23}}$ ($M_{R,\beta=0}$) observable has 
$\as^3 L^4$ and $\as^4 L^6$ terms.
The fact that this observable has one additional logarithm makes its
analytical calculation somewhat easier (cf.\
section~\ref{sec:super-leading-logs} below).
Note that the toy shower is not, in general, suitable for evaluating
single-logarithmic ($\as^n L^n$) terms for non-global observables.
However, once again, the existence of such terms in the toy model
signals their existence also in the full shower.

It is natural to ask why we do not see the impact of super-leading
logarithms in our all-order results.
This will be easier to discuss below, once we have explained their
origin in detail.

\begin{figure}
  \centering
  \includegraphics[width=0.32\textwidth,page=7]{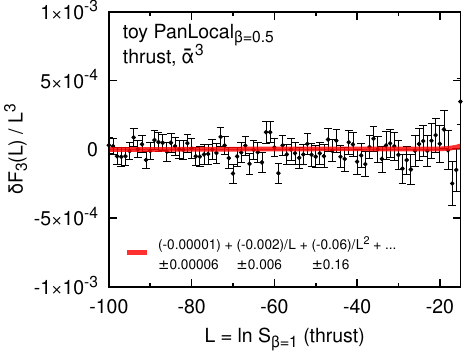}
  \includegraphics[width=0.32\textwidth,page=8]{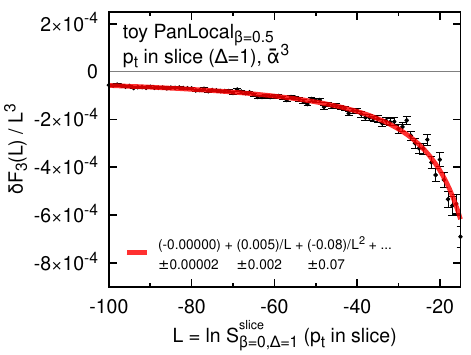}
  \includegraphics[width=0.32\textwidth,page=9]{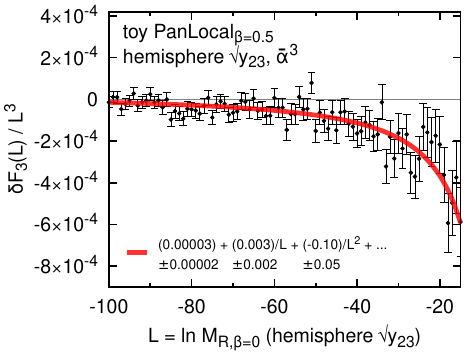}
  \\
  \includegraphics[width=0.32\textwidth,page=10]{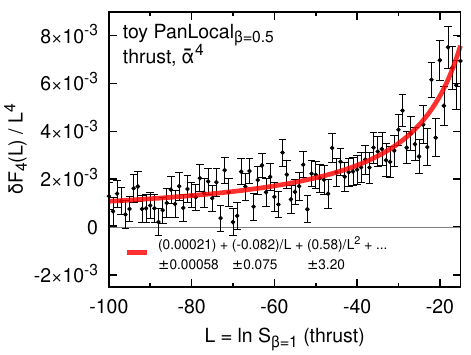}
  \includegraphics[width=0.32\textwidth,page=11]{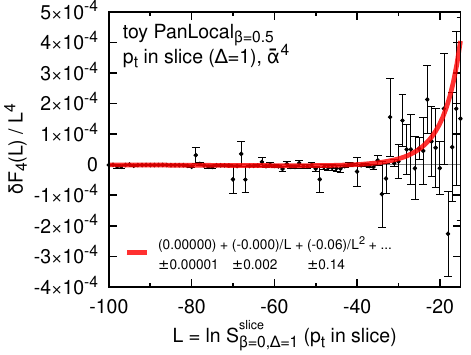}
  \includegraphics[width=0.32\textwidth,page=12]{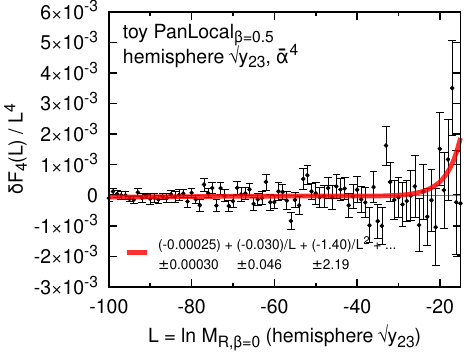}
  \caption{Analogue of Fig.~\ref{fig:super-leading-logs},
    demonstrating the absence of NLL (or super-leading) issues at
    fixed order in the toy version of the PanLocal $\beta=0.5$ shower.
    At order $\abar^4$, we include fit terms down to $L^{-4}$.
  }
  \label{fig:no-super-leading-logs-PSb05}
\end{figure}

We close this section by illustrating the kind of result that one
expects to obtain for a NLL-correct shower.
This is shown in Fig.~\ref{fig:no-super-leading-logs-PSb05} for the
PanLocal $\beta=1/2$ shower, again at third and fourth order.
In all cases, $\delta F_n/ L^n$ tends to zero for large negative $L$,
as required for NLL correctness.

%----------------------------------------------------------------------
\subsection{Super-leading logarithms}
\label{sec:super-leading-logs}

\subsubsection{Hemisphere max $p_\perp$}
\label{sec:hemi-max-pt}

The simplest observable with which to understand super-leading
logarithms is $M_{R,\beta=0}$ (or just $M_{R,0}$ for brevity), the
maximum $p_\perp$ of emissions in the right hemisphere, because the
super-leading terms are visible already from order $\as^3$.
In the toy-model approach involving only soft primary
emissions and fixed-coupling, the correct all-orders result for this
observable is $\Sigma = e^{-\abar L^2/2}$, amounting to
exponentiation of the leading-order result.
The full QCD NLL
result would additionally require inclusion of
hard-collinear single-logarithmic corrections, a resummation of
non-global logarithms, and running coupling effects.
Such contributions are not included in our toy shower and so do not
enter into our discussion below.

We are interested here in the fixed-order results for the difference,
$\delta F$, between dipole showers and the correct NLL
result (i.e.\ the CAESAR result from section~\ref{sec:toy-shower}).
Recall that $F$ was defined in Eq.~(\ref{eq:F}) and that for $n \le 3$
$\delta F_n$ is equivalent to $\delta (\ln \Sigma)_n$.
A NLL discrepancy between the dipole shower and the correct result
would reveal itself via single-logarithmic terms $\alpha_s^n L^n$ in
$\delta F$, while for a NLL correct shower $\delta F$ should contain
terms that are at most $\order{\alpha_s^n L^{n-1}}$.

We start by examining order $\as^2$, where $M_{R,0}$
already reveals a NLL discrepancy,
due to the recoil issue, i.e.\ one finds an $\as^2 L^2$ term in the
difference between the dipole shower and the correct NLL result.  To
calculate the coefficient of this $\as^2 L^2$ term we use the method
we introduced in Ref.~\cite{Dasgupta:2018nvj} and examine dipole
showers with evolution variable $v =p_{\perp}$. We consider two
emissions with values of the shower evolution variable $\hat p_{\perp,1}$
and $\hat p_{\perp,2}$ respectively, with
$\hat p_{\perp,1} > \hat p_{\perp,2}= \zeta \hat p_{\perp,1}$. The NLL order $\as^2$
discrepancy arises from a situation where the first emission has
$\hat \eta_1>0$, i.e.\ is emitted in the right hemisphere $\cal{H}_R$, and has
its transverse momentum modified by receiving recoil from the second emission
also emitted in the right hemisphere. (The contribution from $p_2$ in
the left-hand hemisphere vanishes after azimuthal integration.)
The second emission then must have rapidity 
$\hat \eta_2<\frac{1}{2}(\hat  \eta_1-\ln \hat p_{\perp 1})$ (cf.\ Eq.~(\ref{eq:midpoints})),
and the transverse momentum 
for the first emission is modified so that $\hat p_{\perp,1} \to p_{\perp,1} = \hat p_{\perp,1}
\sqrt{1+\zeta^2-2\zeta \cos\phi}$.
We then obtain, for the difference
between the dipole shower and the correct result,
\begin{multline}
\label{eq:deltasigma2}
\abar^2 \delta F_2 = {\bar{\alpha}}^2 \int_0^1\frac{d \hat p_{\perp,1}}{\hat p_{\perp,1}}  \int_0^{\ln 1/\hat p_{\perp,1}} d\hat \eta_1 \int_0^1\frac{d\zeta}{\zeta } \int_0^{\frac{1}{2} \left(\hat\eta_1+\ln 1/\hat p_{\perp,1} \right)} d\hat \eta_2  \int_0^{2\pi}\frac{d\phi}{2\pi} \times \\
\times \left [ \Theta \left (v-{\max}(p_{\perp,1}, \zeta \hat  p_{\perp,1} \right))- \Theta \left(v-\hat p_{\perp,1} \right) \right ],
\end{multline}
where  $\bar{\alpha}=2 C_F \as/\pi$ and $v$ is the maximum allowed
value of $M_{R,0}$.
There are also configurations where $\hat p_1$ is close to the
hemisphere boundary, and gets pulled in/out by an emission $\hat p_2$
in the unmeasured hemisphere.
This could contribute at $\as^2 L^2$, but after azimuthal averaging
yields zero.
Defining $f_{\text{max}} = \max \left(\zeta,\sqrt{1+\zeta^2-2\zeta \cos\phi}\right)$ and $L=\ln v$ and carrying out the required integrals gives the result 
\begin{equation}
\label{eq:as2mr0}
\abar^2 \delta F_2 =  - \frac{3}{4}  \bar{\alpha}^2 L^2 \int \frac{d\phi}{2\pi} \frac{d\zeta}{\zeta} \ln f_{\text{max}}  + \mathcal{O} \left(\bar{\alpha}^2 L\right) =  -0.0685389 \, \bar{\alpha}^2 L^2+ \mathcal{O} ({\bar{\alpha}}^2 L).
\end{equation}
\begin{figure*}
    \includegraphics[width=0.7 \textwidth]{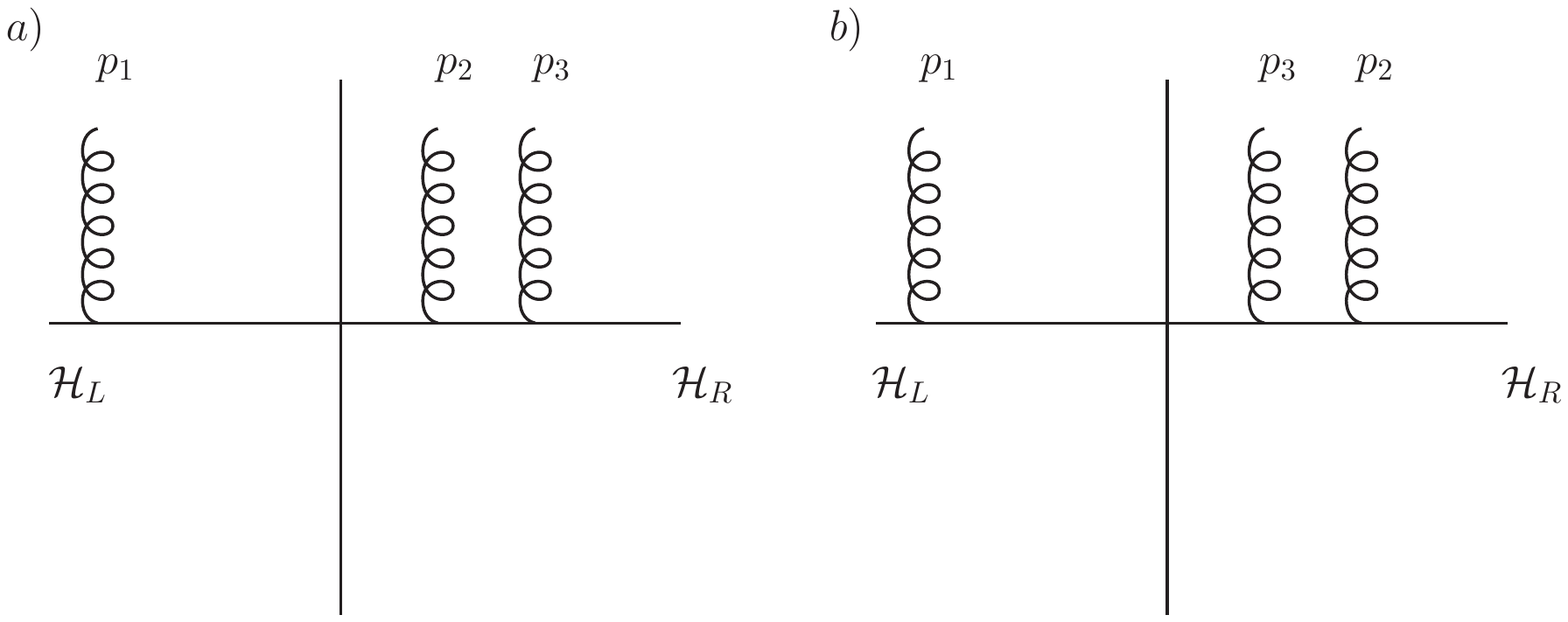}%
        \vspace*{-12cm}
        \caption{Figure showing emission $p_1$ in the left hemisphere
          and emissions $p_2$ and $p_3$ in the right hemisphere. On
          the left we show region a) where $\eta_3>\eta_2$ while on
          the right we show region b) where $\eta_2>\eta_3>0$. In
          region b) the condition for $p_3$ to give recoil to $p_2$
          depends on $p_{\perp,1}$ and $\eta_1$, which gives rise to a
          mismatch with the case where $p_1$ is virtual, resulting in
          a super-leading logarithm as explained in the text.}
      \label{fig:real}
\end{figure*}%
Now we turn to order $\as^3$.
To obtain $\delta \ln \Sigma_3$ we need to consider $p_{\perp}$
ordered parton configurations with three real emissions as well as
those with two real emissions and a virtual emission. At this order
we have
\begin{equation}
\label{eq:dlnsigma}
\abar^3 \delta F_3 = \abar^3(\delta \Sigma_3-\Sigma_1 \, \delta F_2)
= \abar^3 \delta \Sigma_3  - \frac{3}{8} \bar{\alpha}^3 L^4 \int \frac{d\zeta}{\zeta} \frac{d\phi}{2\pi} \ln f_{\mathrm{max}}+ \mathcal{O} \left(\bar{\alpha}^3 L^3 \right),
\end{equation}
where we note the presence of an $\bar{\alpha}^3 L^4$ term which
arises from the product of the double-logarithmic leading-order term,
$\abar \Sigma_1 = -\frac{1}{2} \bar{\alpha} L^2 +\mathcal{O}
\left(\bar{\alpha} L\right)$,
and the recoil discrepancy $\abar^2 \delta \Sigma_2$ in
Eq.~\eqref{eq:as2mr0}. For the discrepancy between dipole showers and
the NLL result to be at most single-logarithmic, we should see that
this $\bar{\alpha}^3 L^4$ term cancels against a similar term in
$\abar^3 \delta \Sigma_3$.

In order to obtain the $\bar{\alpha}^3 L^4$ term from
$\abar^3 \delta \Sigma_3$ we first consider the same two-parton configuration
that gives the $\bar{\alpha}^2 L^2$ term for $\abar^2 \delta \Sigma_2$,
alongside an additional virtual emission.
We label the virtual
emission $p_1$, the two real emissions $p_2$ and $p_3$ and consider
the ordering $\hat p_{\perp,1} \gg \hat p_{\perp,2} > \hat p_{\perp,3}$ so that $p_2$
absorbs recoil from the emission of $p_3$. Integrating over the full
rapidity range $\ln 1/\hat p_{\perp,1} > \hat \eta_1 > - \ln 1/\hat p_{\perp,1}$ and
over $\hat p_{\perp,1}$ in the range $1> \hat p_{\perp,1} >\hat p_{\perp,2}$,
produces a factor $-\ln^2 \hat p_{\perp,2}$, with the minus sign accounting
for unitarity. It is simple to repeat the calculation that yields
$\delta \Sigma_2$, using emissions $p_2$ and $p_3$ and multiplied by
this additional factor accounting for the virtual emission, to obtain
\begin{equation}
  \abar^3 \delta \Sigma^{p_1 \, \text{virtual}}_3 = -\abar L^2 \delta \Sigma_2
 = \frac{3}{4}  \bar{\alpha}^3 L^4 \int \frac{d\phi}{2\pi} \frac{d\zeta}{\zeta} \ln f_{\text{max}} + \mathcal{O} \left(\bar{\alpha}^3 L^3  \right).
\end{equation}
Next we consider the situation where emission $p_1$  is real.
To produce the $\bar{\alpha}^3 L^4$ term we require that emission
$p_1$, which has the largest transverse momentum and is emitted first
in the shower, should not affect the observable, which is the case
when $p_1 \in \mathcal{H}_L$ i.e.\ $\hat \eta_1<0$.
We also require that
$p_3$ gives recoil to $p_2$ rather than to $p_1$, since we are again
examining situations that lead to a difference between dipole showers
and the correct result.

There are then two distinct regions to be
considered. In region a) we have the situation shown on the left in
Fig.~\ref{fig:real}, where $\hat \eta_3> \hat \eta_2$. In this situation $p_3$
can only give recoil to $p_2$ and one is free to integrate over
$\hat \eta_1$ and $\hat p_{\perp,1}$ such that $p_1$ is in $\mathcal{H}_L$.  The
integral over $\hat \eta$ of all emissions and over $\hat p_{\perp,1}$ gives
\begin{equation}
\label{eq:volumea}
\mathcal{V}_{\text{a}} = \int_{\hat p_{\perp,2}}^1 \frac{d \hat p_{\perp,1}}{\hat p_{\perp,1}} \int_{-\ln 1/\hat p_{\perp,1}}^{0}  d\hat \eta_1 \int_{0}^{\ln 1/\hat p_{\perp,2}} d\hat \eta_2 \int_{\hat \eta_2}^{\frac{1}{2}\left(\hat \eta_2 + \ln 1/\hat p_{\perp,2}\right)} d\hat \eta_3 = \frac{1}{8} \ln^4 \hat p_{\perp,2}.
\end{equation}
On the other hand for region b) shown on the right of
Fig.~\ref{fig:real}, the condition that $p_2$ receives recoil from $p_3$
is affected by the presence of $p_1$.
In this region the
rapidity constraints are more involved and the integral over
$\hat \eta_1, \hat p_{\perp,1}$ cannot be performed independently as was the
case in region a) and also for the virtual correction.  The integral
over $\hat \eta$ of all emissions and over $\hat p_{\perp,1}$ now gives
\begin{equation}
\label{eq:volumeb}
\mathcal{V}_{\text{b}} = \int_{\hat p_{\perp,2}}^1 \frac{d \hat p_{\perp,1}}{\hat p_{\perp,1}} \int_{-\ln 1/\hat p_{\perp,1}}^{0}  d\hat \eta_1 \int_{0}^{\ln 1/\hat p_{\perp,2}} d\hat \eta_2 \int_{0}^{\hat \eta_2} d \hat \eta_3 \, \Theta \left( \hat \eta_3 -\frac{1}{2} \left(\hat \eta_1+\hat \eta_2+  \ln\frac{\hat p_{\perp,2}}{\hat p_{\perp,1}}\right)\right) = \frac{11}{48} \ln^4 \hat p_{\perp,2}.
\end{equation}
Combining the contributions from Eqs.~\eqref{eq:volumea} and \eqref{eq:volumeb}  we can perform the remaining integrals over $\hat p_{\perp,2}$, $\zeta$ and $\phi$ to obtain
\begin{equation}
\abar^3 \delta \Sigma_3^{\text{all real}} =\frac{17}{48} \int \frac{d\phi}{2\pi} \frac{d\zeta}{\zeta}\int_0^1 \frac{d \hat p_{\perp,2}}{\hat p_{\perp,2}} \ln^4{\hat p_{\perp,2}} \left[ \Theta \left(v-\hat p_{\perp,2} f_{\mathrm{max}} \right)-  \Theta  \left(v-\hat p_{\perp,2}\right)  \right] = - \frac{17}{48}  \bar{\alpha}^3 L^4 \int \frac{d\phi}{2\pi} \frac{d\zeta}{\zeta} \ln f_{\text{max}} + \mathcal{O} \left(\bar{\alpha}^3 L^3  \right).
\end{equation}
Adding together real and virtual contributions, neglecting any $\mathcal{O} \left(\bar{\alpha}^3 L^3 \right)$ terms, we then have $\delta  \Sigma_3 = \delta \Sigma_3^{\text{all real}}+\delta \Sigma^{p_1 \, \text{virtual}}_3$ and using Eq.~\eqref{eq:dlnsigma} we are left with a surviving super-leading logarithmic contribution:
\begin{equation}
\abar^3 \delta F_3 =  \frac{1}{48} \bar{\alpha}^3 L^4 \int \frac{d\phi}{2\pi} \frac{d\zeta}{\zeta} \ln f_{\text{max}} = 0.00190386 \, \bar{\alpha}^3 L^4 + \order{\abar^3L^3},
\end{equation}
in agreement with the fit in Fig.~\ref{fig:super-leading-logs}, to
within errors.
At next order in $\as$, one needs to consider up to two emissions in
the unobserved hemisphere, each of which brings a factor $\as L^2$,
leading to the $\as^4 L^6$ term seen in
Fig.~\ref{fig:super-leading-logs}. 

%......................................................................
\subsubsection{Considerations at all orders}
\label{sec:SLL-all-orders}

\begin{figure}
  \centering
  \includegraphics[width=0.48\textwidth,page=1]{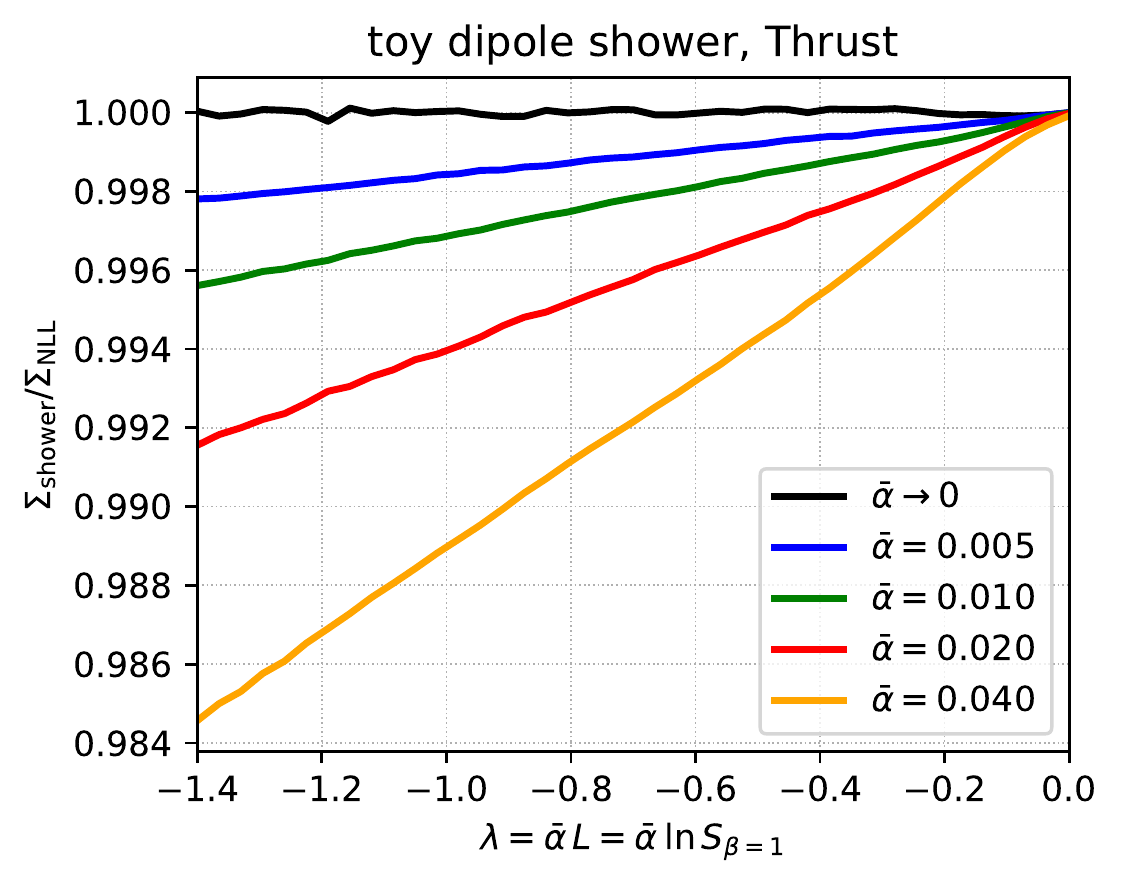}
  \includegraphics[width=0.48\textwidth,page=2]{plots/plot_sll.pdf}
  \caption{Toy-shower all-order result for the thrust ($S_{\beta=1}$,
    Eq.~(\ref{eq:FC-S-toy})).
    Left: $\Sigma_\text{dipole}/\Sigma_\text{NLL}$, where the NLL
    result is given by running the CAESAR version of the shower.
    Four values of $\abar$ are shown, together with the extrapolation
    to $\abar =0$, showing that the all-order dipole-shower result (in
    our usual limit of fixed $\abar L$ and $\abar\to0$) is
    consistent with the NLL result, despite the super-leading logarithmic terms
    that are visible in Fig.~\ref{fig:super-leading-logs}.
    Right: $(\Sigma_\text{dipole}/\Sigma_\text{NLL}-1)/\abar$, again
    for three values of $\abar$ and the extrapolation to $\abar=0$.
    The fact that these curves converge is a sign that the all-order
    (toy) dipole-shower discrepancy with respect to NLL behaves as a term
    that vanishes proportionally to $\abar$, i.e.\ as an NNLL term.
    The results here involve fixed coupling, i.e.\ they do not include
    a correction of the form of Eq.~(\ref{eq:7}).
  }
  \label{fig:toy-thrust-all-order}
\end{figure}
It is perhaps surprising that observables with super-leading
logarithms in the toy shower should resum to give apparent good
agreement with NLL in the full shower (cf.\ amber triangles in
Fig.~\ref{fig:limits-y3} right).
Fig.~\ref{fig:toy-thrust-all-order} shows the toy-shower all-order
result for one of these observables, the thrust, so that we can
compare fixed-order and all-order results within a single framework
(using the same underlying shower insertion code, as described in
section~\ref{sec:toy-shower}).
The left-hand plot shows the usual
$ \Sigma_\text{dipole} / \Sigma_\text{NLL}$ ratio as a function of
$\abar L$ for several $\abar$ values, together with its extrapolation
to $\abar=0$.
The extrapolation is consistent with unity, i.e.\ apparent all-order
agreement with the NLL result, as in the full shower.
Note that for $\abar L = -1.5$, the $\abar^3 L^3$  and
$\abar^4 L^5$ terms from Fig.~\ref{fig:super-leading-logs} would
respectively contribute $-0.054$ and $-0.87$ to the $\abar=0.01$
($L=-150$) line in Fig.~\ref{fig:toy-thrust-all-order} (left).
The all-order results undoubtedly have the statistical power to
resolve such effects, yet do not show any sign of them. 

The right-hand plot of the same figure shows
$(\Sigma_\text{dipole} / \Sigma_\text{NLL} -1)/\abar$ and its
extrapolation to $\abar=0$.
This serves as a verification that in this specific limit (i.e.\
$\abar L$ fixed and $\abar \to 0$, implying $\abar L^2 \to \infty$)
any all-order discrepancies with respect to the NLL result mimic a
standard NNLL, or even higher order, correction.

The presence of super-leading logarithms that evade detection at all
orders is a particularly unpleasant characteristic of dipole showers,
because it risks giving a false sense of security as to the validity
of the underlying logarithmic structure.
An analytic study of the all-order resummation of the
super-leading logarithms is beyond the scope of this manuscript.
However, a reader wishing to understand how an apparently large effect
at fixed order seemingly vanishes at all orders, could consider the
following argument.
For all the amber triangles in
Fig.~\ref{fig:limits-y3}, one contribution to the super-leading
logarithms comes from an $\abar^2 L$ (NNLL) contribution promoted
by additional factors of $\abar L^2$.
The $\abar^2 L$ term arises when a first emission $a$, contributing a
factor $\abar$, absorbs recoil from a second, unresolved, emission $b$
with commensurate $p_\perp$.
Integrating over the rapidity of the second emission yields a factor
$\abar L$, giving the overall $\abar^2 L$.
The $\abar L^2$ enhancement factor that arises at next order comes
about because there is a double logarithmic region for an emission $c$
with $\hat p_{\perp,c} \gg \hat p_{\perp,a}$ that alters
whether $b$ can induce recoil for $a$ (for example if $\hat \eta_c <
\hat \eta_b$, then one has a dipole chain $(a-c-b)$ and $a$ will not
receive recoil from $b$).
At all orders, the typical rapidity extent
($|\hat\eta_a - \hat\eta_b|$) in which one can have an $a-b$ dipole
without any other higher-$p_\perp$ particles in between can become of
order either $1/\sqrt{\abar}$ or $1$, depending on the context.
This causes the original $\abar^2 L$ factor to have $L$ replaced at
all orders by
$1/\sqrt{\abar}$ or $1$ respectively, giving $\abar^{3/2}$ or $\abar^2$,
i.e.\ even smaller than NNLL (which itself can arise from a multitude of
sources).\footnote{
  Note that in the hemisphere maximum $p_\perp$ case ($M_{R,\beta=0}$),
  studied at fixed order in section~\ref{sec:hemi-max-pt}, since the
  second-order result for $\delta F$ behaves as $\abar^2 L^2$, with part of
  each $L$ factor coming from an observed $p_{\perp}$ boundary, the
  result after all-order resummation does not vanish and instead mimics
  an NLL effect.
  }

%----------------------------------------------------------------------
\subsection{Lund-plane declustering for $e^+e^-$ collisions and
  $\Delta \psi$ resummation}
\label{sec:ee-lund}
In this section we introduce the definition of the azimuthal
separation between two Lund-plane declusterings~\cite{Dreyer:2018nbf}
in $e^+e^-\to \text{ jets}$ events.
This observable has been used in
the letter to test the azimuthal dependence of the effective
double-soft strongly angular-ordered squared amplitude
in different showers.
A proper definition of the azimuthal angle $\psi$ of a declustering
$p_i \to p_j + p_k$ requires the introduction of a dynamic reference
axis such that $\psi$ is always defined with respect to the direction
of $p_i$. We introduce the reference axes $\vec{n}_x = (1,0,0)$,
$\vec{n}_y = (0,1,0)$, and $\vec{n}_z = (0,0,1)$. We denote a generic
rotation matrix $R$, which rotates the $z$ axis to a direction
$\vec d$ that has an inclination $\theta$ and an azimuth $\phi$,
\begin{equation}
  \label{eq:1}
  R(\vec d\,)\equiv R(\theta,\phi) = 
  \left(
  \begin{array}{ccc}
    \cos(\phi) & -\sin(\phi) & 0\\
    \sin(\phi) &  \cos(\phi) & 0\\
    0 &   0 & 1
  \end{array}
  \right)
  \left(
  \begin{array}{ccc}
    \cos(\theta) & 0  & \sin(\theta)\\
    0 & 1 & 0\\
    -\sin(\theta) & 0 &  \cos(\theta)\\
  \end{array}
  \right)
  \left(
  \begin{array}{ccc}
    \cos(\phi) & +\sin(\phi) & 0\\
    -\sin(\phi) &  \cos(\phi) & 0\\
    0 &   0 & 1
  \end{array}
  \right)\,.
\end{equation}
We start with a configuration with two hard jets $j_1$ and $j_2$ in
the event, and we set the initial reference direction as
$\vec d_\text{ev} \propto \vec j_1 - \vec j_2$, with $j_1$ being the
jet with the larger (positive) value of $p_z$, and we normalise
$|\vec d_\text{ev}| = 1$. We define the initial rotation matrix
$\mathcal{R}_0 = R(\vec d_\text{ev})$, and we assign a sign $s=1$ to
the direction of the jet $j_1$ (with the larger $z$ component), and a
sign $s=-1$ to the direction of $j_2$.

Consider a declustering $p_i \to p_j + p_k$, primary or secondary,
which we label as being step $n$ in the declustering sequence
constructed with the Cambridge algorithm~\cite{Dokshitzer:1997in}
(four momenta are combined according to the standard $E$ scheme, i.e.\
a $4$-vector sum~\cite{Cacciari:2011ma}).
We introduce the reference directions of the declustering,
\begin{equation}
\vec u_i = \frac{\vec p_i}{ |\vec p_i|}\,,\qquad
\delta \vec u_{kj} = \vec u_{k} - \vec u_j.
\end{equation}
We then determine an updated rotation matrix as
\begin{equation}
  \label{eq:3}
  \mathcal{R}_{n} = R( \mathcal{R}_{n-1}^{-1}(s u_i)) \cdot \mathcal{R}_{n-1},
\end{equation}
where $\mathcal{R}_{n-1}$ is the rotation matrix relative to the
previous declustering stage. We finally evaluate $\psi$ for this
declustering as
\begin{equation}
  \label{eq:4}
  \psi = \text{atan2}(\delta \vec u_{kj} . \mathcal{R}_n . \vec n_y,\,
                      \delta \vec u_{kj} . \mathcal{R}_n . \vec n_x)\,.
\end{equation}
The Lund-plane declustering technology allows the inspection of the
structure of the primary and secondary radiation in the event. We
consider the two highest-$p_\perp$ primary Lund declusterings as an IRC
safe proxy for the two leading (i.e.\ highest-$p_\perp$) primary emissions,
which are the main element of NLL resummations for global
observables. The azimuthal separation $|\Delta \psi_{12}|$ used in
Fig.~\ref{fig:limits-dpsi} is simply defined as the absolute value of
the difference between the $\psi$ angles of the two leading
declusterings.

Specifically, we show the $|\Delta \psi_{12}|$ distribution for a given bin
of the leading declustering ($k_{t1}$), with an additional constraint
on the next highest $p_\perp$ primary declustering ($k_{t2}$) of the form
\begin{equation}
  k_{t,12}^{\rm min} < \frac{k_{t2}}{k_{t1}} < k_{t,12}^{\rm
    max};\quad k_{t,12}^{\rm min} =0.3\,,\quad k_{t,12}^{\rm max} = 0.5 \,.
\end{equation}
We consider the distribution normalised to the NLL vetoed cross
section $\Sigma(k_{t1})$ given in
refs.~\cite{Banfi:2001bz}.
The NLL prediction for this observable reads
\begin{equation}
\Sigma(\Delta \psi_{12}, k_{t2} | k_{t1}) \equiv
    \frac{\Sigma(k_{t1},k_{t2},|\Delta\psi_{12}|)}{\Sigma(k_{t1})} =
    \frac{1}{\pi} \left(e^{ R'(\lambda) \ln  k_{t,12}^{\rm max}} - e^{ R'(\lambda) \ln  k_{t,12}^{\rm min}}\right)\,,\qquad
R'(\lambda) \equiv \frac{4 C_F}{\pi} \frac{\lambda}{1-2 \beta_0 \lambda}\,,
\end{equation}
where $\lambda = \alpha_s \ln(Q/k_{t1})$, and
$\beta_0= (11 C_A - 2 n_F)/(12\pi)$.

%----------------------------------------------------------------------
\subsection{Subjet multiplicities}
\label{sec:multiplicities}

The scaling of particle multiplicities with centre-of-mass energy is
one of the few observables that has long been included among the
logarithmic structure tests
routinely carried out for parton showers~\cite{Marchesini:1983bm}.
Relative to the infrared collinear-safe global (shape-like)
observables and the non-global slice observable that we have
concentrated on for NLL accuracy tests here, it has the interesting
feature of being sensitive to the full nested soft-collinear
multiple branching structure.
Multiplicity itself is not infrared and collinear safe, however subjet
multiplicities for specific jet algorithms are.
We therefore concentrate on the latter.
They are known to NLL accuracy~\cite{Catani:1991pm} for the $e^+e^-$
Durham ($k_t$) jet algorithm~\cite{Catani:1991hj}.
Whereas for global shape-like observables NLL implies control of terms
$\as^n L^n$ in $\ln \Sigma$, NLL for multiplicities implies control of
terms $\as^n L^{2n-1}$ in the multiplicity itself.
Accordingly we take the limit $\as \to 0$ for fixed
$\as L^2$ (rather than fixed $\as L$).
Therefore, for an N$^p$LL result, we should control terms suppressed
by $\as^{p/2}$ relative to the LL result.
With $L = \frac12 \ln
y_\text{cut}$, we consider
\begin{equation}
  \label{eq:6}
  \frac{N^\text{subjet}_\text{shower}(\as, \as L^2 )/
    N^\text{subjet}_\text{NLL}(\as, \as L^2 ) - 1}{\sqrt{\as}}\,
\end{equation}%
\begin{figure}
  \centering
  \includegraphics[height=0.35\textwidth,page=1]{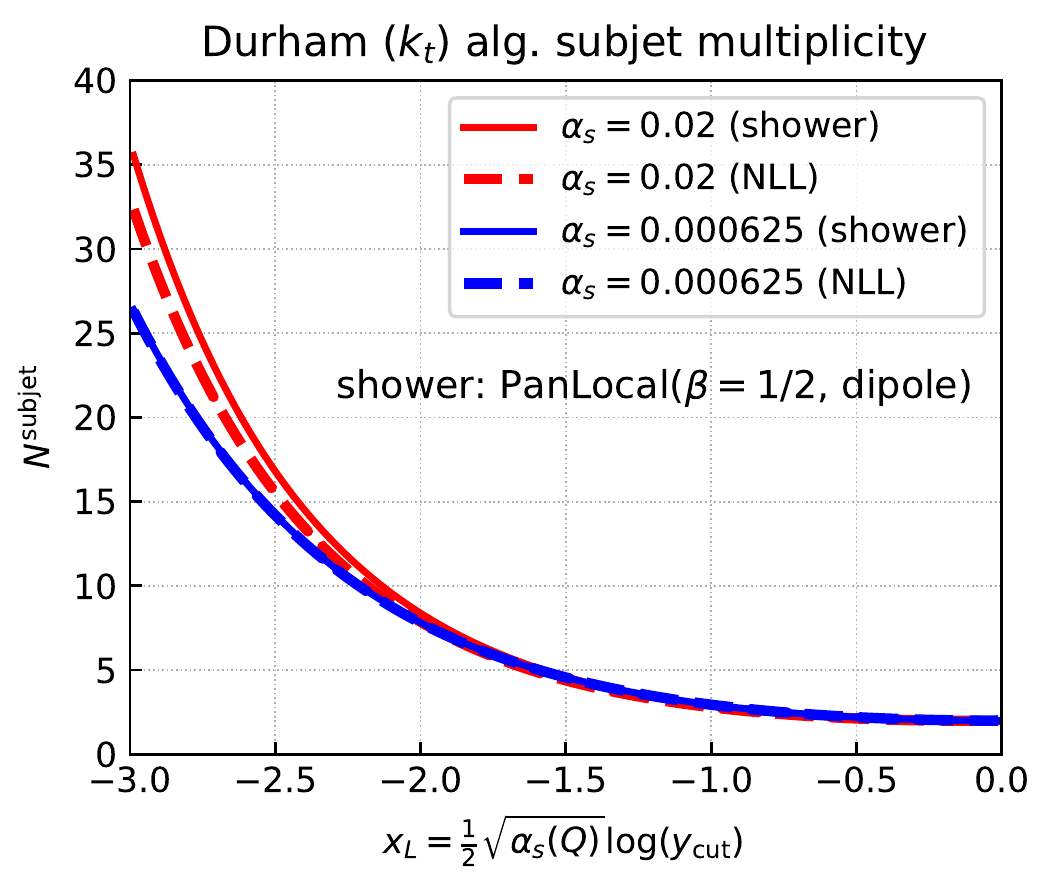}
  \includegraphics[height=0.35\textwidth,page=2]{plots/plot-multiplicity-paper.pdf}
  \caption{Checks of the $k_t$ algorithm subjet multiplicity.
    Left: the multiplicity as a function of
    $\frac12 \sqrt{\as(Q)}\ln y_{cut}$, comparing the PanLocal
    $\beta=0.5$ shower (dipole variant) with the NLL prediction, for two choices of
    $\as$.
    Right: Eq.~(\ref{eq:6}) for the same shower, for several $\as$
    values, together with the $\as \to 0$ limit.
  }
  \label{fig:multiplicity-checks}
\end{figure}%
and in particular take its $\as \to 0$ limit while keeping $\as L^2$ fixed.
Eq.~(\ref{eq:6}) should vanish as $\sqrt{\as}$ if the shower is NLL
accurate.
Fig.~\ref{fig:multiplicity-checks} (left) illustrates the
multiplicity as a function of $\frac12\sqrt{\as(Q)}\ln y_{cut}$ for
two values of $\as$, comparing the PanLocal $\beta=0.5$ shower (in its
dipole variant) to the
NLL result.
The right-hand plot shows Eq.~(\ref{eq:6}) for the same shower, for
three values of $\as$ as well as the extrapolation to $\as=0$.
It illustrates the good agreement across the range of
$\frac12\sqrt{\as(Q)}\ln y_{cut}$ values (with the usual exception of
the region close to $0$, which is not sufficiently asymptotic).
Other showers give similar results.
Note that the $\as \to 0$ extrapolation is more delicate for the
multiplicity than for other observables, because of the effective
expansion in powers of $\sqrt{\as}$ rather than $\as$.
This implies a need for a broader range of $\as$ values in order to
obtain reliable results.

%----------------------------------------------------------------------
\subsection{Considerations for $\as \to 0$ limits of showers}
\label{sec:taking-limits}

To reach a conclusion about NLL correctness of showers, it has been
crucial for us to be able to disentangle NLL terms from NNLL and yet
higher-order contributions.
This was achieved by considering the $\as \to 0$ limit of
\begin{equation}
  \label{eq:5}
  \frac{\Sigma_\text{shower}^\text{obs}(\as,\as L)}{\Sigma_\text{NLL}^\text{obs}(\as,\as L)}\,,
\end{equation}
for each given observable at fixed $\as L$.
The requirement of small $\as$ implies large $L$.
Other than for the subjet multiplicity studies discussed in the previous
section, the smallest $\as$ values used in producing
Fig.~\ref{fig:limits-y3} were either $0.005$ or $0.01$, depending on
the specific observable and shower.

Consider for now the smallest value, $\as=0.005$.
To achieve $ \as L = -0.5$, one needs $L = -100$.
In practice we typically add an event generation buffer of $B=18$
units of the logarithm below the value of interest for the observable.
The limit on the precision on the observable distribution is then
expected to be proportional to $e^{-B}$.
The logarithm of the span of scales in the event generation, roughly
$-118$, takes us into a regime that is far beyond that needed for
parton showers in normal phenomenological contexts and introduces a
number of challenges.
In what follows, we outline some of those challenges and the main
solutions we have adopted.

%......................................................................
\subsubsection{The challenges}
\label{sec:challenges}

\textbf{Particle multiplicities.}
%\label{sec:part-mult}
%
The \emph{logarithm} of the particle multiplicity $N$ in an event
scales as $\sqrt{\as L^2}$.
As we take $\as \to 0$ with $\as L$ fixed, $\as L^2 = (\as L)^2/\as$ grows large and
particle multiplicities become huge.
%
%% qd real run of
%%
%% ./example-framework -shower dire -out a -alphas 0.005 -g2gg -g2qq \
%%                     -no-obs -nloops 2 -lnvmin -118 -lnkt-cutoff -118 
%%
%% results in:
%% logbook/2020-02-06-brief-checks-for-paper/qdreal-mult+timing-as0.005.dat
%%
%% N = 200k, 30s/event 
With $\as=0.005$ and generation down to
$\ln k_t/Q = -118$, the average multiplicity, $\langle N \rangle$, is of
the order of $10^5$.
For computational strategies in event generation and subsequent
analysis that scale as $N$ this is at the edge of accessibility.
Where they scale as $N^2$ (or worse) it can well be beyond it.

%......................................................................
\textbf{Numerical precision.}
Covering a range of momenta that stretches from $1$ (in units of the
centre-of-mass energy $Q$) down to $e^{-118} \simeq 6\cdot10^{-52}$
poses challenges.
This is particularly the case for the calculation of the dot products
that arise, for example, in
Eqs.~\eqref{eq:evolution-variable},\eqref{eq:IIc-ak-bk-Sudakov-coeffs}.
Normal double-precision evaluations can fail already when
$k_t/Q \sim \sqrt{\epsilon}$, where $\epsilon \sim 10^{-16}$ is
machine precision.

%......................................................................
\textbf{Size of cross sections.}
The cross section $\Sigma(L)$ for (say) the $\sqrt{y_{23}}$ observable to
be below $e^{L}$ scales as
$ \Sigma(L) \simeq \exp[ - 2C_F \as L^2 /\pi ]$ at fixed coupling.
Including full LL
and NLL terms with running coupling, for $\as = 0.005$ and $\as L = -0.5$ one obtains
$\Sigma(L) \sim 10^{-29}$.
If one uses unweighted events, then a precision determination of
$\Sigma(L)$ requires that one needs to generate substantially more
than $1/\Sigma(L)$ events.
This is simply not feasible.

%......................................................................
\subsubsection{Solutions adopted}
\label{sec:solutions}

We have taken a hybrid approach to these challenges.
Firstly, insofar as possible, we have attempted to limit the
challenge.
For example we use techniques similar to those of
Ref.~\cite{Kleiss:2016esx} supplemented with a
Roulette Wheel~\cite{DBLP:journals/corr/abs-1109-3627} choice of 
dipole to
eliminate the $N^2$ scaling for all showers except the PanGlobal
one. (We believe $N^2$ scaling for the PanGlobal showers could perhaps
also be eliminated with suitable caching, but so far this wasn't a
critical necessity).

To address the precision issue, in the small-angle limit we use
cross-products rather than dot products, which generically remain
viable down to angles of $\epsilon$ rather than $\sqrt{\epsilon}$.
A further improvement is to align the initial $q\bar q$ event along
the $z$ axis.
Then there is almost no limit on the smallest angle that can be
calculated with respect to the original $q\bar q$ direction, and when
the shower produces an emitter at an angle $\theta$ with respect to
the $z$ axis, the smallest angle that can subsequently be calculated with respect
to that emitter is $\epsilon \theta$.
When these techniques reach their limit, we switch to the use of
\texttt{dd\_real} and \texttt{qd\_real} types~\cite{hida2000quad},
though thanks to other advances below, this was necessary mainly for
code testing and for the runs used to determine average
multiplicities.
One could also envisage techniques to handle arbitrarily small angles
fully in double precision.

A second element of our approach is to avoid generating emissions in
regions that are of no relevance to a given study.
We make an estimate for each emission of the most it could contribute
to that observable
($L_\text{approx} = \ln k_t - \beta_\text{obs} |\eta|$ with
$\eta = \bar \eta + \frac{1}{\beta}\ln\rho$ for the PanScales showers,
ignoring distinctions between primary and secondary emissions) and
then avoid generating emissions whose contributions would be below
$e^{-B}$ times the largest contribution so far in the shower.
This dynamic cutoff reduces multiplicities to a manageable level and
also keeps us in a regime where we can carry out calculations in double
precision rather than \texttt{dd\_real} and \texttt{qd\_real}. 

One potential concern about the dynamic cutoff is that while it is
safe from the point of view of the global observables (because they
are rIRC safe), it does cut out phase-space regions associated with
the generation of super-leading logarithms in dipole showers.
However the fact that our full-shower results (with the dynamic
cutoff) agree with the toy-shower results (which use the full primary
phase space, i.e.\ no dynamic cutoff) gives us some reassurance that
the dynamic cutoff is not generating misleading all-order conclusions
in Fig.~\ref{fig:limits-y3}.
Ultimately part of the reason that there is no issue here is that the
super-leading logarithms are anyway strongly reduced by all-order
resummation effects.
This parametrically suppresses their effect in the $\as\to 0$ limit
and restricts the relevant phase-space region to a band of width
$1/\sqrt{\as}$ below the observable boundary, a width that we
account for by the generation buffer $B$.

The final element that we bring to the problem is weighted event
generation. 
When $\beta_\text{PS} = \beta_\text{obs}$, this is fairly
straightforward (to avoid ambiguity in this paragraph, we add a
``PS'' subscript to the $\beta$ used in the parton shower ordering
variable).
We split event generation into bins of the first-emission $\ln v$.
For a given bin we include an event weight proportional to the Sudakov
above the upper edge of the $\ln v$ bin.
When $\beta_\text{PS} = \beta_\text{obs}$ there is a strong
correlation between the shower variable $\ln v$ for the first emission
and the value of the observable and so the weight distribution for a
given observable bin is relatively compact.
This approach works less well when $\beta_\text{PS} \neq
\beta_\text{obs}$ and in some challenging cases (notably
$\beta_\text{PS}=\frac12$, $ \beta_\text{obs} = 0$) further
refinements are needed.
There, we choose a bin for the highest $L_\text{approx}$ that we will
accept, start the shower with the least negative $\ln v$ that can be
in that bin (with an appropriate Sudakov weight) and then immediately
reject an event if at any stage in its showering one produces an
emission with an $L_\text{approx}$ that is larger than the upper edge
of the bin.
This rejection tends to happen in the early stages of the generation,
so there is a limited cost to generating the (many) events that
ultimately get rejected.
It is conceivable that there might be more sophisticated approaches,
but this one was sufficient in order to reach the few-per-mil
precisions needed on the $\as \to 0$ extrapolations at fixed $\as L$ for
Fig.~\ref{fig:limits-y3}, though it only just allowed us to
explore the region down to $\as L = -0.5$ for $\beta_\text{PS}=0.5$ showers.

Note that extrapolations to $\as=0$ are performed using a polynomial
of degree $n-1$ for $n$ values of $\alpha_s$.
%

%----------------------------------------------------------------------

\setcounter{figure}{0}
% For PRL only
%\setcounter{page}{1}

%%% Local Variables:
%%% mode: latex
%%% TeX-master: "panscales-NLL"
%%% End:

% LocalWords:  Eq Lund PanLocal PanGlobal NLL Jacobian Pythia FC Eqs
% LocalWords:  exponentiate unitarity

\end{document}